\documentclass{elsart}

\usepackage[square,comma,numbers]{natbib}
\usepackage{graphicx}
\usepackage{pxfonts}
\usepackage{lineno}

\usepackage{amssymb}
\usepackage{footnote}
\usepackage{multirow}
\usepackage{varwidth}
\usepackage{array}
\usepackage{epstopdf}
\usepackage{url}

\journal{}

\begin{document}
\thispagestyle{empty}
\begin{Large}
	\textbf{DEUTSCHES ELEKTRONEN-SYNCHROTRON}
	
	\textbf{\large{Ein Forschungszentrum der Helmholtz-Gemeinschaft}\\}
\end{Large}

DESY 18-113

August  2018

\begin{eqnarray}
	\nonumber &&\cr \nonumber && \cr \nonumber &&\cr
\end{eqnarray}
\begin{eqnarray}
	\nonumber
\end{eqnarray}
\begin{center}
	\begin{Large}
		\textbf{Relativity and Synchrotron Radiation: Critical Reexamination of Existing Theory}
	\end{Large}
	\begin{eqnarray}
		\nonumber &&\cr \nonumber && \cr
	\end{eqnarray}
	
	%
		
		Evgeny Saldin
	%
	\textsl{\\Deutsches Elektronen-Synchrotron DESY, Hamburg}
	\begin{eqnarray}
		\nonumber
	\end{eqnarray}
	\begin{eqnarray}
		\nonumber
	\end{eqnarray}
	ISSN 0418-9833
	\begin{eqnarray}
		\nonumber
	\end{eqnarray}
	\begin{large}
		\textbf{NOTKESTRASSE 85 - 22607 HAMBURG}
	\end{large}
\end{center}
\clearpage
\newpage

\begin{frontmatter}



\title{Relativity and Synchrotron Radiation: Critical Reexamination of Existing  Theory}


\author[DESY]{Evgeny Saldin}
\address[DESY]{Deutsches Elektronen-Synchrotron (DESY), Hamburg, Germany}

\begin{abstract}

Maxwell's equations are valid only in Lorentz frame i.e. in inertial frame where the Einstein synchronization procedure is used to assign values of the time coordinate. Einstein time order must be applied and kept in consistent way in both dynamics and electrodynamics.
However, the usual for accelerator engineering non-covariant treatment of  relativistic particle dynamics in a constant magnetic field looks precisely the same as in non-relativistic Newtonian dynamics. According to both  treatments, the magnetic field  is only capable of altering the direction of motion, but not the speed of an electron. However, the non-covariant trajectory does not include relativistic kinematics effects. The covariant  electron trajectory is viewed from the Lorentz lab frame as a result of successive infinitesimal Lorentz transformations. Like it happens with  the Galilean boosts composition, collinear Lorentz boosts also commute.  Therefore, for the rectilinear motion,  non-covariant and covariant approaches produce the same trajectory. But this result was incorrectly extended to arbitrary trajectory. In fact, Lorentz boosts in different directions do not commute and the composition of non-collinear boosts will result in a Wigner rotation  which has no non-covariant analogue. As one of the consequences of non-commutativity of non-collinear Lorentz boosts, we find an  unusual momentum-velocity relation, which also has no  non-covariant analogue. 
The theory of relativity shows us that unusual momentum-velocity relation and Wigner rotation have to do with the effects of acceleration in curved trajectories. 
We point out  that both these effects can be regarded as the two sides of the same coin: they are manifestations of the relativity of simultaneity that is expressed as a  mixture of positions and time.
One of the consequences of non-commutativity  of non-collinear Lorentz boosts is a difference between  covariant and non-covariant  single particle trajectories in a constant magnetic field. One can see that this essential point has never received attention in the physical community. As a result a correction of the conventional radiation theory is required.
In this paper we present a critical reexamination of existing  synchrotron radiation theory.
The main emphasis of this paper is on spontaneous synchrotron radiation from bending magnets and undulators.

\end{abstract}

%
%

%
\end{frontmatter}



\section{ Introduction }

The general approach to the determination of the motion of the particle is the following:
at any instant a particle has a well-defined velocity $\vec{v}$  as measured in a laboratory frame of reference. How is a velocity of a particle found? The velocity  is determined once the coordinates in the  lab frame are chosen, and is then  measured at appropriate time intervals along the particle's trajectory. But how to measure a time interval between events occurring at different points in space? In order to do so, and  hence measure  the velocity of a particle within a single inertial lab frame, one first has to synchronize distant clocks.  The concept of synchronization is a key concept in the understanding of special relativity. It is possible to think of various methods to synchronize the distant clocks.
To quote Moeller \cite{M}: "All methods for the regulation of clocks meet with the same fundamental difficulty. 
The concept of simultaneity between two events in different places obviously has no exact objective meaning at all, since we cannot specify any experimental method by which this simultaneity could be ascertained.
The same is therefore true also for concept of velocity." Following Einstein, the theory of relativity offers a procedure of clock synchronization based on the constancy of the speed of light in all inertial framers. Covariant particle tracking is based on the use this synchronization convention.

Relativistic effects start to be important when velocities of objects get closer to the speed of light. However, up to recently there were no man-made macroscopic objects possessing relativistic velocities: usually, in experiments, only microscopic particles can travel at velocities close to that of light. But with the operation X-ray free electron lasers (XFELs) this situation changes. An X-ray free electron laser is the best, exciting example of an engineering system where improvements in accelerator technology makes it possible to develop ultrarelativistic macroscopic objects with an internal fine structure, and the special theory of relativity plays an essential role in their description. An ultrarelativistic electron bunch modulated at nanometer-scale in XFELs has indeed a macroscopic finite-size of order of 10 $\mu$m. Its internal, collective structure is characterized in terms of a wave number vector. 

Let us suppose that a  modulated electron beam moves along the $z$-axis of a Cartesian $(x,y,z)$ system in the lab frame. As an example, suppose that the modulation wavefront is perpendicular to the velocity $v$. How to measure this orientation? A moving electron bunch changes its position with time. The natural way to do this is to answer the question: when does each electron cross the $x$-axis of the reference system? 
If we have adopted a method for timing distant events (i.e. a synchronization convention), we can also specify a method for measuring the orientation of the modulation wavefront: if electrons located  at the position with maximum  density cross the $x$-axis simultaneously at certain position $z$,  then the modulation wavefront is perpendicular to $z$-axis. In other words, the modulation wavefront is defined as a plane of simultaneous events (the events being the arrival of particles located at maximum density): in short, a plane of simultaneity.

It is generally believed that the modulation wavefront orientation has objective meaning.
If the modulation wavefront is tilted  of an angle $\theta$ with respect to $x$ axis, one usually just concludes that electrons move at constant speed $v_p = v\theta$ along the plane of simultaneity (i.e. along the modulation wavefront). 
When the trajectories of the particles calculated in the Lorentz reference frame (i.e. in inertial frame where Einstein synchronization procedure is used to assign values to the time coordinate) they must include such relativistic kinematics effect as relativity of simultaneity. 
It is very important to point out that the relativity of simultaneity is dictated by the finiteness of the velocity of light. 
In the ultrarelativistic asymptote, the orientation of the modulation wavefront , i.e the orientation of the plane of simultaneity is always perpendicular to the electron beam velocity when the evolution of the modulated electron beam is treated using Lorentz coordinates.

We should remark that Maxwell's equations are valid only in Lorentz reference frames.
Einstein's time order should obviously be applied and kept in consistent way in both dynamics and electrodynamics. It is important at this point to emphasize that the theory of relativity dictates that a modulated electron beam in ultrarelativistic asymptote has the same kinematics in Lorentz coordinates as a laser beam. According to Maxwell's equations, the wavefront of the laser beam is always orthogonal to the propagation direction. In other words, in ultrarelativistic limit we have for  modulated electron beam massless particle limit which is the same as for instance in the photon case.

What does this wavefront readjustment  mean in terms of measurements? In classical physics the simultaneity of a pair of events has an absolute character. The absolute character of the temporal coincidence of two events is a consequence of the as well absolute classical concept of time. 
However, according to the theory of relativity we establish a criterion for the simultaneity of events, which is based on the invariance  of the speed of light. It is immediately understood that, as a result of the motion of electrons along  the tilted wavefront (i.e. along the plane of simultaneity) with the velocity $v\theta$, the simultaneity of different events is no longer absolute, i.e. independent of the tilt angle $\theta$. This reasoning is in analogy with Einstein's train-embankment thought experiment.   
The time $t$ under the Einstein synchronization in the lab frame is readily obtained  by introducing the time shift $\Delta t = d_w(v\theta)/c^2$, where $d_w$ is the distance along the wavefront in the plane of rotation.
This time shift has the effect of rotation the plane of simultaneity (that is modulation wavefront) on the angle $\Delta \theta = v\Delta t/d_w = v(v\theta/c^2)$.
As a consequence of this, the modulation wavefront rotates in the lab frame. In ultrarelativistic limits, $\Delta \theta = \theta$, and wavefront is readjusted along the new direction of motion of the beam.

The angle of wavefront tilt has no exact objective meaning, because the constancy of the speed of light in all inertial frames takes place. The statement that the wavefront orientation has objective meaning to within a certain accuracy can be visualized by the picture of wavefront in the proper orientation with approximate angle extension (blurring) given by $\Delta \theta \simeq v(v\theta/c^2)$. This relation specifies the limits within which the classical (non relativistic ) theory can be applied. 
In fact, it follows that for a very classical electron beam for which $v^2/c^2$ is very small, the angle "blurring" becomes very small too. In this case angle of wavefront tilt $\theta$ is practically sharp $\Delta \theta/\theta \simeq v^2/c^2 \ll 1$. This is a limiting case of classical (i.e. non-relativistic) kinematics. Classical kinematics holds for non-relativistic particles; the angle "blurring" is a peculiarity of relativistic beam motion.  In ultrarelativistic asymptotics when $v \simeq c$, the  wavefront tilt has no exact objective meaning at all  since due to the finiteness of  the speed of light, we cannot specify any experimental method by which this tilt could be ascertained.

In existing literature a theoretical analysis of XFELs driven by an electron beam with wavefront tilt was presented in \cite{TKS,BH,ML, ML2},  based on the use the usual Maxwell's equations and standard simulation codes. We state that this approach is conceptually incorrect.  In the XFEL case  we deal  with an ultrarelativistic electron beam  and within the Lorentz lab frame (i.e. within the validity of the Maxwell's equations) the tilted modulation wavefront is at odds with the special theory of relativity.

\subsection{A non-covariant approach to relativistic particle dynamics}

We would like to make some further remark about wavefront tilt. When considering the conventional particle tracking universally used for particle accelerator physics, there are several cases 
where a wavefront tilt can occur in XFELs, mainly through introduction of a deliberate angular trajectory error (or "kick").  
As well-known result of conventional particle tracking states that after the electron beam is kicked by a weak dipole magnet there is a change in the trajectory of the electron beam, while the orientation of the modulated wavefront remains as before. In other words, the kick results in a difference between the directions of the electron motion and the normal to the modulation wavefront (i.e. in a wavefront tilt). In XFEL simulations it is generally accepted that coherent radiation from the undulator placed after the kicker is emitted ( in accordance with Maxwell's electrodynamics) along the normal to the modulation wavefront. Therefore, when the angular kick exceeds the divergence of the output radiation, emission in the electron beam motion is suppressed.  
An angular kick is often an essential part of many XFEL related diagnostic or experimental procedures. The standard gain length measurement procedure in XFELs employs such kicks. Other applications include "beam-splitting" schemes where different polarization components  are separated by means of an angular kick to the modulated electron beam \cite{L,NUHN}.

We know that, in the ultrarelativistic asymptote, the orientation of the modulation wavefront is always perpendicular to the electron beam velocity when the evolution of the modulated electron beam is treated using Lorentz coordinates.
So we must conclude that for the accelerated motion in a constant magnetic field the covariant trajectory of the particle $\vec{x}(t)_{cov}$ and result from conventional (non-covariant) particle tracking $\vec{x}(t)$ differ from each other.

It is generally accepted that in order to describe the dynamics of relativistic particles in the lab reference frame one only needs to take into account the relativistic dependence of the particles momenta on the velocity. In other words, the treatment of relativistic particle dynamics involves only a corrected Newton's second law. Note that this solution of the dynamics problem in the lab frame makes no reference  to Lorentz transformations. Conventional particle tracking treats the space-time continuum  in a non-relativistic format, as (3+1) manifold.
In other words, in this approach, introducing as only modification the relativistic mass, time differ from space. In fact, we have no mixture of positions and time. 

For the rectilinear acceleration  the non relativistic Newtonian trajectory $x(t)_{classic}$  and the result of conventional particle tracking $x(t)$ differ from each other. In Newtonian dynamics the particle keeps picking up speed. In relativistic dynamics, the particle keeps picking up, not speed, but momentum.   

Now  let us discuss the accelerated motion in a constant  magnetic field. 
According to both (classical and relativistic) approaches, the magnetic field  is only capable of altering the direction of motion, but not the speed (i.e. mass) of an electron. 
This usual for accelerator engineering study of relativistic particle motion in a constant magnetic field looks precisely the same as in nonrelativistic Newtonian dynamics and kinematics. 
The trajectory of the electron, which follows from the solution of the corrected Newton's second law, does not include  relativistic effects and  Galilean vectorial law of addition of velocities is actually used. What is surprising and what we should understand is the origin of the identity between classical $\vec{x}(t)_{classic}$ and  non covariant $\vec{x}(t)$ trajectories in a constant magnetic field.

A non-covariant (3+1) approach to relativistic particle dynamics has been used in particle tracking calculations for about seventy years. However, the type of clock synchronization which provides the time coordinate $t$ in the corrected Newton's equation has never been discussed in literature. It is clear that without answer to the question about the method of synchronization used, not only the concept of velocity, but also the dynamics law has no physical meaning. 
A  non-covariant (3+1) approach to relativistic particle dynamics is forcefully based on a definite synchronization assumption but this is actually hidden assumption.  
The dynamical evolution in the lab frame is based on the use of the lab frame time $t$ as an independent variable, independent in the sense that $t$ is not related to the spatial variables. Such  approach to relativistic particle dynamics   is actually based on the use of a not standard (not Einstein) clock synchronization assumption in the lab frame. 

The trajectory of the particle  $\vec{x}(t)$, which follows from the solution of the corrected Newton's second law by integrating from initial conditions does not include relativistic kinematics effects. In particular, think of the algorithm that one actually uses while updating the velocity from one moment in time to the next in conventional particle tracking: one just uses the Galilean law of addition of velocities , not Einstein's one, and this is direct result following from the hidden assumption on non-standard clock synchronization.

In contrast to this, in the case of Einstein's synchronization convention relativistic kinematics effects arise and the covariant trajectory $\vec{x}_{cov}(t)$ is viewed from the lab frame as a result of successive Lorentz transformations. Under the Einstein's synchronization convention the lab frame time $t$ in the equation of motion cannot be independent from the space variables. This is because Lorentz transformations lead to a mixture of positions and time.

We should underline that we claim there is a difference between $\vec{x}(t)$ and $\vec{x}_{cov}(t)$.
We state that it depends on the choice of a convention, namely the synchronization convention of clocks in the lab frame.  Whenever we have a theory containing an arbitrary convention, we should examine what parts of the theory depend on the choice of that convention and what parts do not. We may call the former convention-dependent, and the latter convention-invariant parts. Clearly, physically meaningful  results must be convention-invariant.

Consider, for example, the motion of two charged particles in a given magnetic field which is used to produce special particle trajectories. Suppose there are two apertures at point $A$ and at point $A'$. Particle trajectories may be found  according to conventional particle tracking by integrating from initial conditions. From this solution of the corrected Newton's equation of motion we conclude, for example, that simultaneously first particle gets through  the aperture  at $A$ and second particle gets through the aperture at $A'$. These two events at point $A$ and point $A'$  have exact objective meaning i.e. convention-invariant. In contrast to this, simultaneity of these two events is convention-dependent and has no exact objective meaning. In particular, in the case of Einstein synchronization convention relativity of simultaneity arises and according to covariant particle tracking generally there may be some time shift between these two events.

Consistently with the conventionality of simultaneity, also the value of the velocity of particle is a matter of convention and has no exact objective meaning. 
Even  for a single particle we are able to demonstrate the difference between conventional and covariant particle tracking results. In fact, we use Einstein's rule for adding velocities to track the particle motion in a covariant way. But in the conventional particle tracking the velocity summation is curried out differently. In accelerator physics the dynamical evolution in the lab frame is based on the usual Galileo (vectorial) rule which is in agreement with velocity summations of Newtonian mechanics.

\subsection{Error in standard coupling fields and particles}

There is a common mistake made in accelerator and plasma physics connected with the difference between $\vec{x}(t)$ and $\vec{x}_{cov}(t)$. Let us look at this difference from the point of view of electrodynamics of relativistically moving charges. To evaluate fields arising from external sources we need to know their velocity and positions as a function of the lab frame time $t$. Suppose one wants to calculate properties of synchrotron radiation. Given our previous discussion the question arises, whether one should solve the usual Maxwell's equations in the lab frame with current and charge density created by particle moving along non-covariant trajectories like $\vec{x}(t)$. We claim that the answer to this question is negative.
In our previous publications \cite{OURS1,OURS2,OURS3,OURS4,OURS5,OURS6,OURS7} we argued that
this algorithm for solving usual Maxwell's equations in the lab frame, which is considered in all standard treatments as relativistically correct, is at odds with the principle of relativity. 
This essential point has never received attention in the physical community. Only the solution of the dynamics equations in covariant form gives the correct coupling between the usual Maxwell's equations and particle trajectories in the lab frame. We conclude that previous theoretical and experimental results in accelerator and plasma physics should be reexamined 
in the light of the pointed difference between conventional and covariant particle tracking.
In particular, a correction of the conventional synchrotron radiation theory is required. One can see that the difference between conventional particle trajectory and covariant particle trajectory seems to have been entirely overlooked using the usual Maxwell's equations and $\vec{x}(t)$, instead of $\vec{x}_{cov}(t)$, simply because this difference has never been considered before.

In this paper we present a critical reexamination of existing  synchrotron radiation theory.
The main emphasis of this paper is on  spontaneous synchrotron radiation from bending magnets and undulators.
But before the discussion of the main topic it would be well to illustrate  error in standard coupling fields and particles in accelerator and plasma physics by considering the relatively simple example, wherein the essential physical features are not obscured by unnecessary mathematical difficulties.
This illustrative example is mainly addressed to readers with limiting knowledge of accelerator and synchrotron radiation physics. Fortunately, the error in standard coupling fields and particles can be explained in a very simple way. 

\subsection{ An illustative example}

There is a realistic configuration encountered in practice, which involves the production of coherent undulator radiation. Perhaps the most interesting applications of the theory of relativity  concern X-ray free electron lasers (XFELs). Let us consider an ultrarelativistic electron beam, modulated by the FEL process in the main XFEL undulator, kicked by a weak dipole field before entering a downstream undulator radiator. We want to study the process of emission of coherent undulator radiation from such setup. This problem gives, in fact, a first idea of the influence of the difference between  $\vec{x}(t)$ and $\vec{x}_{cov}(t)$ on the radiation by relativistic charged particles.

It would be well to begin with a bird's-eye view of some of the main results.
According to non-covariant particle tracking,  after the beam is kicked there is a trajectory change, while the orientation of the modulation wavefront remains as before. In other words, the kick results in a difference between the direction of the electron motion and the normal to the wavefront. In standard Maxwell's electrodynamics, coherent radiation is emitted in the direction normal to the modulation wavefront. Therefore, according to the conventional coupling of fields and particles \footnote{This means: according to usual algorithm for solving Maxwell's equations in the lab frame with charge and current density created  by particles moving along the trajectories calculated by using non covariant particle tracking}, which we claimed incorrect, when the angular kick exceeds the divergence of the output coherent radiation, emission in the direction of the electron beam motion is strongly suppressed. We have shown that our coupling of fields and particles predicts an effect in complete contrast to the conventional treatment.
Namely, when the evolution of the electron beam modulation is treated according to covariant particle tracking, the orientation of the modulation wavefront in the  ultra-relativistic asymptotic is always perpendicular to the electron beam velocity. In other words, relativistic kinematics shows the surprising effect that after the kick the orientation of the modulation wavefront is readjusted along the new direction of the electron beam. As a result, using standard electrodynamics we predict strong emission of coherent undulator radiation  from the  modulated electron beam in the kicked direction. It should have been made clear that in our example even the direction of emission of coherent undulator radiation is beyond the predictive power of the conventional synchrotron radiation theory.

Let us now go back and consider quantitatively the problem  which concerns the kick of modulated ultrarelativistic electron beam. 
Let us suppose that the ultrarelativistic  modulated electron beam  is kicked by a weak dipole magnetic field  before entering downstream undulator and study the process of emission of coherent radiation with and without kick. Suppose that a modulated electron beam moves, initially, at the ultrarelativistic velocity $v$ parallel to the $z$-axis upstream the kicker, assuming for simplicity that the kick angle $\theta_k \simeq v_x/v$ is small compare with $1/\gamma$, where $\gamma = 1/\sqrt{1-v^2/c^2}$ is the relativistic factor. This means that we take the limit $\gamma \gg 1$, $\gamma v_x/v \ll 1$ and that the speed $v$ is close to the speed of light, $v \simeq c$. It is necessary to mention that  in XFEL engineering we deal indeed with an ultrarelativistic electron beam ( $c - v  \ll 10^{-8}c$) and  with a transverse velocity after the kick, which is much smaller than speed of light ($(v_x/c)^2  \ll 10^{-8}$), so that our studies of this simplistic model nevertheless yields a correct quantitative description in large variety of practical problems.

\subsubsection{Kicker setup. Treatment according to  non-covariant (3+1) approach}

Let us first discuss the results from usual particle tracking. 
We will solve the dynamics problem of motion of a relativistic electron in the force field of a weak dipole magnet by working only up to the order $\gamma\theta_k$. Even under this approximation we will be able to demonstrate the difference between conventional and covariant particle trajectories. Suppose that the modulation wavefront is perpendicular to the velocity upstream the kicker.
After the kick, the beam velocity components are $(v_x,0,v_z)$, where $v_z = \sqrt{v^2 - v_x^2}$. The velocity component along the $z$-axis remains unchanged in our first order approximation i.e. $v_z \simeq v$. Assuming further that the magnetic field in the setup does not depend on the transverse coordinates, which is typically justified for kicker setup in XFELs, after the beam is kicked the propagation axis of the electron beam is deflected , while the wavefront orientation is preserved.  

We note that the configuration under study in this section is of interest to the XFEL designers.
This discrepancy between directions of the electron motion and wavefront normal after the kick have been discussed previously (see, for example, Fig. 1 in \cite{TKS}). 
One particular consequence that received attention following the \cite {TKS} is the effect of the trajectory error (single kick error) on the XFEL amplification process.  It was pointed out that coherent radiation is emitted towards the wavefront normal of the beam modulation. Thus, according to conventional coupling of  fields and particles (which we claimed incorrect), the discrepancy between the two directions decreases the radiation efficiency \cite{TKS}.

Note that we started with the formulation of the initial conditions upstream of the kicker in terms of wavefront orientation and particle velocities. However, in order to measure  those, one first has to synchronize distant clocks within the lab frame upstream of the kicker. 
We already mentioned that the type of clock synchronization which results in time coordinate $t$ in corrected Newton's equation is never discussed in accelerator and plasma physics. The question now arises
how to assign synchronization in the lab frame upstream of the kicker. We need to give an "operational" answer to this question. Suppose that  clocks are synchronized by light signals. The synchronization procedure that follows  is the usual Einstein synchronization procedure. After this at least our initial conditions have experimental interpretation. 

The convention chosen  for clock synchronization is nothing more than a definite choice of coordinate system in an inertial frame of reference. Upstream of the kicker in the lab inertial frame we selected a special type of coordinate system, a Lorentz coordinate system to be precise.
Within a Lorentz frame (i.e. an inertial frame with Lorentz coordinates), Einstein's synchronization of distant clocks and Cartesian space coordinates are enforced.

\subsubsection{Kicker setup. Treatment under Einstein's time order}

Now let us see  what happens if we  keep Lorentz coordinates system in the lab frame downstream of the kicker. Using Einstein synchronization procedure in the lab frame downstream of the kicker we automatically assume that different Lorentz frames are related by  Lorentz transformations. 
Now let us try to get a better understanding of the relativistic kinematics,
which is, in fact, a comparative study between different coordinate frames. It requires two relativistic observers and two coordinate systems. Consider downstream of the kicker a Lorentz reference frame $S'$ moving with uniform motion at speed $v_x$  along the $x$-axis  of the Lorentz lab frame $S$. In the inertial frame $S'$, the wavefront normal  and electron motion have the same direction  along $z$ axis. A setup in the  inertial frame $S'$ downstream of the kicker reproduces the situation upstream of the kicker. 
Theory of relativity states  that in the Lorentz lab frame, after an electron beam is kicked, there is a change in the trajectory of the beam which is viewed from the lab frame as a result of Lorentz transformation. 
It is immediately understood that the simultaneity of events, and consequently orientation of wavefront,  is no longer absolute (i.e. independent of the kick), as a result of the invariance of the speed of light. 

Suppose, in fact, two electrons cross the $x$-axis simultaneously at certain position $z$ upstream of the kicker. Two events are simultaneous in a Lorentz reference frame if they are coincident with the arrival of light signals previously emitted from the position at equal distance from both events. Before the kick, light signals are emitted from a place equidistant from the positions along $x$-axis where the events happened. After the kick, instead, the place where the light signals is emitted is not equidistant to the positions where the events happened. Light signals do not arrive simultaneously at each electron in the Lorentz lab frame downstream of the kicker: the electrons have time to move from their positions equidistant from the source because the signal propagates with finite speed. This reasoning is analogy  with Einstein's train-embankment thought experiment. Finally the time $t$ under standard synchronization in the lab frame is readily obtained by introducing the offset factor $xv_x/c^2$ and substituting $t' = t - xv_x/c^2$. This expression forms the Lorentz transformation for time in the first order approximation. This time shift has the effect of  rotation the plane of simultaneity (that is modulation wavefront) on the angle $v_x/c$ in the first order approximation. As a consequence of this, the modulation wavefront rotates in the lab frame. 
This rotation is simply a consequence of the relativity of simultaneity between the two Lorentz frames $S$ and $S'$. In ultrarelativistic limits, $v \simeq c$, and the wavefront rotates exactly as the velocity vector $\vec{v}$, i.e. wavefront is readjusted along the new direction of motion of the kicked beam.

Now we need to give an "operational" answer to the question how to assign Lorentz coordinates to the inertial lab frame in the case when the electron beam is accelerated by the kicker. Upstream of the kicker one picks a Lorentz coordinate system. Then, after the kick, the beam velocity changes of an small value $v_x$ along the $x$-axis. In order to keep a Lorentz coordinate system in the lab frame downstream of the kicker, one needs to perform a clock resynchronization by introducing an infinitesimal time shift $t' = t - xv_x/c^2$. This form of the Lorentz transformation  is justified by the fact that we are dealing  with first order approximation. Therefore, $v_x/c$ is so small that $v_x^2/c^2$ can be neglected and one arrives at $x' =x - v_xt$, $t' = t - xv_x/c^2$. This infinitesimal Lorentz transformation just described differs from Galilean transformation  only by the inclusion of the relativity of simultaneity, which is only relativistic effect that appearing in the first order in $v_x/c$.

\subsubsection{Hidden synchronization assumption in the non-covariant (3+1) approach}

Let us now return to the conventional particle tracking.  In this (3+1) approach we have no mixture of positions and time. In conventional particle tracking, the simultaneity along the $x$ direction has an absolut character, meaning that it is independent of the kick. When a kick is introduced, electrons move at constant speed $v_x$ along the plane of simultaneity (i.e. along the wavefront), while the orientation of the plane of simultaneity stays unvaried. The trajectories of the particles, which follows from the solution of the corrected Newton's second law by integrating from initial conditions, does not include such relativistic effects as relativity of simultaneity. Therefore, 
conventional particle tracking is based on the use of a non-standard and unusual synchronization convention within the theory of relativity.

Now we are ready to investigate how the synchronization assumption  is hidden in the non-covariant (3+1) approach to relativistic dynamics. Let us return to kinematics and try to get an understanding of the relationship between two inertial frames $S$ and $S'$ downstream of the kicker in the case of conventional particle tracking.

Consider downstream of the kicker a Lorentz reference frame $S'$ moving with uniform motion at speed $v_x$  along the $x$-axis  of the lab frame $S$.  A setup in the  Lorentz frame $S'$ downstream of the kicker reproduces the situation upstream of the kicker i.e. the wavefront normal  and electron velocity with the same direction, along the $z$ axis. According to conventional particle tracking, a kick along the $x$ direction is equivalent to a coordinate transformation as $x' = x - v_xt$. This transformation is completed with the invariance of the simultaneity; in other words, if two electrons arrive simultaneously at the certain position $z$ upstream of the kicker, then after the transformation downstream of the kicker the same two electrons reach position $z' = z$ one more simultaneously i.e. $\Delta t' = \Delta t$. The absolute character of temporal simultaneity between two events is a consequence of the identity $t' = t$. As a result,  the hidden synchronization convention has the form of absolute time convention. In this situation the lab observer actually sees electron trajectories after the kick as a result of Galilean boost rather than a Lorentz boost.

The question now arises how to operationally interpret this absolute, global time convention i.e. how one should change the rule-clock structure of the lab reference frame after the kick. This actually correspomds to the simplest method of synchronization, which consists in keeping without changing the same set of synchronized  clocks used for experimental interpretation of the initial conditions in conventional particle tracking. Such trivial synchronization convention preserves simultaneity  and is actually based on the absolute time convention. This choice is usually the most convenient one from the viewpoint of connection to laboratory reality. When time coordinate is assigned in the lab frame, non-covariant particle trajectories can be experimentally interpreted by a laboratory observer. Due to the particular choice of synchronization convention, relativistic kinematics effects such as relativity of simultaneity do not exist in the lab frame. 
As matter of fact this hidden synchronization convention is used, in practice, in accelerator and plasma physics. Particle tracking calculations usually become much simpler if the particle beam evolution is treated in terms of absolute time (or simultaneity). This time synchronization convention is self-evident and this is the reason why this subject is not discussed in relativistic engineering.

In non covariant particle tracking, time differ from space and the particle trajectory in a constant magnetic field can be seen from the lab frame as a result of successive Galilean boosts that track the acceleration motion.

The use of Galilean transformations within the theory of relativity requires some special discussion. Many physicists still tend to think of Galilean transformations as old, incorrect transformations between spatial coordinates and time. It is simply not true in physics.
The special theory of relativity is the theory of four-dimensional space-time with pseudo-Euclidean  geometry. From this viewpoint, the principle of relativity is a simple consequence of the space-time geometry, and the space-time continuum can be described in arbitrary coordinates. In the process of transition to arbitrary coordinates, the geometry of the four-dimensional space-time does not change.  Therefore, contrary to the view presented in many textbooks,  Galilean transformations are actually compatible with the principle of relativity although, of course, they alter the form of Maxwell's equations.

This illustrative example is mainly addressed to reader with limiting knowledge of the theory of relativity and here we do not want to go through the detail of this subject, which is conceptually subtle. Because of our using Galilean transformations within the theory of relativity, we  have some apparent paradoxes, which we will gradually reduce one by one  in the following sections and will demonstrate that there is in fact no  difficulty with the (3+1) non covariant approach in relativistic dynamics and electrodynamics. It is perfectly satisfactory. It does not matter which convention and hence transformation  is used to describe the same reality. What matter is that, once fixed, such convention should be applied and kept in a consistent way in both dynamics and electrodynamics.

\subsubsection{Discussion}

This is a good point to make a general remark about Lorentz coordinates.
Obviously, it is convenient to describe the dynamics in the lab frame based on the use of the absolute time convention. In fact, in this case things looks precisely the same as in Newtonian  kinematics. 
In the case of Einstein synchronization convention, in contrast to the absolute time convention, we have a mixture of positions and time. As a consequence of this, kinematics effects are not what intuitively expected. Nevertheless, there is a reason to prefer Lorentz coordinates within the framework of electrodynamics. We are better off using covariant trajectories when we want to solve the electrodynamics problem based on Maxwell's equations in their usual form. One might choose to use  non-covariant trajectories, but the price to pay would be a change in the form of Maxwell's equations. 
In fact, the use of non-covariant trajectories also implies the use of much more complicated electromagnetic field equations. 

To solve the electrodynamics problem with minimal efforts we need to pick Lorentz coordinates.
As just discussed, the problem of assigning Lorentz coordinates to the lab frame in the case of an acceleration motion is complicated even in our very idealized situation. We already found that, in order to keep  a Lorentz coordinate system in the lab frame one needs to perform a clock resynchronization  by introducing a time shift after the kick. It should be clear that Lorentz coordinate systems are only mental construct, but manipulations with non existing clocks are an indispensable prerequisite for the application of the usual Maxwell's equations for moving light sources. 

It is interesting to note that we can interpret manipulations with rule-clock structure in the lab frame  simply as a change of the time variable according to the transformation $t \to t + xv_x/c^2$. The overall combination of Galileo transformation and time variable changes actually yields the infinitesimal (in our case of interest) Lorentz transformation in the (3+1) space and time, but in this context this transformation
are only to be understood as useful mathematical device, which allow one to solve the electrodynamics problem  in the (3+1) space and time with minimal effort. 
We state that this variable change has no intrinsic meaning. One can see the connection between the time shift and the issue of clock synchrony. The convention-independent results of calculations are precisely the same in the new variables. As a consequence, we should not care to transform the results of the electrodynamics problem solution into the original (3+1) variables.

An idea of studying dynamics and electrodynamics in (3+1) space and time using technique involving a change of variables is useful from a pedagogical point of view. It is worth remarking that the absent of a dynamical explanation for wavefront rotation has disturbed some physicists. It should be clear from the preceding discussion that a good way to think of the wavefront rotation is to regard it as a result of transformation to a new time variable.

\subsubsection{Wigner rotation}

Above we demonstrated that if the velocity of our modulated electron beam is close to velocity of light, Lorentz transformations work out in such a way that the rotation angle of the modulation wavefront coincides with the angle of rotation of the velocity. 
As known, a composition of noncollinear Lorentz boosts does not results in a different boost but in a Lorentz transformation involving a boost and  a spatial rotation, the Wigner rotation  \cite{WI, WI1,WI2}. The rotation of the modulation wavefront after the beam kicking is one concrete example of Wigner rotation.

Suppose the beam velocity is perpendicular to the wavefront of the modulation upstream of the kicker. As seen from the lab frame, the wavefront of the beam modulation rotates relative to  the Cartesian axes of the lab Lorentz frame when a modulated electron beam is accelerated in the kicker's field. 
Our calculations are performed in ultrarelativistic limit. In the case  of an arbitrary electron beam velocity, expression for the Wigner rotation is given by \cite{Rit}

\begin{eqnarray}
\vec{\delta \Phi} = 	\left(1 - \frac{1}{\gamma} \right)\frac{\vec{v}\times d\vec{v}}{v^2}	
=	\left(1 - \frac{1}{\gamma} \right) \vec{\delta \theta} ~.
\label{RT}
\end{eqnarray}

where $d\vec{v}$ is the vector of  small  velocity change due to acceleration, $\Phi$  is the Wigner rotation angle of the  wavefront, and $\theta$ is the orbital angle of the particle in the lab frame. From Eq. (\ref{RT}) follows that in the ultra relativistic limit $\gamma \longrightarrow \infty$, the wavefront rotates exactly as the velocity vector $\vec{v}$. Above we demonstrated  that in ultrarelativistic asymptotic the Wigner rotation results directly from the relativity of simultaneity \footnote{Above we worked out a simple case. The result involved the assumption that the orbital angle of the particle $\delta\theta$ is smaller than $1/\gamma$. this corresponds to a lag smaller than $1/\gamma^2$, which  is zero with respect to  (ultrarelativistic) approximation accuracy}. 

Thomas precession is a particular case of Wigner rotation corresponding to an infinitely small change in the velocity vector. Eq.(\ref{RT}) written in terms of the angular velocity $\Omega_\mathrm{T} = d\Phi/dt$, where $t$ is the time in the lab Lorentz frame, is represented as $\Omega_\mathrm{T} = (1-1/\gamma)\omega_0$, where $\omega_0 = d\theta/dt$ is the angular velocity of orbiting measured in the lab frame. In deriving expressions for the Thomas precession, the majority of authors were supposedly guided by the incorrect expression for Thomas precession from  Moeller's monograph \cite{M}. The expression obtained by Moeller is given by  $\vec{\delta \Phi} =  (1 - \gamma) \vec{v}\times d\vec{v}/v^2 =	
(1 - \gamma) \vec{\delta \theta}$ (and subsequently $\Omega_\mathrm{T} =  (1 - \gamma)\omega_0 $).
It should be note that, in his monograph, Moeller stated several times  that this expression valid in the lab Lorentz frame. Clearly, this expression and  Eq. (\ref{RT}) differ both in sign and in magnitude. 

It is important at this point to emphasize that    
the theory of relativity dictates that a modulated electron beam in the ultrarelativistic asymptote has the same kinematics in Lorentz coordinates, as a laser beam. In other words, in the limit $\gamma \longrightarrow \infty$ we have a limit where our modulated electron beam approaches a beam of massless particles. In contrast, 
according to  Moeller's expression for Wigner rotation in the lab frame  the modulation wavefront rotates in opposite direction  and $\Phi = (1- \gamma\theta) \to -\infty$ in the limit $\gamma \longrightarrow \infty$.

An analysis of the reason why Moeller obtained an incorrect expression for the Wigner rotation in the lab frame is the focus of Ritus paper \cite{Rit}. As shown in \cite{Rit}, the Moeller's mistake is not computational, but conceptual in nature.  In review \cite{MA} it is shown that the correct result was obtained in the works of several authors, which were published more than half century ago but remained unnoticed against the background of numerous incorrect works.

\subsubsection{Undulator radiation setup}

The most elementary of the effect that represents a crucial test of the correct coupling fields and particles is a problem involves the production of coherent undulator radiation by modulated ultrarelativistic electron beam kicked by a weak dipole field before entering a downstream undulator. We want to study the process of emission of coherent undulator radiation from such setup.

The key element of a XFEL source is the udulator, which forces the electrons to move along curved periodical trajectories. There are two popular undulator configurations: helical and planar. To understand the basic principles of undulator source operation, let us consider the helical undulator. The magnetic field on the axis of the helical undulator is given by $ \vec{H}_w = \vec{e}_xH_w\cos(k_wz) -\vec{e}_y H_w\sin(k_wz)$, where $k_w = 2\pi/\lambda_w$ is the undulator wavenumber and $\vec{e}_{x,y}$ are unit vectors directed along the $x$ and $y$ axes. We neglected the transverse variation of the magnetic field. It is necessary to mention that in XFEL engineering  we deal with a very high quality of the undulator systems, which have a sufficiently wide good-field-region,  so that our studies, which refer to a simple model of undulator field nevertheless yields a correct quantitative description in large variety of practical problems.
The Lorentz force $\vec{F} = -e\vec{v}\times \vec{H}_w/c$ is used to derive the equation of motion      
of electrons with charge $-e$ and mass $m$ in the presence of magnetic field. The explicit expression for the electron velocity in the field of the helical undulator has the form $c\theta_w[\vec{e}_x \cos(k_wz) - \vec{e}_y\sin(k_wz)]$, where $\theta_w = K/\gamma$ and $K = eH_w/(k_w mc^2)$ is the undulator parameter.
This means that the reference electron in the undulator moves along the constrained helical trajectory parallel to the $z$ axis. As a rule, the electron rotation angle $\theta_w$ is small and the longitudinal electron velocity $v_z$ is close to the velocity of light, $v_z = \sqrt{v^2 - v^2_{\perp}} \simeq v(1-\theta_w^2/2) \simeq c$.

Let us consider a modulated ultrarelativistic electron beam moving alone the $z$ axis in the field of the helical undulator. In the present study we introduce the following assumptions. First,
without kick the electrons move along constrained helical trajectories in parallel with the $z$ axis. Second, 
electron beam density at the undulator entrance is simply $n = n_0(\vec{r}_{\perp})[1 + a\cos \omega(z/v_z - t)]$, where $a = \mathrm{const.}$ In other words we consider the case in which there are no variation in amplitude and phase of the density modulation in the transverse plane.
Under these assumptions the transverse current density may be written in the form $\vec{j}_{\perp} = -e\vec{v}_{\perp}(z)n_0(\vec{r}_{\perp})[1+a\cos \omega(z/v_z -t)]$. Even through the measured quantities are real, it is generally more convenient to use complex representation, starting with real $\vec{j}_{\perp}$, one defines the complex transverse current density: $j_x+ij_y = -ec\theta_wn_0(\vec{r}_{\perp})\exp(-ik_wz)[1+a\cos \omega(z/v_z -t)]$. The transverse current density has an angular frequency $\omega$ and two waves traveling in the same direction with variations $\exp i(\omega z/v_z - k_wz -\omega t)$ and  $\exp - i(\omega z/v_z + k_wz -\omega t)$ will add to give a total current proportional to 
$\exp(-ik_wz)[1+a\cos \omega(z/v_z -t)]$. The factor  $\exp  i(\omega z/v_z - k_wz -\omega t)$ indicates a fast wave, while the factor  $\exp - i(\omega z/v_z + k_wz -\omega t)$ indicates a slow wave. The use of the word "fast"  ("slow") here implies a wave with a phase velocity faster (slower) than the beam velocity.

Having defined the sources, we now should consider the electrodynamics problem. 
Maxwell equations can be manipulated
mathematically in many ways in order to yield derived equations more
suitable for certain applications.  For example, from Maxwell
equations we can obtain an equation which depends only on the
electric field vector $\vec{E}$ (in Gaussian units):
$c^2 \nabla^2 \vec{E} - \partial^2 \vec{E}/\partial t^2 = 4
	\pi c^2 \vec{\nabla} \rho + 4 \pi \partial \vec{j}/\partial t$.
Once the charge and current densities  $\rho$ and $\vec{j}$ are specified as a function of time and position, this equation allows one to calculate the electric field $\vec{E}$ at each point of space and time. Thus, this nonhomogeneous wave equation is the complete and correct formula for radiation. However we want to apply it to still simpler circumstance in which second term  (or, the current term) in the right-hand side provides the main contribution to the value of the radiation field. It is relevant to remember that our case of interest is the coherent undulator radiation and the divergence of this radiation is much smaller compared to the angle $1/\gamma$. It can be shown that when this condition is fulfilled the gradient term, $4\pi c^2 \vec{\nabla} \rho$, in the right-hand side of the nonhomoheneous wave equation can be neglected.
Thus we consider the wave equation $c^2 \nabla^2 \vec{E} - \partial^2 \vec{E}/\partial t^2 =   4 \pi \partial \vec{j}_{\perp}/\partial t$.

We wish to examine the case when the phase velocity of the current wave is close to the velocity of light. This requirement may be met under resonance condition $\omega/c = \omega/v_z - k_w$. This is the condition for synchronism between the transverse electromagnetic wave and the fast transverse current wave with the propagation constant $\omega/v_z - k_w$. With the current wave traveling with the same phase speed as electromagnetic wave, we have the possibility of obtaining a spatial resonance between electromagnetic wave  and electrons. If this the case, a cumulative interaction between modulated electron beam and transverse electromagnetic wave in empty space takes place. We are therefore justified in considering the contributions of all the waves except the synchronous one to be negligible as long as the undulator is made of a large number of periods.

Here follows an explanation of the resonance condition which is elementary in the sense that we can see what is happening physically. 
The field of electromagnetic wave has only transverse components, so the energy exchange between the electron and electromagnetic wave is due to transverse component of the electron velocity. For effective energy exchange between the electron and the wave, the scalar product $-e\vec{v}_{\perp}\cdot\vec{E}$ should be kept nearly constant along the whole undulator length. We see that required synchronism  $k_w + \omega/c - \omega/v_z = 0$ takes place when the wave advances the electron beam by the wavelength at one undulator period $\lambda_w/v_z = \lambda/(c -v_z)$, where $\lambda = 2\pi/\omega$ is the radiation wavelength. This tells us that  the angle between the transverse velocity of the particle $\vec{v}_{\perp}$ and the vector of the electric field $\vec{E}$ remains nearly constant.
Since $v_z \simeq c$ this resonance condition may be written as $\lambda \simeq \lambda_w/(2\gamma_z^2) = \lambda_w (1 + K^2)/(2\gamma^2)$.

We will use an adiabatic approximation that can be taken  advantage of, in all practical situations involving XFELs, where the XFEL modulation wavelength  is much shorter than the electron bunch length $\sigma_b$, i.e. $\sigma_b\omega/c \gg 1$. Since we are interested in coherent emission around the modulation wavelength the theory of coherent undulator radiation  is naturally developed in the space-frequency domain. In fact, in this case one is usually interested into radiation properties at fixed modulation frequency.

We first apply a temporal Fourier transformation  to the inhomogeneous wave equation to obtain the inhomogeneous  Helmholtz equation
$c^2 \nabla^2 \vec{\bar{E}} + \omega^2 \vec{\bar{E}} =  - 4 \pi i \omega \vec{\bar{j}}_{\perp}$,
where 
$\vec{\bar{j}}_{\perp}(\vec{r},\omega)$ is the Fourier transform of the current density
$\vec{j}_{\perp}(\vec{r},t)$.
The solution  can be represented as a weighted superposition of solutions corresponding to a unit point source located at $\vec{r}'$. The Green function for the inhomogeneous Helmholtz equation is given by (for unbounded space and outgoing waves) $4\pi G(\vec{r}, \vec{r'}, \omega) =   \exp\left[i \omega
|\vec{r} - \vec{r'}|/c\right]/|\vec{r} - \vec{r'}|$, with $|\vec{r} - \vec{r'}| = \sqrt{(x' - x )^2 + (y' - y)^2 + (z' - z)}^2$.
With the help of this Green function we can write a formal solution for
the field equation as: $\vec{\bar{E}} = \int d\vec{r'} ~G(\vec{r}, \vec{r'}) \left[- 4 \pi i \omega \vec{\bar{j}_{\perp}}/c^2\right]$.

This is just a mathematical description of the concept of Huygens' secondary sources and waves, and is of course  well-known, but we still recalled how it follows directly from  the Maxwell's equations. We may consider the amplitude  of the beam radiated by plane of oscillating electrons
as a whole to be the resultant of radiated spherical waves. This is because Maxwell's theory has no intrinsic anisotropy \footnote{This property of the electromagnetic field theory only holds in an inertial frame with Lorentz coordinates}. The electrons lying on the plane of simultaneity gives rise to spherical radiated wavelets, and these combine according to  Huygens' principle to form what is effectively a radiated wave. If the plane of simultaneity is  the $xy$-plane (i.e. beam modulation wavefront is perpendicular to the $z$- axis),  then the Huygens' construction shows that plane wavefronts will be emitted along the $z$-axis. 

In summary: according to Maxwell's electrodynamics, coherent radiation is always emitted in the direction normal to the modulation wavefront. We already stressed  that Maxwell's equations are valid only in a Lorentz reference frame, i.e. when an inertial frame where the Einstein synchronization procedure is used to assign values to the time coordinates. Einstein's time order should be applied and kept in consistent way in both dynamics and electrodynamics. Our previous description implies quite naturally that Maxwell's equations in the lab frame are compatible only with covariant trajectories 
$\vec{x}_{cov}(t)$, calculated by using Lorentz coordinates and, therefore, including relativistic kinematics effects.  

Let us go back to the modulated electron beam, kicked transversely with respect to the direction of motion, that was discussed before. 
Conventional particle tracking shows that while the electron beam direction changes after the kick, the orientation of the modulation wavefront stays unvaried. In other words, the electron motion and the wavefront normal have different directions. Therefore, according to conventional coupling of fields and particles that we deem incorrect, the coherent undulator radiation in the kicked direction produced in a downstream undulator is expected to be dramatically suppressed as soon as the kick angle is larger than the divergence of the output coherent radiation.

In order to estimate the loss in radiation efficiency in the kicked direction  according to the conventional coupling of fields and particles, we make the assumption that the spatial profile of the modulation  is close to that of the electron beam and has a Gaussian shape with standard deviation $\sigma$. A modulated electron beam in an undulator can be considered as a sequence of periodically spaced oscillators. The radiation produced by these oscillators always interferes coherently at zero angle with respect to the undulator axis. When all the oscillators are in phase there is, therefore,  strong emission in the direction $\theta = 0$. If we have a triangle with a small altitude $r \simeq \theta z $ and long base $z$, than the diagonal $s$ is longer than the base. The difference is $\Delta = s - z\simeq z\theta^2/2$. When $\Delta$ is equal to one wavelength, we get a minimum in the emission. This is because in this case the contributions of various oscillators are  uniformly distributed in phase from $0$ to $2\pi$.  
In the limit for a small size of the electron beam, $\sigma \to 0 $, the interference will be constructive within an angle of about $\Delta\theta \backsimeq \sqrt{c/(\omega L_w)} = 1/(\sqrt{4\pi N_w}\gamma_z) \ll 1/\gamma$, where $L_w = \lambda_w N_w$ is the undulator length. In the limit for a large size of the electron beam,  the angle of coherence is about $\Delta\theta \backsimeq c/(\omega\sigma)$ instead. The boundary between these two asymptotes is for sizes of about $\sigma_\mathrm{dif} \backsimeq \sqrt{cL_w/\omega}$. \footnote{The parameter $\omega\sigma^2/(cL_w)$ can be referred to as the electron beam Fresnel number} 
It is worth noting that, for XFELs, the transverse size of electron beam $\sigma$ is  typically much larger  than $\sigma_\mathrm{dif}$ (i.e electron beam Fresnel number is large).  
Thus, we can conclude that the angular distribution of the radiation power in the far zone has a Gaussian shape with standard deviation $\sigma_\mathrm{\theta} \backsimeq c/(\sqrt{2}\omega\sigma)$. 
However, still according to the conventional treatment, after the electron beam is kicked we have
the already-mentioned discrepancy between direction of the electron motion and wavefront normal. Then, the radiation intensity along the new direction of the electron beam can be approximated as $I \backsimeq I_0  \exp[- \theta_k^2/(2\sigma_\mathrm{\theta}^2)]$, where $I_0$ is the  on-axis intensity without kick and $\theta_k$ is the kick angle.  The exponential suppression factor is due to the tilt  of the modulation wavefront with respect to the direction of motion of the electrons. 

We presented a study of very idealized situation  for illustrating the difference between conventional and covariant coupling of fields and particles. We solved the dynamics problem of the motion of a relativistic electrons in the prescribed force field of weak kicker magnet by working only up to the order of $\gamma\theta_k$. This approximation is of particular theoretical interest because it is relatively simple and at the same time forms the basis for understanding relativistic kinematic effects such as Wigner rotation \footnote{We used this restriction in order to understand all the physical principles very clearly. We considered an ultrarelativistic electron beam, meaning  that we already have small problem parameter $1/\gamma^2 \ll 1$.  In small kick angle  approximation we also have a second small problem parameter $v_x^2/v^2 = v_x^2/c^2 \ll  1$. 
It would have been difficult  for us in this illustrative example to discuss the interdependence of these two small parameters, so we studied only a situation where all velocities are non relativistic even in the initial frame where the beam was at rest upstream of the kicker.
Let the $S$ be a lab frame of reference and $S'$ a comoving with velocity $\vec{v}$ relative to $S$. Upstream of the kicker, the modulated beam is at rest in the frame $S'$. 
One can study what happens in $S'$ before the kick. Our modulated beam  is at rest and the kicker is running towards it with velocity $-\vec{v}$. The moving magnetic field of the kicker produces an electric field orthogonal to it. When the kicker interacts with the particle in $S'$ we thus deal with an electron moving in the combination of perpendicular electric and magnetic fields.  It is easy to see that the acceleration in the crossed fields yields an electron velocity  $v'_x =\gamma v_x$ parallel to the $x$-axis and $v'_z = - v(\gamma v_x/c)^2/2$ parallel to the $z$-axis. We assumed that $(\gamma v_x/c)^2 \ll 1$. If we neglect terms in $(\gamma v_x /c)^2$, second relativistic correction does not appear in this approximation. In other words, even in the $S'$ frame, the transverse motion of the beam is non relativistic }.   
Let us discuss the region of validity of our small kick angle approximation $\theta_k\gamma \ll 1$. Since  in XFELs the Fresnel number is rather large,  we  can always consider a kick angle which is relatively large compared to the divergence of the output coherent radiation, and, at the same time, it is relatively small compared to the angle $1/\gamma$. In fact, from $\omega\sigma^2/(cL_w) \gg 1$, with some rearranging, we obtain $\sigma_\theta^2 \simeq c^2/(\omega^2\sigma^2)\ll c/(\omega L_w)$. Then we recall that  $\sqrt{c/(\omega L_w)} = 1/(\sqrt{4\pi N_w}\gamma_z) \ll 1/\gamma$.
Therefore, the first order approximation used to investigate the kicker setup in this section is of practical interest in XFEL engineering.

\subsubsection{Results of experiment}

Above we have shown that our covariant coupling of fields and particles predicts an effect in complete contrast to the conventional treatment. Namely, in the ultrarelativistic limit, the plane of simultaneity, that is wavefront orientation of the modulation,  is always perpendicular to  the electron beam velocity. As a result, we predict strong emission of coherent undulator radiation  from the  modulated electron beam in the kicked direction.

From a pragmatic viewpoint, physical theories should be able to predict experimental results in agreement with measurements, i.e. they should "work". The fact that our theory  predicts reality in a satisfactory way is well-illustrated by comparing the prediction we just made with the results of an experiment involving "X-ray beam splitting" of a circularly-polarized XEL pulse from the linearly-polarized XFEL background pulse, a technique used in order to maximize the degree of circular polarization at XFELs\cite{NUHN}.
The "X-ray beam splitting" experiment at the LCLS \cite{NUHN}  apparently demonstrated that after a modulated electron beam is kicked on a large angle compared to the divergence of the XFEL radiation  \footnote{ The tuning limit of deflection angle  was set at $\sim$ 5 rms of XFEL radiation divergence by beamline aperture}, the modulation wavefront is readjusted along the new direction of motion of the kicked beam, see Fig. 14 in \cite{NUHN} . This is the only way to justify coherent radiation emission from the short undulator placed after the kicker and along the kicked direction.

The authors of \cite{NUHN} found that coherent undulator radiation was produced in the kicked direction. These results came unexpectedly, but from a practical standpoint, the  "apparent wavefront readjusting"  immediately led to the realization that
the unwanted, linearly-polarized radiation background could be fully eliminated without extra-hardware. In other words a single corrector,  already part of the baseline installations in the intersection between undulator segments, effectively worked as the complex and expensive bending system designed according to the theory of conventional particle tracking in \cite{L}. The results of the "beam splitting" experiment at the LCLS, demonstrated that even the direction of emission of coherent undulator radiation is beyond the predictive power of the conventional theory.

We showed that the authors of \cite{NUHN} actually witnessed an apparent wavefront readjusting due to the phenomenon of Wigner rotation, but they never drew this conclusion. We are actually first in considering the idea that results of the conventional theory of radiation by relativistically moving charges are not consistent with the principle of relativity. In previous literature, identification of the trajectories in the source part of the usual Maxwell's equations with the trajectories calculated by conventional particle tracking in the lab frame has always been considered obvious. The impact of \cite{NUHN} on our studies was immediate. 
Now everything fits together, and our theory, albeit shocking, shows the existence of  coherent radiation in the kicked direction.


\

\section{What is special relativity?} 

The laws of physics are invariant with respect to Lorentz transformations. This is a restrictive principle and does not determine the exact form of the dynamics in question. Understanding the postulates of the theory of relativity is similar to understanding energy conservation: at first we learn this as a principle and later on  we study  microscopic interpretations that must be consistent with this principle. 
For any system to which the energy conservation principle can be applied, a  deeper theory should exist which yields insight into the detailed physical processes involved. Of course, this deeper theory must lead to energy conservation. 

The principle of conservation of energy is very useful in making analyses without knowing all the formulas of the fundamental theory.  A methodological analogy with the postulates of the special relativity emerges by itself. 
Suppose we do not know why a muon disintegrates, but we know the law of decay in the Lorentz rest frame. This law would then be a phenomenological law. The relativistic generalization of this law to any Lorentz frame allows us to make a prediction on the average distance traveled by a muon. In particular, when a Lorentz transformation of the decay law is tried, one obtains the prediction that after the travel distance $\gamma v\tau_0$, the population in the lab frame  would be reduced to 1/2 of the origin population. We may interpret this result by saying that, in the lab frame, the characteristic lifetime of a particle has increased from $\tau_0$ to $\gamma\tau_0$. 

However, the theory of relativity is necessary incomplete. Constructive (microscopic) theories like electrodynamics or quantum field theory provide more insight into the nature of things than restrictive theories like special relativity.
Relativistic kinematics  is only an interpretation of the behavior of the dynamical matter fields in the view of different observers. 
The point is that one can, in principle compute any relativistic quantity directly from the underlying theories of matter without involving relativity at all. 
For example, muons in motion behave relativistically because the field forces that are responsible for the muon disintegration satisfy quantum field equations that are Lorentz covariant.  
Of course, in the "microscopic" approach to relativistic phenomena,  Lorentz covariance of all the fundamental laws of physics remains, similarly to energy conservation, an unexplained fact, but all explanation must stop somewhere.

\section{Different approaches to special relativity}

In literature, three approaches to special relativity are discussed: Einstein's approach, the usual covariant approach, and the space-time geometric approach (see e.g. \cite{IV} and references therein).  

Einstein formulation is based on postulates: the principle of relativity and the constancy of the velocity of light. 
The usual covariant approach mainly deals with the components of 4-tensors in a specific basis, i.e. when Lorentz coordinates are chosen in an inertial frame of reference. 
In space-time geometric approach, primary importance is attributed to the geometry of space-time; it is supposed that the geometry of space-time is a pseudo-Euclidean geometry in which only 4-tensors quantities do have real physical meaning.

In this most general approach the principle of relativity in contrast to Einstein formulation of the special relativity is a simple consequence of the space-time geometry. 
Since the space-time geometric approach deals with all possible choices of coordinates of the chosen reference frames, the second Einstein postulate referred to the constancy of the coordinate velocity of light does not hold in this formulation of the theory of relativity. Only in Lorentz coordinates, when Einstein's synchronization of distant clocks and Cartesian space coordinates are used, the coordinate speed of light is isotropic and constant. 
Thus the basic elements of the space-time geometric formulation of the special relativity  and the usual Einstein,s formulation, are quite different.

\subsection{The usual Einstein's approach}

Traditionally, the special theory of relativity is built on the principle of relativity and on a second additional postulate concerning the velocity of light:

1. Principle of relativity. The laws of nature are the same  (or take the same form) in all inertial frames

2. Constancy of the speed of light. Light propagates with constant velocity $c$ independently of the direction of propagation, and of the velocity of its source.

The constancy of the light velocity in all inertial systems of reference is not a fundamental statement of the theory of relativity. The central principle of special relativity is the Lorentz covariance of all  the fundamental laws of physics. It it important to stress at this point that the second ”postulate”, contrary to the view presented in textbooks, is not a separate physical assumption, but a convention that cannot be the subject of experimental tests. In fact, in order to measure the one-way speed of light one has first to synchronize the infinity of clocks assumed attached to every position in space, which allows us to perform time measurements. Obviously, an unavoidable deadlock appears  if one synchronizes the clocks by assuming a-priori that the one-way speed of light is $c$. In fact, in that case, the one-way speed of light measured with these clocks (that is the Einstein speed of light) cannot be anything else but $c$: this is because the clocks have been set assuming that particular one-way speed in advance. Therefore, it can be said that the value of the one-way speed of light is just a matter of convention without physical meaning. In contrast to this, the two-way speed of light, directly measurable along a round-trip, has physical meaning, because round-trip experiments rely upon the observation of simultaneity or non-simultaneity of events at a single point in space.

Assuming postulate 2 on the constancy of the speed of light in all inertial frames we also automatically assume Lorentz coordinates, and that different inertial frames are related by Lorentz transformations. In other words, according to such limiting understanding of the theory of relativity it is assumed  that only Lorentz transformations must be used to map the coordinates of events between inertial observers.

\subsection{The usual covariant approach}

In the usual covariant approach the special of relativity is understood as the theory of space-time with pseudo-Euclidean geometry. Quantities of physical interest are represented by tensors in a four-dimensional space-time, i.e. by covariant quantities, and the laws of physics are written in manifestly covariant way  as four-tensor equations.  

Any event in the usual covariant approach is mathematically represented by a point in space-time, called world-point. The evolution of a particle is, instead, represented by a curve in space-time, called world-line. If $ds$ is the infinitesimal displacement along a particle world-line, then

\begin{eqnarray}
&& ds^2 =  c^2 dT^2 - dX^2 - dY^2 - dZ^2~ ,\label{MM1}
\end{eqnarray}
where we have selected a special type of coordinate system (a Lorentz coordinate system),  defined by the requirement that Eq. (\ref{MM1}) holds.

To simplify our writing we will use, instead of variables $T, X, Y, Z$,  variables $X^{0} = cT, ~ X^{1} = X,~ X^{2} = Y,~ X^{3} = Z$. Then, by adopting the tensor notation, Eq. (\ref{MM1}) becomes $ds^2 = \eta_{ij}dX^{i}dX^{j}$, where Einstein summation is understood. Here $\eta_{ij}$ are the Cartesian components of the metric tensor and by definition, in any Lorentz system, they are given by $\eta_{ij} = \mathrm{diag}[1,-1,-1,-1]$, which is the metric canonical, diagonal form. As a consequence of the space-time geometry, Lorentz coordinates systems are connected by Lorentz transformations, which form the Lorentz group. Since the metric is invariant under Lorentz transformations the Lorentz group is also called the stability group of the metric.

The usual covariant formulation of the theory of relativity deals with the pseudo-Eucledian space-time  geometry and with the invariance of $ds$, but it is understood only in a limited sense when the metric is strictly diagonal. As a matter of fact, a widespread argument used to support the incorrectness of Galilean transformations is that they not preserve the diagonal form of the metric. To quote L. Landau and E. Lifshitz \cite{LL}:  "This formula is called the Galileo transformation. It is easily to verify that this transformation, as was to be expected, does not satisfy the requirements of the theory of relativity; it does not leave the interval between events invariant.".  This statement is obviously incorrect, because the space-time continuum can be described  equally well from the point of view of any coordinate system, which cannot possibly change $ds$. Assuming diagonality of the metric we also automatically assume Lorentz coordinates, and that different inertial frames are related by Lorentz transformations. In other words, according to such limiting  understanding of the covariant approach, it is assumed  that only Lorentz transformations must be used to map the coordinates of events between inertial observers. 

Physical quantities are represented by space-time geometric (tensor) quantities. When some basis is introduced, the representation of a tensor as geometric quantity comprise both components and basis. 
In usual covariant approach, one only deals  with the basis components of tensors in the Lorentz coordinates i.e. with the case when the basis four-vectors are orthogonal. As a result one deals only with four-tensor equations of physics written out in the component form. 

However, the concept of a tensor in the usual covariant approach is given in terms of the transformation properties of its components.  
For example in the usual covariant approach the electromagnetic "tensor" $F^{\mu\nu}$ is actually not a tensor since $F^{\mu\nu}$ are only components implicitly taken in standard (orthogonal) basis. The components are coordinate quantities and they do not contain the whole information about the physical quantity, since a basis of the space-time is not included. This is no problem only in the limiting case when transformations from one orthogonal basis to another orthogonal basis are selected i.e. only assuming that Lorentz transformations must be used to map the coordinates of events. According to the usual covariant approach, another  transformations  from standard to non standard (not orthogonal) basis, like Galilean transformations, are "incorrect".

It should be note that usual formulation  of the theory of relativity is limited but absolutely correct if Lorentz coordinates are  applied and kept in consistent way in both dynamics and electrodynamics. The common mistake, discussed above, made in accelerator and plasma physics is connected with the incorrect algorithm for solving the electromagnetic field equations. Only the solution of the dynamics equations in covariant form (i.e. in Lorentz coordinates)  gives the correct coupling between the usual Maxwell's equations and particle trajectories in the lab frame.

\subsection{The space-time geometric approach}

Common textbook presentations of the special theory of relativity use the Einstein approach or, as generalization, the usual covariant approach which deals, as discussed above, only with components of the 4-tensors in specific (orthogonal) Lorentz basis. 
The fact that in the process of transition to arbitrary coordinates the geometry of the space-time does not change, is not considered in textbooks. 
As a consequence there is a widespread belief among experts that a Galilean transformation ( which is actually a transformation from an orthogonal Lorentz basis to a non orthogonal basis) is incorrect, while a Lorentz transformation (which is a transformation from an orthogonal Lorentz basis to another orthogonal Lorentz basis) is correct. This is not true. We can describe physics in any arbitrary coordinates system. The different transformations of coordinates  only correspond  to a change in the way of components of 4-tensors are written, but not influence of 4-tensors themselves.    
Although the Einstein synchronization i.e. Lorentz coordinates choice, is preferred by physicists due to its simplicity and symmetry, it is nothing more "physical" than any other. A particularly  very unusual choice of coordinates, the absolute time coordinate choice, will be considered and exploited in this paper. 

The reason why in this paper the understanding of Galilean transformation in terms of the theory of relativity is given so much attention  is that our accelerator engineering colleagues have been using the non covariant (3+1) approach to relativistic particle dynamics in particle tracking calculations for about  seventy years. However, the type of clock synchronization which provides the time coordinate $t$ in the corrected Newton's equation has never been discussed in literature. We claim, and this claim is quite central for our reasoning, that in conventional particle tracking in accelerator and plasma physics the description of the dynamical evolution of charged particles in the lab frame is based on the use of the absolute time convention.
Much unusual as this choice may seem in the theory of relativity, it is actually the most convenient one in relativistic engineering.
In this kind of non-covariant particle tracking,  time differs from space and  particle's trajectory can be seen from the lab frame view as a result of successive Galilean boosts that track the motion of the accelerated particle. The usual Galileo (vectorial) rule for addition of velocities is used to fix Galileo boosts tracking a particular particle along its motion.

\subsubsection{General form of pseudo-Eucledian metric}

The space-time continuum, determined by the interval  Eq. (\ref{MM1}) can be described in arbitrary coordinates and not only in Lorentz coordinates. In the transition to arbitrary coordinates, the geometry of four-dimensional space-time obviously does not change, and in the special theory of relativity we are not limited in any way in the choice of a coordinates system. The space coordinates $x^1, x^2, x^3$ can be any quantities defining the position of particles in space, and the time coordinate $x^0$ can be defined by an arbitrary running clock.
The components of the metric tensor in the coordinate system $x^i$ can be determined by performing the transformation from the Lorentz coordinates  $X^{i}$ to the arbitrary variables $x^{j}$, which are fixed as $X^{i} = f^{i}(x^{j})$. One then obtains

\begin{eqnarray}
	&& ds^2 = \eta_{ij}dX^{i}dX^{j} = \eta_{ij}\frac{\partial X^{i}}{\partial x^{k}}\frac{\partial X^{j}}{\partial x^{m}}dx^{k}dx^{m} = g_{km}dx^{k}dx^{m} ~ ,\label{MM3}
\end{eqnarray}

This expression represents the general form of the pseudo-Euclidean metric.
In textbooks and monographs, the special theory of relativity is generally presented in relation to an interval $d s$ in the Minkowski form Eq.(\ref{MM1}), while  Eq.(\ref{MM3}) is ascribed to the theory of general relativity. 

However, in the space-time geometric approach, special relativity is understood as a theory of  four-dimensional space-time with pseudo-Euclidean geometry. In this formulation of the theory of relativity
the space-time continuum can be described  equally well from the point of view of any coordinate system, which cannot possibly change $ds$. At variance, the usual formulation of the theory of relativity also deals with the invariance of $ds$, but it is understood only in a limited sense when the metric is strictly diagonal.

\subsubsection{Pseudo-Eucledian metric and Galilean transformations}

Absolute simultaneity can be introduced in special relativity without affecting neither the logical structure, no the  (convention-independent) predictions of the theory. Actually, it is just a simple effect related with a particular parametrization.  
We begin with the Minkowski metric as the true measure of space-time intervals for an inertial observer $S'$ with coordinates $(t',x')$. Here we neglect the two perpendicular space components that do not enter in our reasoning. We transform coordinates $(t,x)$ that would be coordinates of an inertial observer $S$ moving with velocity $-v$ with respect to the observer $S'$, using a Galilean transformation: we substitute $x' =x-vt$, while leaving time unchanged $t' = t$ into the Minkowski metric $ds^2 = c^2 dt'^2 - d x'^2$ to obtain

\begin{eqnarray}
	&& ds^2 = c^2(1-v^2/c^2)dt^2 + 2vdxdt - dx^2  ~ .\label{GGG3}
\end{eqnarray}
Inspecting Eq. (\ref{GGG3}), or using transforming the Minkowski metric using the Galilean transformation above we can find the components of the metric tensor $g_{\mu\nu}$ in the coordinate system $(ct,x)$ of $S$. We obtain $g_{00} = 1-v^2/c^2$, $g_{01} = v/c$, $ g_{11} = - 1$. Note that the metric in Eq. (\ref{GGG3}) is not diagonal, since, $g_{01} \neq 0$, and this implies that time is not orthogonal to space.

The velocity of light in the coordinate system $(t', x')$ for $S'$, defined above as "at rest", is $c$. In the coordinate system $(t, x)$ \footnote{Transformation is interpreted in the passive sense}, however, the speed of light cannot be equal $c$ anymore because $(t, x)$ is related to $(t',x')$ via a Galilean transformation. As a result, the speed of light in the direction  parallel to the $x$ axis is equal to $c + v$ in the positive direction, and $-c +v$  in the negative direction. This is readily verified if one recalls that the velocity of light in the reference system "at rest" is equal to $c$. If $ds$ is the infinitesimal displacement along the world line  of  a ray of light, then  $ds^2 = 0$ and we obtain $c^2 = (dx'/dt')^2$. In the moving reference system, since $x' = x - vt$ and $t' = t$, this expression takes the form $c^2 = (dx/dt -v)^2$, which can be seen by a trivial change of variable, or setting $ds^2 =0$ in Eq. (\ref{GGG3}). This means that in the moving reference system of coordinates $(ct,x)$ the velocity of light parallel to the x-axis,  is $dx/dt = c + v$ in the positive direction, and $dx/dt = -c + v$  in the negative direction as stated above.

We conclude that the speed of light emitted by a  moving source measured in the lab frame $(t, x)$ depends on the relative velocity of source and observer, in our example $v$. In other words, the speed of light is compatible with the Galilean law of addition of velocities. The reason why it is different from the electrodynamics constant $c$
is due to the fact that the clocks are synchronized following the absolute time convention, which is fixed because $(t, x)$ is related to $(t',x')$ via a Galilean transformation. 
Note that from what we just discussed follows the statement that the difference between the speed of light and the electrodynamics constant $c$ is convention-dependent and has no direct physical meaning. 

\subsubsection{Galilean transformations and wave equation for electromagnetic fields}

We used the four-geometric arguments to show that,  due to the absolute time synchronization convention in the lab frame, the speed of light is compatible with the Galilean law of addition of velocities.  We are now ready to study the same outcome in terms of the properties of the dynamical fields. In fact, light propagation  can  be explained in the framework of the electromagnetic field theory.

In the comoving frame, fields are expressed as a function of the independent variables $x', y', z'$, and $t'$. Let us consider Maxwell's equations in free space.  The electric field $\vec{E}'$ of an electromagnetic wave satisfies the equation $\Box'^2\vec{E}' =  \nabla'^2\vec{E}' - \partial^2\vec{E}'/\partial(ct')^2  = 0$.
However, the variables $x',y',z',t'$ can be expressed in terms of the independent variables $x, y, z, t$ by means of a Galilean transformation, so that fields can be written in terms of  $x, y, z, t$. From the Galilean transformation $x' = x - vt, ~ y' = y, ~ z' = z, ~ t' = t $, after partial differentiation, one obtains $\partial/{\partial t} = \partial/{\partial t'} - v\partial/{\partial x'}$, $\partial/{\partial x} = \partial/{\partial x'}$.
Hence the wave equation transforms into

\begin{eqnarray}
&& \Box^2\vec{E} = \left(1-\frac{v^2}{c^2}\right)\frac{\partial^2\vec{E}}{\partial x^2}  - 2\left(\frac{v}{c}\right)\frac{\partial^2\vec{E}}{\partial t\partial x}
+ \frac{\partial^2\vec{E}}{\partial y^2} + \frac{\partial^2\vec{E}}{\partial z^2}
- \frac{1}{c^2}\frac{\partial^2\vec{E}}{\partial t^2} = 0 ~ , \label{GGT2}
\end{eqnarray}

where coordinates and time are transformed according to a Galilean transformation. The solution of this equation $F[x - (c +v)t] + G[x + (-c +v)t]$ is the sum of two arbitrary functions, one of argument $x - (c +v)t$ and the other of argument $x + (-c +v)t$.
Here we obtained the solution for waves which move in the $x$ direction by supposing that the field does not depend on $y$ and $z$. The first term represents a wave traveling forward in the positive $x$ direction, and the second term a wave traveling backwards in the negative $x$ direction. This result agrees with what we would have found more rapidly using the metric Eq.(\ref{GGG3}).  However, in this way we have  provided a dynamical underpinning for our previous discussion of the behavior of the speed of light under a Galilean transformation.

We would like to make some further remarks about kinematic relativistic effects. As discussed above, the Galilean transformation connecting the reference frame $S'$,  moving with velocity $v$ relative to the lab frame $S$, is given by  $x' =  x - vt, ~ t' = t$. This transformation implies a particular choice of synchronization convention  in the lab frame, which we called the "absolute time convention", so that the motion of particles looks precisely the same as predicted by Newtonian kinematics:  relativistic effects like Wigner rotation, time dilation,  Lorentz-Fitzgerald contraction and relativistic corrections in the law of composition of velocities  do not exist in this description.
In agreement with the principle of relativity, the usual Maxwell's equations can be used in a moving inertial frame where a charge is instantaneously at rest. However, the transformation connecting any co-moving frame to the lab frame in the case of the absolute time convention is a Galilean transformation, and Maxwell's equations do not remain invariant with respect to Galilean transformation. When a Galilean transformation of Maxwell's equations is tried, the new terms that have to be added into  Maxwell's equations lead to those relativistic phenomena that were left out from the description of dynamics in terms of Newtonian kinematics. It does not matter which convention and hence transformation is used to describe the same reality. What matters is that, once fixed, such convention should be applied and kept in a consistent way in both dynamics and electrodynamics.

\subsubsection{Lorentz and Galilean transformations in the non-relativistic limit}

It is generally believed that a Lorentz transformation reduces to a Galilean transformation in the non-relativistic limit. We state that this is incorrect. As discussed, kinematics is a comparative study which requires two coordinate systems, and one needs to assign time coordinates to the two systems. Different types of clock synchronization provide different time coordinates.
The convention on the clock synchronization amounts to nothing more than a definite choice of the coordinate system in an inertial frame of reference in Minkowski space. Pragmatic arguments for choosing one coordinate system over another may therefore lead to different choices in different situations. Usually, in practice, we have a choice between absolute time coordinate and Lorentz time coordinate. The space-time continuum can be described equally well in both coordinate systems. This means  that  for arbitrary particle speed,  the Galilean coordinate transformations well characterize a change in the reference frame from the lab inertial observer to a co-moving inertial observer in the context of the theory of relativity. Let us consider the non relativistic limit. The Lorentz transformation, for $v/c$ so small that $v^2/c^2$ is neglected can be written a $x' = x - vt$, $t' = t - xv/c^2$. This infinitesimal Lorentz transformation differs from the infinitesimal Galilean transformation $x' = x - vt$, $t' = t$. The difference is in the term $xv/c^2$ in the Lorentz transformation for time, which is a first order term. If lab frames $S$ and comoving frame $S'$ have coordinates in a non standard (absolute time) configuration, we need to transform Maxwell's equations according to a Galilean transformation, and we obtain Eq.(\ref{GGT2}). We can see that the wave equation in the lab frame after the Galileo boost has non-diagonal form even in the non-relativistic limit $v/c \ll 1$, $\gamma \sim 1$. The difference consists in the crossed term $\partial^2/\partial t\partial x$ which arises when applying the Galileo boost.

\subsubsection{Metric diagonalization}

The Galilean transformation connecting Lorentz coordinates $(t', x', y', z')$ with diagonal metric, Eq.(\ref{MM1}), to coordinates $(t,x, y, z)$ with non diagonal metric, Eq.(\ref{GGG3}), is equivalent to a rotation in the plane $x',t'$ to non-orthogonal axes $t,x$. In the coordinates system\footnote{As before, we can neglect $y$ and $z$.} $(t,x)$  we therefore have, as already discussed, much more complicated field equations. To get around this difficulty we observe that the non-diagonal metric can always be simplified. In fact, the space-time line-element in Eq.(\ref{GGG3}) can be separated in a temporal part $dt_d$ and a spatial part $dx_d$ as

\begin{eqnarray}
	&& ds^2  = c^2 dt_d^2 - dx_d^2 ~ ,\label{GG7}
\end{eqnarray}

with $dx_d^2 = dx^2/(1 - v^2/c^2)$ and $c^2dt_d^2 = \left[ \sqrt{1-v^2/c^2}cdt +  (v/c)dx/\sqrt{1-v^2/c^2}\right]^2$.

In practice we are "diagonalizing" the metric by completing the square and collecting terms in $dx$. Obviously, transforming to new variables leads to the usual Minkowski form of the metric. From Eq. (\ref{GG7}) we find $dx_d/dt_d = c$. As expected, in the new variables the velocity of light is constant in all directions, and equal to the electrodynamics constant $c$. The overall combination of Galilean transform and variable changes specified above actually yields to the Lorentz transformation $dx_d = \gamma(dx' + vdt')~ , dt_d = \gamma(dt' + vdx'/c^2)$.

As we already discussed in Introduction (see paragraph 1.4.4), the idea of studying dynamics and electrodynamics in (3+1) space and time using a technique involving Galilean transformations and a change of variables is useful from a pedagogical point of view.
In the non-covariant (3+1) approach, there are no relativistic kinematics effects. 
One might think that the relativistic kinematics effects like Wigner rotation, time dilation,  Lorentz-Fitzgerald contraction and relativistic corrections in the law of composition of velocities are results of transformation Eq.(\ref{GG7}) to new variables.

\subsubsection{Way to solve the electromagnetic field equations  in the (3+1) space and time}

We already found that, starting from the diagonal form of the metric tensor in the rest frame and applying a Galilean transformation we obtain the non-diagonal metric Eq.(\ref{GGG3}). We observed that this non-diagonal metric can always be simplified. In particular, we could transform it to the usual Minkowski form by changing variables. Let us take the dynamical field viewpoint and use it to understand this change of variables.

After properly transforming the d'Alembertian through a Galileo boost, which changes the initial coordinates $(x',y',z',t')$ into $(x,y,z,t)$, we can see that the homogeneous wave equation for the field in the lab frame  has nearly but not quite  the usual, standard form that takes when there is no uniform translation in the transverse direction with velocity $v$. The main difference consists in the crossed term $\partial^2/\partial t\partial x$, which complicates the solution of the equation. To get around this difficulty, we observe that simplification is always possible. The trick needed here is to further make a change of the time variable according to the transformation $t' = t - x v_x/c^2$. In the new variables in i.e. after the Galilean coordinate transformation and the time shift we obtain the  d'Alembertian in the following form

\begin{eqnarray}
\Box^2 =	\left(1-\frac{v_x^2}{c^2}\right)\frac{\partial^2}{\partial x^2} + \frac{\partial^2}{\partial y^2} + \frac{\partial^2}{\partial z^2}
- \left(1-\frac{v_x^2}{c^2}\right)\frac{1}{c^2}\frac{\partial^2}{\partial t^2} ~.
\end{eqnarray}
A further change of a factor $\gamma$  in the scale of time and of the coordinate along the direction of uniform motion leads to the usual Maxwell's equations. In particular, when coordinates and time are transformed according to a Galilean transformation followed by the  variable changes specified above, the  d'Alembertian  $\Box'^2 = \nabla'^2 - \partial^2/\partial(ct')^2$  transforms into  $\Box^2 = \nabla^2 - \partial^2/\partial(ct)^2$ . The overall combination of Galileo transformation and variable changes actually yields the Lorentz transformation in the "3+1" space and time. 
Since the Galilean transformation, completed by the introduction of the new variables,  is mathematically equivalent to a Lorentz transformation, it obviously follows that transforming to new variables leads to the usual Maxwell's equations.

\section{Relativistic mechanics of a particle}

\subsection{Usual four-dimensional  covariant representation. Equation of motion} 

Dynamics equations can be expressed as tensor equations in Minkowski space-time. When coordinates are chosen, one may work with components, instead of  geometric objects. Relying on the geometric structure of Minkowski space-time, one can define the class of inertial frames and can adopt a Lorentz frame with orthonormal basis vectors for any given inertial frame. Within the chosen Lorentz frame, Einstein's synchronization of distant clocks and Cartesian space coordinates are then automatically enforced, the metric tensor components are the usual $g_{\mu\nu} = \mathrm{diag}(1, -1, -1, -1)$, and any two Lorentz frames are related by a Lorentz transformation that preserves the metric tensor components, so that in any Lorentz coordinate system the law of motion becomes

\begin{eqnarray}
&& m\frac{d^2 x_{\mu}}{d\tau^2} = e F^{\mu\nu}\frac{dx_{\nu}}{d\tau}~ ,\label{DDE}
\end{eqnarray}
Here the electromagnetic field is described by the second-rank, antisymmetric tensor with components $F^{\mu\nu}$. The coordinate-independent proper time $\tau$ is a parameter describing the evolution of physical system under the relativistic laws of motion, Eq. (\ref{DDE}).

The covariant equation of motion for a relativistic charged particle under the action of the four-force $K_{\mu} = e F^{\mu\nu}dx_{\nu}/d\tau$ in the Lorentz lab frame, Eq.(\ref{DDE}), is a relativistic "generalization" of the Newton's second law. 
The three-dimensional Newton second law $md\vec{v}/dt = \vec{f}$  can always be used in the instantaneous Lorentz comoving frame. Relativistic "generalization" means that the previous three independent equations expressing Newton second law are be embedded into the four-dimensional Minkowski space. In Lorentz coordinates there is a kinematics constraint $u^{\mu}u_{\mu} = c^2$ for the four-velocity $u_{\mu} = dx_{\mu}/d\tau$. Because of this constraint, the four-dimensional dynamics law, Eq.(\ref{DDE}),  actually includes only three independent equations of motion.  

\subsection{Non-covariant particle tracking}

As discussed in Introduction, it is generally accepted that in order to describe the dynamics of relativistic particles in the lab reference frame, which we assume inertial, one can take into account the relativistic dependence of the particle momentum on the velocity. The treatment of relativistic particle dynamics involves a version of Newton's second law corrected by the relativistic factor $\gamma$. In a given lab frame, there is an electric field $\vec{E}$ and magnetic field $\vec{B}$. They push on a particle in accordance with

\begin{eqnarray}
&& \frac{d\vec{p}}{dt} = e\left(\vec{E} + \frac{\vec{v}}{c}\times \vec{B}\right) ~,\cr &&
\vec{p} = m\vec{v}\left(1 - \frac{v^2}{c^2}\right)^{-1/2}~ ,\label{NN}
\end{eqnarray}

where here the particle's mass, charge, and velocity are denoted by $m$, $e$, and $\vec{v}$  respectively.
In other words, aside for a straightforward correction in the relativistic mass, these three equations looks precisely the same as in Newtonian dynamics. 
The Lorentz force law, plus measurements on the components of acceleration of test particles, can be viewed as defining  the components of the electric and magnetic fields. Once field components are known from the acceleration of test particles, they can be used to predict the accelerations of other particles.

Once a prescribed force field is independently specified, the particle trajectory may be found by integration from initial conditions. The conventional study a relativistic particle motion in a prescribed force field can thus be framed, mathematically, as a well-defined initial value problem.
This study of relativistic particles motion looks precisely the same as in non relativistic Newtonian dynamics. Conventional particle tracking treats the space-time continuum in a non-relativistic format, as a (3+1) manifold. In other words, in the lab frame, Minkowski space-time "splits up" into three dimensional space and one dimensional time. 
Indeed, in conventional particle tracking in accelerator and plasma physics time and space are treated differently. This approach does not require the introduction of Minkowski space-time and is widely used in the study of relativistic particle motion in prescribed force fields, since it is a well-defined initial value (Cauchy) problem.


\subsection{The common view on the relation between  covariant and (3+1) approaches}

Having written down the motion equation in a 4-vector form, Eq.(\ref{DDE}), and determined the components of the 4-force, we satisfied the principle of relativity for one thing, and, for another, we obtained the four components of the equation of motion. This is covariant relativistic generalization of the usual three dimensional Newton's equation of motion which is based on particle proper time as the evolution parameter.

We next wish to describe the motion in the Lorentz lab frame using the lab time $t$ as the evolution parameter. Let us determine the first three spatial components of the 4-force.
We consider for this the spatial part of the dynamics equation, Eq.(\ref{DDE}): $\vec{K} = (dt/d\tau) d(m\gamma\vec{v})/dt = \gamma d(m\gamma\vec{v})/dt$. The prefactor $\gamma$ arises from the change of the evolution variable from the proper time $\tau$, which is natural since $\vec{K}$ is the space part of a four-vector, to the lab frame time $t$, which is  needed to introduce the usual force three-vector $\vec{f}$: $\vec{K} = \gamma\vec{f}$. Written explicitly,  the relativistic form of the three-force is

\begin{eqnarray}
&& \frac{d(m\vec{v}/\sqrt{1-v^2/c^2})}{dt} = e\left(\vec{E} + \frac{\vec{v}}{c}\times \vec{B}\right)~ .\label{DDDE1}
\end{eqnarray}

The time component is obtained as follows

\begin{eqnarray}
&& \frac{d(m c^2/\sqrt{1-v^2/c^2})}{dt} = e\vec{E}\cdot\vec{v} ~ .\label{DFE1}
\end{eqnarray}

The evolution of the particle is subject to these four equations, but also to the constraint

\begin{eqnarray}
&& \mathcal{E}^2/c^2 - |\vec{p}|^2 = mc^2  ~ .\label{DCFE1}
\end{eqnarray}

According to non-covariant (3+1) approach we seek for the initial value solution to these equations.
Using explicit expression for Lorentz force we find that the three  equations Eq.(\ref{DDDE1}) automatically imply the constraint  Eq.(\ref{DCFE1}),
once this is satisfied initially at $t = 0$.
In the (3+1) approach, the four equations of motion "split up" into (3+1) equations and we have no mixture of space and time parts of the dynamics equation Eq.(\ref{DDE}). 
This approach to relativistic particle dynamics relies on the use of three independent equations of motion  Eq.(\ref{DDDE1}) for three independent coordinates and velocities, "independent" meaning that  equation Eq.(\ref{DFE1}) (and constraint Eq.(\ref{DCFE1})) is automatically satisfied. 

The previous commonly accepted derivation of the equations for the particle motion in the three dimensional space from the covariant equation Eq.(\ref{DDE}) has one delicate point. In Eq.(\ref{DDDE1}) and Eq.(\ref{DFE1}) the restriction  $\vec{p}  = m\vec{v}/\sqrt{1 - v^2/c^2}$ has already been imposed. One might  well wonder why, because, equation  Eq.(\ref{DDE}) tells us that the force is the rate of change of the momentum $\vec{p}$, but does  not tell us how momentum varies with speed. 
The components of momentum four-vector $p_\mu = (E/c,\vec{p})$ behave under transformations from one Lorentz frame to another, exactly in the same manner as the component of the four-vector event $x = (x_0,\vec{x})$.  Surprises can surely be expected when we return from the four-vectors language  to the three-dimensional velocity vector $\vec{v}$, which can be represented in terms of the components of four-vector as $\vec{v} = d\vec{x}/dx_0$.  
In contrast with the pseudo-Euclidean four-velocity space, the relativistic three-velocity space is a three-dimensional space with constant negative curvature, i.e. three-dimensional space with Lobachevsky geometry. 
It is well-known that for a rectilinear motion, the  restriction  $\vec{p}  = m\vec{v}/\sqrt{1 - v^2/c^2}$ holds. We claim that this restriction does not hold when we dealing with a particle accelerating along a curved trajectory.
In this section we will investigate in detail the reason why this is the case.

\subsection{Mistake in commonly used method of covariant particle tracking}

In the  non-covariant (3+1) approach, the solution of the dynamics problem in the lab frame makes no reference to Lorentz transformations. This means that, for instance, within the lab frame the motion of particles in constant magnetic field looks precisely  the same as predicted  by Newtonian kinematics: relativistic effects do not have a place in this description. In conventional particle tracking  a particle trajectory  $\vec{x}(t)$ can be seen from the lab frame as the result of successive Galileo boosts that track the motion of the accelerated (in a constant magnetic field) particle. The usual Galileo rule for addition of velocities is used to determine the Galileo boosts tracking a particular particle, instant after instant, along its motion along the curved trajectory.

In order to obtain relativistic kinematics effects, and in contrast to conventional particle tracking, one actually needs to solve the dynamics equation in manifestly covariant form by using the coordinate-independent proper time $\tau$ to parameterize the particle world-line in space-time. Relying on the geometric structure of Minkowski space-time, one defines the class of inertial frames and adopts a Lorentz frame with orthonormal basis vectors. 
Within the chosen Lorentz frame, Einstein's synchronization of distant clocks and Cartesian space coordinates are enforced. In the Lorentz lab frame (i.e. the lab frame with Lorentz coordinate system) one thus has a coordinate representation of a particle world-line as ($t(\tau), x_1(\tau), x_2(\tau), x_3(\tau)$). These four quantities basically are, at any $\tau$, components of a four-vector describing an event in space-time. Therefore, if one  chooses the lab time $t$ as a parameter for the trajectory curve, after inverting the relation $t = t(\tau)$, one obtains that the space position vector  of a particle in the Lorentz lab frame has the functional form $\vec{x}_{cov}(t)$. 
The trajectory $\vec{x}_{cov}(t)$ is viewed from the lab frame as the result of successive Lorentz transformations that depend on the proper time. In this  case relativistic kinematics effects arise. 
In view of the Lorentz transformation composition law, one will experience e.g.
the Einstein's rule of addition of velocities applies.

Attempts to solve the dynamics equation  Eq.(\ref{DDE}) in manifestly covariant form can be found in literature  (see e.g. \cite{RM,GF,CT}). 
It is general believe that the  integration from initial conditions of the four-dimensional covariant equation of motion Eq.(\ref{DDE})  gives the covariant  particle trajectory.
However, in such  approach the four equations of motion "split up" into (3+1) equations and similar to non-covariant approach we have no mixture of space and time parts of the dynamics equation Eq.(\ref{DDE}). 
The trajectory which is found in this way does not include relativistic kinematics effects.
Therefore, it cannot be identified with $\vec{x}_{cov}(t)$ even if, at first glance, it appears to be derived following covariant prescription.

Consider, for example, the motion of a particle in a given electromagnetic field. 
The simplest case, of great practical importance, is that of a uniform electromagnetic field meaning that  $F^{\mu\nu}$ is constant on the whole space-time region of interest. In particular we consider the motion of a particle in a constant homogeneous magnetic field, specified by tensor components $F^{\mu\nu} = B(e^{\mu}_2e^{\nu}_3 - e^{\nu}_2 e^{\mu}_3)$  where $e^{\mu}_2$ and $e^{\mu}_3$ are orthonormal space like basis vectors $e^2_2 = e^2_3 = - 1$, $e_2\cdot e_3 = 0$. In the lab frame of reference where $e^{\mu}_0$ is taken as the time axis, and   $e^{\mu}_2$ and $e^{\mu}_3$ are space vectors the field is indeed purely magnetic, of magnitude $B$ and parallel to the $e_1$ axis. Let us set the initial four-velocity $u^{\mu}(0) = \gamma c e^{\mu}_0 + \gamma v e^{\mu}_2$, where $v$ is the initial particle's velocity relative to the lab observer along the axis $e_2$ at the instant $\tau = 0$, and $\gamma = 1/\sqrt{1-v^2/c^2}$. The components of the equation of motion are then $du^{(0)}/d\tau = 0$, $du^{(1)}/d\tau = 0$, $du^{(2)}/d\tau = - eBu^{(3)}/(mc)$, $du^{(3)}/d\tau =  eBu^{(2)}/(mc)$. We seek for the initial value solution to these equations as  done in the existing literature. A distinctive feature of the initial value problem in relativistic mechanics, is that the dynamics is always constrained. In fact, the evolution of the particle is subject to $md u^{\mu}/d\tau = eF^{\mu\nu}u_{\nu}$, but also to the constraint $u^2 = c^2$. However, such a condition can be weakened requiring its validity at certain values of $\tau$ only, let us say initially, at $\tau = 0$. To prove this,
we calculate the scalar product between both sides of the equation of motion and $u_{\mu}$. Using the fact that  $F^{\mu\nu}$ is antisymmetric (i.e. $F^{\mu\nu} = - F^{\nu\mu}$), we find $u_{\mu}d u^{\mu}/d\tau = eF^{\mu\nu}u_{\mu}u_{\nu} = 0$. Thus, for the quantity $Y = (u^2 - c^2)$ we find $dY/d\tau = 0$. Therefore, if $Y(\tau)$ vanishes initially, i.e. $Y(0) = 0$, then $Y(\tau) = 0$ at any $\tau$. In other words, the differential Lorentz-force equation implies the constraint $u^2 = c^2$ once this is satisfied initially. Integrating with respect to the proper time we have  $u^{\mu}(\tau) = \gamma e^{\mu}_0 + \gamma v[e^{\mu}_2\cos(\omega\tau)
+ e^{\mu}_3\sin(\omega\tau)]$ where $\omega = eB/(mc)$. We see that $\gamma$ is constant with time, meaning that the energy of a charged particle moving in a constant magnetic field is constant. After two successive integrations we have $X^{\mu}(\tau) = X^{\mu}(0) + \gamma c\tau e^{\mu}_0 + R[e^{\mu}_2 \sin(\omega\tau) - e^{\mu}_3\cos(\omega\tau)]$ where $R = \gamma v/\omega$.  This enables us to find the time dependence $[0,X^{(2)}(t), X^{(3)}(t)]$ of the particle's position since $t/\gamma = \tau$.  From this solution of the equation of motion we conclude that the motion of a charged particle in a constant magnetic field is a uniform circular motion \cite{RM,GF,CT}.

One could expect that the particle's trajectory in the lab frame, following from the previous reasoning $[0, X^{(2)}(t), X^{(3)}(t)]$, should be identified with $\vec{x}_{cov}(t)$. However, paradoxical result are obtained by doing so. In particular, the trajectory $[0,X^{(2)}(t), X^{(3)}(t)]$  does not include relativistic kinematics effects.  In fact, the calculation carried out above shows that  $t/\gamma = \tau$  and one can see the connection between this dependence  and the absolute simultaneity convention. Here we have a situation where the temporal coincidence of two events has the absolute character:  $\Delta \tau = 0$ implies $\Delta t = 0$.

We found that the usual integration of the four-dimensional covariant equation of motion Eq.(\ref{DDE})  gives   particle trajectory which looks precisely the same as in Newton dynamics and kinematics. The trajectory of the electron does not include relativistic effects and the Galilean vectorial law of addition of velocities is actually used. The old kinematics is especially surprising, because we are based on the use of the covariant approach. So we must have made a mistake. We did not make a computational mistake in our integrations, but rather a conceptual one. 
We must say immediately that there is no objection to the first integration of Eq.(\ref{DDE}) from initial conditions over proper time $\tau$. With this, we find the four-momentum. The momentum has exact objective meaning i.e. it is convention-invariant. What must be recognized is that the concept of velocity  is only introduced in the second integration step. However,  in accepted covariant approach, the solution of the dynamics problem  for the momentum in the lab frame makes no reference to three-dimensional velocity.
In fact, the initial condition which we used is
$u^{\mu}(0) = \gamma c e^{\mu}_0 + \gamma v e^{\mu}_2$ and includes $\gamma c$ and  $\gamma v$, which are actually notations for the time and space parts of the initial four-momentum. 
The three-dimensional trajectory and respectively velocity, which are convention-dependent, are only found after the second integration step. Then,  where does the old kinematics comes from? 
The second integration was performed using the relation  $d\tau =dt/\gamma$.
It is only after we have made those replacement for $d\tau$ that we obtain the usual formula for conventional (non-covariant) trajectory for an electron in a constant magnetic field.

We should then expect to get results similar to those obtained in the case of the (3+1) non-covariant particle tracking. In fact, based on the structure of the four components of the equation of motion Eq.(\ref{DDE}),  we can  arrive to another mathematically identical formulation of the dynamical problem. The fact that the evolution of the particle in the lab frame is subject to a constraint has already been mentioned. This means that the mathematical form of the dynamics law includes only three independent equations of motion. It is easy to see from the initial set of four equations, $du^{(0)}/d\tau = 0$, $du^{(1)}/d\tau = 0$, $du^{(2)}/d\tau = - eBu^{(3)}/(mc)$, $du^{(3)}/d\tau =  eBu^{(2)}/(mc)$,
that the presentation of the time component simply as  the relation $d\tau = dt/\gamma$ between proper time and coordinate time is just a simple parametrization that yields the corrected Newton's equation Eq.(\ref{NN}) as another equivalent form of these four equations in terms of absolute time $t$ instead of proper time of the particle. This approach to integrating dynamics equations from the initial conditions relies on the use of three independent spatial coordinates and velocities without constraint and is intimately connected with old kinematics. 
The presentation of the time component simply as  the relation $d\tau = dt/\gamma$ between proper time and coordinate time is based on the hidden assumption that the type of clock synchronization, which provides the time coordinate $t$ in the lab frame, is based on the use of the absolute time convention.

\subsection{Covariant particle tracking}

We now want to describe the machinery of the covariant particle tracking.
We will consider a relativistic particle accelerating in the lab inertial frame, and we will analyze its evolution within the framework of special relativity. We will use the usual covariant approach.
The problem of assigning Lorentz coordinates to the lab frame in the case of accelerated  motion is complicated. The  permanent rest frame of the particle is obviously not inertial and any transformation of observations in the lab frame, back to the rest frame, cannot be made by means of Lorentz transformations. To get around that difficulty in the usual covariant approach one introduces an infinite sequence of co-moving frames. At each instant, the rest frame is a Lorentz frame  centered on the particle and moving with it. As the particle velocity changes to its new value at an infinitesimally later instant,  a new Lorentz frame centered on the particle and moving with it at the new velocity is used to observing the particle. All reference frames are assumed to be orthogonal. This ensemble of comoving coordinate systems or tetrads can be constructed by choosing, for each value of $\tau$ along the world line $\sigma$ of the particle, an inertial  system  whose origin coincides with $\sigma(\tau)$ and whose $x'_0$-axis is tangent to $\sigma$ at $\sigma(\tau)$. The zeroth basis vector $e'_0$ is therefore directed as the 4-velocity $u$.
In the tetrad basis $e'_i(\tau)$, the particle has four velocity $u = (c, 0, 0, 0)$ and four acceleration $a = (0,a_1,a_2,a_3)$.
The basis vectors of the tetrad  $e'_0(\tau), e'_1(\tau), e'_2(\tau), e'_3(\tau)$ at any proper time $\tau$ are then related to the basis vectors $e_0, e_1, e_2, e_3$ of some given inertial lab frame by a Lorentz transformation $e'_\mu(\tau) = \Lambda^\nu_\mu(\tau)e_\nu$. Therefore, the basis vectors at two successive instants must also be related to each other by a Lorentz transformation.

In the lab frame one thus has a coordinate representation of the world-line as  $\sigma(\tau) = (t(\tau), x_1(\tau), x_2(\tau), x_3(\tau))$. The covariant particle trajectory $\vec{x}_{cov}(t)$ is calculated by projecting world line  to the lab frame basis  and using the lab time $t$ as a parameter for the trajectory curve.
In this paper we claimed many times that there is a difference between the non-covariant particle trajectory $\vec{x}(t)$, calculated by solving the corrected  Newton's equations and the covariant particle trajectory $\vec{x}_{cov}(t)$, calculated by projecting the world line onto the  lab frame Lorentz basis. There is a fundamental reason for this difference. The trajectory $\vec{x}_{cov}(t)$ is viewed from the lab frame as the result of Lorentz transformations $\Lambda^\nu_\mu(\tau)$ that depend on the proper time. Therefore, the composition law that follows from the group properties of the Lorentz transformations  is used to express the conditions of co-moving sequence of frames tracking a particle. In contrast to this, $\vec{x}(t)$ follows from solving the corrected Newton's equations and does not include the composition law of Lorentz transformations.

As is known, the composition of non-collinear Lorentz boosts does not result in a different boost but in a Lorentz transformation involving a boost and a spatial rotation, the Wigner rotation. Suppose that our  particle moves along an arbitrary accelerated world line. As just discussed, the basis vectors of the tetrad defining the instantaneously co-moving frames is related to the basis vectors of the lab frame  by a  Lorentz transformation depending on the proper time $e'_\mu(\tau) = \Lambda^\nu_\mu(\tau)e_\nu$. The most general Lorentz transformation  $\Lambda^\nu_\mu(\tau)$  can be  uniquely separated into a pure Lorentz boost followed by spatial rotation. As seen from the lab frame, space vectors of the tetrad (those with indexes $\mu = 1,2,3$) rotate relative to  the Cartesian axes of the lab frame.

\subsection{An illustrative example of covariant single particle tracking}

Let us try out our algorithm for reconstructing $\vec{x}_{cov}(t)$ on some example, to see how it works.
An electron kicker setup is a practical case of study for illustrating the difference between covariant and non-covariant trajectories.
We have already discussed the kicker setup in Introduction. We assumed before that the kick angle 
was small compared to $1/\gamma$ and evaluated the transformations up to first order $\theta_k\gamma$.
Let us see what happens if we increase our accuracy.
We consider now the small expansion parameter $\gamma v_x/c \ll 1$, neglecting terms of order $(\gamma v_x/c)^3$, but not of order $(\gamma v_x/c)^2$. In other words, we use the second-order kick angle approximation.

\subsubsection{Demonstration that the equality $\vec{v}_{cov} = \vec{v}$  does not hold in general}

Let us start with non-covariant particle tracking calculations. The trajectory of the electron, which follows from the solution of the corrected Newton's second law under the absolute time convention, does not include  relativistic effects. Therefore, as usual for Newtonian kinematics, Galilean vectorial law of addition of velocities is actually used. Non-covariant particle dynamics shows that the electron direction changes after the kick, while the speed remains unvaried. According to non-covariant particle tracking, the magnetic field $B\vec{e}_y$ is only capable of altering the direction of motion, but not the speed of the electron. This is clearly true when considering the equations of motion for a single electron  $dv_x/dt = \omega_c v_z$, $dv_z/dt = -\omega_c v_x$, where the characteristic "cyclotron" frequency $\omega_c$ is defined by $\omega_c = eB/(m\gamma )$. This is a well defined initial value problem with initial condition $v_z = v, v_x = 0, v_y = 0$. We recognize the harmonic oscillator differential equation, hence the solution is $v_x = v\sin\omega_ct$, $v_z = v\cos\omega_ct$. 
After the kick, the beam velocity components are $(v_x,0,v_z)$, where $v_z = \sqrt{v^2 - v_x^2}$. 
Taking the ultrarelativistic limit $v \simeq c$ and using the second order approximation we get  $v_z = v [1- v_x^2/(2v^2)] = v [1- v_x^2/(2c^2)] =   v(1- \theta_k^2/2)$.

In contrast, covariant particle tracking, which is based on the use of Lorentz coordinates, yields different results for the velocity of the electron. 
Let us consider a composition of Lorentz transformations that track the motion of the relativistic electron accelerated by the kicker field. Let the $S$ be the lab frame of reference and $S'$ a comoving  frame with velocity $\vec{v}$ relative to $S$. Upstream of the kicker, the particle is at rest in the frame $S'$. In order to have this, we impose that $S'$ is connected to $S$ by the Lorentz boost $L(\vec{v})$, with $\vec{v}$ parallel to the $z$ axis, which transforms a given four vector event $X$ in a space-time into $X' = L(\vec{v})X$. We study what happens in $S'$ before the kick. Our particle is at rest and the kicker is running towards it with velocity $-\vec{v}$. The moving magnetic field of the kicker produces an electric field orthogonal to it. When the kicker interacts with the particle in $S'$ we thus deal with an electron moving in the combination of perpendicular electric and magnetic fields.  It is easy to see that the acceleration in the crossed fields yields an electron velocity  $v'_x =\gamma v_x$ parallel to the $x$-axis and $v'_z = - v(\gamma v_x/c)^2/2$ parallel to the $z$-axis. If we neglect terms in $(\gamma v_x /c)^3$, the relativistic correction in the composition of velocities does not appear in this approximation.

Let $S"$ be a frame fixed with respect to the particle downstream the kicker. As is known, the composition of non collinear Lorentz boosts does not result in a simple boost but, rather, in a Lorentz transformation involving a boost and a rotation. In our second order approximation we can neglect the rotation of the system $S"$  in the plane $(x',z')$ of the system $S'$. Therefore we can use a sequence of two commuting non-collinear Lorentz boosts linking $X'$ in $S'$ to $X''$ in $S''$ as $X"=L(\vec{e}_x v'_x) L(\vec{e}_z v'_z)X' =  L(\vec{e}_z v'_z)L(\vec{e}_x v'_x)X'$ in order to discuss the beam motion in the frame $S'$ after the kick. Here  $\vec{e}_x$ and $\vec{e}_z$ are unit vectors directed, respectively, along the $x$ and $z$ axis. Note that as observed by an observer on $S'$, the axes of the frame $S''$ are parallel to those of $S'$, and the axes of $S'$ are parallel to those of $S$. The relation $X" = L(\vec{e}_x v'_x)L(\vec{e}_z v'_z)L(\vec{e}_z v)X$ presents a step-by-step change from $S$ to $S'$ and then to $S"$. For the simple case of parallel velocities, the addition law is $L(\vec{e}_z v'_z)L(\vec{e}_z v) = L(\vec{e}_z v_z)$. Here $v_z = v(1 - \theta_k^2/2)$ and $\theta_k = v_x/v = v_x/c$ in our (ultrarelativistic) case of interest. The resulting boost composition can be represented as $X" = L(\vec{e}_x v'_x)L(\vec{e}_z v_z)X = L(\vec{e}_z v_z)L(\vec{e}_x v_x)X $. In the ultrarelativistic approximation $\gamma_z^2 = 1/(1 - v_z^2/c^2) \gg 1$, and one finds the simple result $v = v_z$, so that a  Lorentz boost with non-relativistic velocity $v_x$ leads to a rotation of the particle velocity $v_z$ of the angle $v_x/c$.


Note that we discuss particle tracking in the limit of a small kick angle $ \gamma v_x/c \ll 1$. However, even in this simple case and for a single electron we are able to demonstrate the difference between non-covariant and covariant particle trajectories.  The electron speed decreases from $v$ to $v(1-\theta_k^2/2)$. This result is at odds with the prediction from non-covariant particle tracking, because we used Lorentz transforms to track the particle motion. As a result, we track the particle in covariant way.

\subsubsection{Demonstration that the equality $\vec{p}_{cov} = \vec{p}$  holds} 

In our  relativistic but non-covariant study of electron motion in a given magnetic field, the electron has  the same velocity and consequently the same relativistic factor $\gamma$ upstream and downstream of the kicker. Suppose we now put the electron through a bending magnet (i.e. a uniform magnetic field directed along the $x$-axis ).
The motion in the bending magnet we obtained is practically the same as in the case of non-relativistic dynamics, the only difference being the appearance of the relativistic factor $\gamma$ in the determination of cyclotron frequency $\omega_c = eB/(m\gamma)$.  The curvature radius $R$ of the trajectory is derived from the relation $v_{\perp}/R  = \omega_c$, where $v_{\perp} = v(1 - \theta_k^2/2)$ is the component of the velocity normal to the field of the bending magnet $\vec{B} = B\vec{e}_x$. As a result, after the kick, the correction to the radius $R$  is only of order  $\theta_k^2$.

One could naively expect that according to covariant particle tracking, since the total speed of electron in the lab frame downstream of the kicker decreases from $v$ to $v(1 - \theta_k^2/2)$,  this would also lead to a consequent decrease of the three-momentum $|\vec{p}|^2$ from $m\gamma v$  to $m\gamma v(1 - \gamma^2\theta_k^2/2)$ in our approximation. However, such a momentum change would mean a correction to the radius $R$ of order $\gamma^2\theta_k^2$ so that there is a glaring conflict with the calculation of the raqdius according to non covariant tracking. Since the  curvature radius of the trajectory in the bending magnet  has obviously an objective meaning, i.e. it is convention-invariant, this situation seems paradoxical. The paradox is solved taking into account the fact that  in Lorentz coordinates the three-vector of momentum $\vec{p}$  is transformed, under Lorentz boosts, as the space part of the four vector $p_{\mu}$. Let us consider a composition of Lorentz boosts that track the motion of the relativistic electron accelerated by the kicker field. Under this composition of boosts the longitudinal momentum component remains unchanged in our approximation.

Let us verify that this assertion is correct. We have $p_{\mu} = [\mathcal{E}/c,\vec{p}]$. We consider the Lorentz frame $S'$ fixed with respect to the electron upstream the kicker, and in the special case when electron is at rest $p_{\mu}' = [mc , \vec{0}]$. We turn focus on what happens in $S'$. Acceleration in the crossed kicker fields gives rise to an electron velocity $v_x' =  \gamma v_x$ parallel to the $x$-axis and $v_z' = - v(\gamma v_x)^2/2$ parallel to the $z$-axis. Downstream of the kicker the transformed four-momentum is $p_{\mu}' = [mc + m v_x'^2/(2c), mv_x', 0, mv_z']$, where we evaluate the transformation only up to the order $(\gamma v_x/c)^2$, as done above. We note that, due to the transverse boost, there is a contribution  to the time-like part of the four-momentum vector i.e. to the energy of the electron. In fact, the energy increases
from $mc^2$ to $mc^2 + m (\gamma v_x)^2/2$. We remind that  $S'$ is connected to the lab frame $S$ by a Lorentz boost. Now, with a boost to a frame moving at velocity $\vec{v} = - v\vec{e}_z$, the transformation of the longitudinal  momentum component, normal to  the magnetic field of the bend, is $p_z = \gamma(p_z' + vp_0'/c) = \gamma mv$. Therefore we can see that the  momentum component along the $z$-axis remains unchanged in our approximation as it must be. We also have, from  the transformation properties of four-vectors,  that the time component $p_0 = \gamma(p'_0 + vp'_z) = \gamma mc$ .

\subsubsection{Momentum-velocity relation}

Let us now return to our consideration on the covariant  electron trajectory calculation in the Lorentz lab frame when a constant magnetic field is applied. We analyzed a very simple (but very practical) kicker setup and we noticed that, in fact, the three-momentum  is not changed; so we have already verified that  this transformation is  the same as the non covariant transformation for the three-momentum, i.e. $\vec{p}_{cov} = \vec{p}$. We also found that there is a difference between covariant and non covariant output velocities, $v_{cov} < v$.  In these transformations we therefore discovered that  $\vec{p}_{cov}  \neq m\vec{v}_{cov}/\sqrt{1 - v^2_{cov}/c^2}$ for curved  trajectory in ultrarelativistic asymptotic. It is interesting to discuss what it means that there are two different (covariant and non covariant) approaches that  produce the same particle three-momentum.  The point is that both approaches describe correctly the same  physical reality and the curvature radius of the trajectory in the magnetic field (and consequently the three-momentum) has obviously an objective meaning, i.e. is convention-invariant. In contrast to this, the velocity of the particle has objective meaning only up to a certain accuracy, because the finiteness of velocity of light takes place.

From the theory of relativity follows that the equation $\vec{p}_{cov}  = m\vec{v}_{cov}/\sqrt{1 - v^2_{cov}/c^2}$  does not hold for a curved trajectory.
Many experts who learned the theory of relativity using  textbooks will find this statement disturbing at first sight.
First of all, it is  well known that for rectilinear motion the equation $\vec{p}_{cov}  = m\vec{v}_{cov}/\sqrt{1 - v^2_{cov}/c^2}$ holds. How can it be that for  motion along a curved trajectory the usual momentum-velocity relation does not hold? This essential point has never received attention by the physical community.

The situation can be described quite naturally in the following way.
The equations of a particle's motion in three-dimensional space  Eq.(\ref{DDDE1}) and Eq.(\ref{DFE1}) are not a mathematical result, derived from the covariant equation Eq.(\ref{DDE}). In these equations the restriction $m dx_\mu/d\tau  = (\mathcal{E}/c,\vec{p}) =  (c\gamma,v\gamma)$ has already been imposed: it is in the assumption that we are working in three-dimensional momentum representation $\vec{p}_{cov}  = m\vec{v}_{cov}/\sqrt{1 - v^2_{cov}/c^2}$. 
We showed that instant after instant, the trajectory  $\vec{x}_{cov}(t)$ is viewed from the Lorentz lab frame as a result of successive infinitesimal Lorentz transformations. As we see, in Lorentz coordinates the lab time $t$ in the equation of motion  cannot be independent from space variables.  This is because resynchronization of distant clocks according to the relativity of simultaneity in the process of particle acceleration leads to a mixture of positions and time.

It is well known that for the rectilinear motion the combination of the usual momentum-velocity relation and the covariant three-velocity transformation 
(according to Einstein's addition velocity law) is consistent with the covariant three-momentum transformation and both  (non-covariant and covariant) approaches produce the same trajectory\footnote{Let us examine the transformation of the three velocity in the theory of relativity. For a rectilinear motion it is performed in accordance with the following equation:
$v = (v'+V)/(1 + v'V/c^2)$. The "summation" of two velocities is not just the algebraic sum of two velocities, but it is "corrected" by $(1 + v'V/c^2)$. 
The relativistic factor $1/\sqrt{1 - v^2/c^2}$ is given by the following expression:
$1/\sqrt{1 - v^2/c^2} = (1 + v'V/c^2)/(\sqrt{1 - v'^2/c^2}\sqrt{1 - V^2/c^2})$. The new momentum is  then simply $mv$ times the above expression. But we want to express the new momentum in terms of the primed momentum and energy, and we note that $p = (p' + \mathcal{E}'V/c^2)/\sqrt{1 - V^2/c^2}$. Thus, for a rectilinear motion, the combination of  Einstein addition law for parallel velocities and the usual momentum-velocity relation is consistent with the covariant three momentum transformation }. But this result was incorrectly extended to an arbitrary trajectory. 
Like it happens with  the composition of Galilean boosts, collinear Lorentz boosts commute.
This means that the resultant of successive collinear Lorentz boosts is independent of which transformation applies first. On the contrary, Lorentz boosts in different directions do not commute. 
A comparison with the three-dimensional Euclidean space might help here. 
Spatial rotations do not  commute either. However, also for spatial rotations there is a case where the result of  two successive transformations is independent of their order: that is, when we deal  with rotation around the same axis. While the successive application of two Galilean boosts is Galilean boost and the successive application of two rotations is a rotation,  the successive application of two non-collinear Lorentz boosts is not a Lorentz boost. The composition of non-collinear boosts will results to be equivalent to a boost, followed by spatial rotation, the Wigner rotation. The  Wigner rotation is relativistic effect which has no a non-covariant analogue. One of the consequences of non-commutativity of non-collinear Lorentz boosts is the unusual momentum-velocity relation
$\vec{p}_{cov}  \neq m\vec{v}_{cov}/\sqrt{1 - v^2_{cov}/c^2}$, which is also has no a non-covariant analogue.

This is a good point to make a general remark about the unusual momentum-velocity relation discussed above,  and Wigner rotations. The theory of relativity shows that both effects have to do with the effects of acceleration in curved trajectories. But what we can say about relationship between rotation and change in velocity? One could naively expect that a Wigner rotation is a rotation in the ordinary space  and that this would not lead to a change of the three-dimensional velocity vector. In fact, the three-dimensional vector is a geometric object, and it is invariant under rotations in ordinary space. However, it can be shown that this assertion is incorrect. Just to give a slight hint as to how that happens, it should be note that
a Wigner rotation describes the rotation of the axes of a moving reference frame which is observed in the lab frame. But how to measure this orientation? A moving  coordinate system changes its position in time. 
We can only specify a method for measuring the orientation of the axes of a moving reference frame if we have adopted a method of timing events at distance. It has already been pointed that the Wigner rotation results directly from the relativity of simultaneity, which  is related with the time shift $ t \to t+ xv_x/c^2$ along $x$-axis of the lab frame. 
Once we recognize the presence of this time shift, we see that there is also the time shift along the velocity direction after the kick and this  projection is proportional to  $\theta_k^2$.  It is not hard to prove that this extra time shift is equivalent to a velocity change:  $\Delta v/v \sim \theta_k^2$. 
We may point out  that the Wigner rotation and the unusual momentum-velocity relation can be regarded as the two sides of the same coin: they are manifestations of the  mixture of positions and time.

\subsubsection{Trajectory and path}

So far we have considered the motion of a particle in three-dimensional space using the vector-valued function $\vec{x}(t)$. We have a prescribed curve (path) along which the particle moves. The motion along the path is described by $l(t)$, where $l$ is a certain parameter (in our case of interest the length of the arc).  
Note the difference between the notions of path and trajectory \cite{JA}. The trajectory of a particle conveys more information about its motion because every position is described additionally by the corresponding time instant. The path is rather a purely geometrical notion. Complete paths or their parts may consist of, e. g., line segments, arcs, circles, helical curves. If we take the origin of the (Cartesian) coordinate system and we connect the point to the point laying on the path and describing the motion of the particle, then the creating vector will be a position vector $\vec{x}(l)$. The derivative of a vector is the vector tangent to the curve described by the radius vector $\vec{x}(l)$. The sense of the $d\vec{x}(l)/dl$ is determined by the sense of the curve arc $l$.  

We want now to describe  how to determine the position vector   $\vec{x}(l)_{cov}$ in covariant particle tracking. We consider the motion in an uniform magnetic field with zero electric field.
Using the Eq.(\ref{DDE}) we obtain

\begin{eqnarray}
	&& \frac{d\vec{p}}{d\tau} =  e \vec{p}\times \vec{B}, ~ ~\frac{d\mathcal{E}}{d\tau} = 0 ~ ~ .
\end{eqnarray}

From $d\mathcal{E}/d\tau = 0$ and constraint Eq.(\ref{DCFE1})  we have $dp/d\tau = 0$, where $p = |\vec{p}|$.
The unit vector $\vec{p}/p$ can be described by the following equation $\vec{p}/p =  d\vec{x}_{cov}/|d\vec{x}_{cov}| = d\vec{x}_{cov}/dl$,
where $|d\vec{x}_{cov}| = dl$ is the differential of the path length.
From foregoing consideration it follows that

\begin{eqnarray}
&& \frac{d^2\vec{x}_{cov}}{dl^2} =  \frac{d\vec{x}_{cov}}{dl}\times \left (\frac{e\vec{B}}{p}\right)~ .
\end{eqnarray}

These three equations corresponds exactly to the equations for components of the position vector that can be found using the non-covariant particle tracking approach. Then $\vec{x}(l)_{cov}$ is exactly equal to $\vec{x}(l)$ as it must be. The path $\vec{x}(l)$ has exact objective meaning  i.e. it is convention-invariant. In contrast to this, and consistently with the conventionality intrinsic in the velocity, the trajectory $\vec{x}(t)$ of the particle is convention dependent and has no exact objective meaning. 
We should also notice  that a uniform magnetic field can be used in making a "momentum analyzer" for high-energy charge particles. It must be recognized that this method for determining the particle's momentum is convention-independent.

\section{Relativity and electrodynamics}

Going to the electrodynamics problem, the differential form of  Maxwell's equations  describing electromagnetic phenomena in the Lorentz lab frame (in cgs units)  is given by the following expressions:

\begin{eqnarray}
	&& \vec{\nabla}\cdot \vec{E} = 4 \pi \rho~, \cr && \vec{\nabla}\cdot
	\vec{B} = 0~ , \cr && \vec{\nabla}\times \vec{E} =
	-\frac{1}{c}\frac{\partial \vec{B}}{ \partial t}~,\cr &&
	\vec{\nabla}\times \vec{B} = \frac{4\pi}{c}
	\vec{j}+\frac{1}{c}\frac{\partial \vec{E}}{\partial t}~. \label{Max}
\end{eqnarray}

Here the charge density $\rho$ and current density $\vec{j}$ are written as

\begin{eqnarray}
	&&\rho(\vec{x}, t) = \sum_{n} e_n\delta(\vec{x} - \vec{x}_n(t)) ~,\cr &&
	\vec{j}(\vec{x},t) = \sum_{n} e_n\vec{v}_n(t)\delta(\vec{x} - \vec{x}_n(t))~, \label{CD}
\end{eqnarray}

where $\delta(\vec{x} - \vec{x}_n(t))$ is the three-dimensional delta function, while $m_n, e_n, \vec{x}_n(t)$, and $\vec{v}_n =   d\vec{x}_n(t)/dt$ denote respectively the rest mass, charge, position, and the velocity of the $n$th particle involved in the electrodynamic process.
To evaluate  radiation fields arising from an external sources in Eq. (\ref{CD}), we need to know the velocity $\vec{v}_n$ and the position $\vec{x}_n$ as a function of the lab frame time $t$. As discussed above, it is generally accepted  that one should solve the usual Maxwell's equations in the lab frame with current and charge density created by particles moving along non-covariant trajectory like $\vec{x}_n(t)$.
The trajectory   $\vec{x}_n(t)$, which follow from the solution of the corrected Newton's second law under the absolute time convention, does not include, however, relativistic effects.

In our previous publications \cite{OURS1,OURS2,OURS3,OURS4,OURS5,OURS6,OURS7} we argued that
this algorithm for solving usual Maxwell's equations in the lab frame, which is considered in all standard treatments as relativistically correct, is at odds with the principle of relativity. 
However, the usual Maxwell's equations in the lab frame, Eq. (\ref{Max}), are compatible only with covariant trajectories calculated by using Lorentz coordinates, therefore including relativistic kinematics effects.

The covariant particle trajectory $\vec{x}_{cov}(t)$ is calculated by projecting the corresponding world line  to the lab frame basis  and using the lab time $t$ as a parameter for the trajectory curve.
The charge and current densities Eq. (\ref{CD}), must be written as 4-vector current  by representing charge world line in Lorentz lab frame

\begin{eqnarray}
	x_{\mu}(\tau) = [t(\tau), x_1(\tau), x_2(\tau), x_3(\tau)]  ~ ,\label{WL}
\end{eqnarray}

and integrating over proper time with an appropriate additional delta function. Thus

\begin{eqnarray}
	&& j_{\mu}(x) = ec\int d\tau  u_{\mu}(\tau) \delta^4(x - x(\tau)) ~ ,\label{CA}
\end{eqnarray}

where charge 4-velocity $u_{\mu}(\tau) = dx_\mu/d\tau$.
The integration over the proper time of $\tau$ leads to

\begin{eqnarray}
	&& j_{\mu}(\vec{x},t) = eu_{\mu}(t)\delta^3(\vec{x} - \vec{x}_{cov}(t)) ~ ,\label{CL}
\end{eqnarray}

Thus we obtain

\begin{eqnarray}
	&&\rho(\vec{x}, t) =  e\delta(\vec{x} - \vec{x}_{cov}(t)) ~,\cr &&
	\vec{j}(\vec{x},t) = e\vec{v}_{cov}(t)\delta(\vec{x} - \vec{x}_{cov}(t))~, \label{CD1}
\end{eqnarray}

where $\vec{v}_{cov} = d\vec{x}_{cov}/dt$.

It is generally believed that the usual momentum-velocity relation $\vec{p}_{cov}  = m\vec{v}_{cov}/\sqrt{1 - v^2_{cov}/c^2}$ holds for any arbitrary world-line $x(\tau)$. Let us present a typical textbook statement \cite {SCHE} concerning the projection of an arbitrary world line onto the Lorentz lab frame basis: "A charged point particle moving along the world line $x(\tau)$, $\tau$ being proper time, within the framework of Special Relativity has the velocity $u(\tau) = dx(\tau)/d\tau = (\gamma c, \gamma\vec{v})$. The four-velocity is normalized such that its invariant squared norm equals $c^2$, $u^2 = c^2\gamma^2(1-\beta^2) = c^2$. While $x(\tau)$ and $u(\tau)$ are coordinate-free definitions the decomposition $u = (c\gamma, \vec{v}\gamma)$ presupposes the choice of a frame of reference $K$.      
The particle, which is assumed to curry the charge $e$, creates the current density   
$j(x) = ec\int d\tau  u(y) \delta^4(y - x(\tau))$. This is a Lorentz vector. [...] Furthermore, in any frame of reference $K$, one recovers the expected expressions for the charge and current densities by integrating over $\tau$ by means of relation $d\tau = dt'/\gamma$ between proper time and coordinate time and using the formula $\delta(y_0 - x_0(\tau)) = \delta(ct - ct') = \delta(t-t')/c$,    
$j_0(t,y) = ce\delta^{(3)}(y-x(t)) \equiv c\rho(t,y)$, $j^i(t,y) = ev^i(t)\delta^{(3)}(y-x(t))$, $i = 1,2,3$." We state that this incorrect and misleading. In fact, as we have already discussed in the previous section, the four-velocity cannot be decomposed into  $u = (c\gamma, \vec{v}\gamma)$  when we deal with a particle accelerating along a curved trajectory in the Lorentz lab frame.


One of the consequences of non-commutativity of non-collinear Lorentz boosts is the unusual momentum-velocity relation. In this case there is a difference between  covariant and non covariant  particle trajectories. One can see that this essential point has never received attention by the physical community.
As a result, a correction of the conventional radiation theory is required.

\subsection{Radiation emitted by a single electron}

We will be interested in the case of an ultra-relativistic electron going through a certain magnetic system. We will discuss of a bending magnet and undulator in order to illustrate our reasoning, but the considerations in this section are fully general, and apply to any magnetic system. Radiation theory is naturally developed in the space-frequency domain, as one is usually interested in radiation properties at a given position in space and at a certain frequency. In this paper we define the relation between temporal and frequency domain via the following definition of Fourier transform pair:

\begin{eqnarray}
&&\bar{f}(\omega) = \int_{-\infty}^{\infty} dt~ f(t) \exp(i \omega t ) \leftrightarrow
f(t) = \frac{1}{2\pi}\int_{-\infty}^{\infty} d\omega \bar{f}(\omega) \exp(-i \omega t) ~.
\label{ftdef2}
\end{eqnarray}

Suppose we are interested in the radiation generated by an electron and observed far away from it. In this case it is possible to find a relatively simple expression for the electric field \cite{JACK}. We indicate the electron velocity in units of $c$ with $\vec{\beta}$, the electron trajectory in three dimensions with $\vec{R}(t)$ and the observation position with $\vec{R}_0$. Finally, we introduce the unit vector

\begin{equation}
\vec{n} =
\frac{\vec{R}_0-\vec{R}(t)}{|\vec{R}_0-\vec{R}(t)|}
\label{enne}
\end{equation}
%

pointing from the retarded position of the electron to the observer. In the far zone, by definition, the unit vector $\vec{n}$ is nearly constant in time. If the position of the observer is far away enough from the charge, one can make the expansion

\begin{eqnarray}
\left| \vec{R}_0-\vec{R}(t) \right|= R_0 - \vec{n} \cdot \vec{R}(t)~.
\label{next}
\end{eqnarray}
We then obtain the following approximate expression for the the radiation field  in the space-frequency domain\footnote{For a better understanding of the physics involved one can refer to e.g. the textbook \cite{JACK}. A different constant of proportionality in Eq. (\ref{revwied}) is to be ascribed to the use of different units and definition of the Fourier transform.}:

\begin{eqnarray}
\vec{\bar{E}}(\vec{R}_0,\omega) &=& -{i\omega e\over{c
		R_0}}\exp\left[\frac{i \omega}{c}\vec{n}\cdot\vec{R}_0\right]
\int_{-\infty}^{\infty}
dt~{\vec{n}\times\left[\vec{n}\times{\vec{\beta}(t)}\right]}\exp
\left[i\omega\left(t-\frac{\vec{n}\cdot
	\vec{R}(t)}{c}\right)\right] \cr &&  \label{revwied}
\end{eqnarray}
where $\omega$ is the frequency, $(-e)$ is the negative electron charge and we make use of Gaussian units.

\subsection{Multipole expansion}

First we will limit our consideration to the case of sources moving in a non-relativistic fashion.
According to the principle of relativity, usual Maxwell's equations Eq. (\ref{Max})  can always  be used in any Lorentz frame where sources are at rest. The same considerations apply where sources are moving in non-relativistic manner. In particular, when oscillating, charge particles emit radiation, and in the non-relativistic case, when the velocities of oscillating charges $v_n \ll c$, dipole radiation will be generated and described with the help of the Maxwell's equations in their usual form,   Eq. (\ref{Max}). 

Let's examine in a more detail how the dipole radiation term comes about.
The time  $\vec{R}(t)\cdot(\vec{n}/c)$ in the integrands of the expression for the radiation field amplitude, Eq. (\ref{revwied}), can be neglected in the cases where the trajectory of the charge changes little during this time. It is easy to find the conditions for satisfying this requirement.  Let us denote by $a$ the order of magnitude of the dimensions of the system. Then the time  $\vec{R}(t)\cdot(\vec{n}/c) \sim a/c$. In order to ensure that the distribution of the charges in the system does not undergo a significant change during this time, it is necessary that $a \ll \lambda$. Thus, the dimensions of the system must be small compared to radiation wavelength. This condition can be written in the in still another form $v \ll c$, where $v$ is of the order of magnitude of the velocities of the charges. 

We consider the radiation associated with the first order term in the expansion of the  Eq. (\ref{revwied}) in power of $\vec{R}(t)\cdot(\vec{n}/c)$. 
In doing so, we neglected all information about the electron trajectory  $\vec{R}(t)$.
In this dipole approximation the electron orbit scale is always much smaller than the radiation wavelength
and  Eq. (\ref{revwied}) gives fields very much like the instantaneous theory. 

Now we consider the radiation associated with the succeeding terms in the expansion of the field amplitude Eq. (\ref{revwied}) in powers of  $\vec{R}(t)\cdot(\vec{n}/c)$ i.e. in the power of the ratio $v/c$. Since $v/c$ is assumed to be small, these terms are  small compared with the first (dipole) term.
Thus, the total radiation consists of independent parts; they are called dipole, quadrupole, octupole terms etc.

Although this looks rather complicated, the result is easily interpreted. 
In accounting only for the dipole part of the radiation we neglect all information about the electron trajectory.
Therefore, one should not be surprised to find that  there is no influence of the difference between the non-covariant and covariant electron trajectories on the electromagnetic dipole radiation. 
But that is only the first term. The other terms tell us that there are corrections to the dipole radiation approximation. The calculation of this correction  requires detailed information about the electron trajectory. Obviously, in order to calculate the correction to the dipole radiation, we will have to use the covariant trajectory and not be satisfied with the non-covariant approach.

\section{Lorentz and Galilean transformations in  electrodynamics}

\subsection{Operational interpretation of Lorentz and absolute time coordinatizations}

The fundamental laws of electrodynamics are expressed by Maxwell's equations, according to which light propagate with the same velocity $c$ in all directions. This is because Maxwell's theory has no  intrinsic anisotropy. It had been stated that in their original form, Maxwell's equations are valid in any inertial frame. However, Maxwell's equations can be written down only if the space-time coordinate system has already been specified. 

We want to consider a relativistic particle, accelerating in a  lab inertial frame, and we want to analyze its radiation within the framework of special relativity.
The problem of assigning Lorentz coordinates  to the lab frame in the case of accelerated motion is complicated. We would like to start with the simpler question of how to assign space-time coordinates to an inertial lab frame, where a dipole source of light is at rest.  
The theory of relativity offer a procedure of clocks synchronization based on the constancy of light speed in inertial frames (Einstein synchronization).

Suppose we have a dipole radiation source. When the light source is at rest, the fields equations are constituted by the usual Maxwell's equations and Einstein synchronization is defined in terms of light signals emitted by the (dipole) source at rest, assuming that light propagate with the same velocity $c$ in all directions.  Using  Einstein synchronization procedure in the rest frame of  the dipole, we actually select the Lorentz coordinate system. In this coordinate system the metric has Minkowski form.

Now we consider the  acceleration of the light source  in the lab frame from rest up to velocity $v$ along the $x$-axis. The influence  of the uniform translational motion of the source along the  $x$-axis on the (dipole) radiation emission can be described purely kinematically. Most important in the study of a moving emitter of light  is the synchronization of the clocks  at rest in the lab frame. The simplest method of synchronization consists in keeping the same set of uniformly scattered already synchronized clocks without any changes.

It is clear that such  synchronization preserve simultaneity and is actually
based on the absolute time (simultaneity) convention. After a boost along the $x$ axis, the Cartesian coordinates of emitter transform as $x' = x-vt, ~ y' = y, ~ z' = z$. This transformation completes with the invariance in simultaneity, $\Delta t' = \Delta t$. The absolute character of the temporal coincidence between two events is a consequence of the as well absolute concept of time $t' = t$. As a result of the boost, the transformation of the time and coordinates of any event has the form of a Galilean transformation.   That is, applying a Galilean transformation, we obtain the not orthogonal metric Eq.(\ref{GGG3}).
From the above we conclude that the coordinate velocity of light from the moving emitter  in the lab frame  is dependent of the relative velocity between emitter and observer.
The speed of light is compatible with the Galilean  law  of addition  of velocities.
The reason  why the velocity of light is different from the electrodynamics constant $c$
is due to the fact that the clocks are synchronized by the absolute time convention.
The coordinate velocity of a light parallel to the x-axis  $dx/dt$ is given as follows: $dx/dt = c + v$ in the positive direction, $dx/dt = -c + v$  in the negative direction.

In agreement with the principle of relativity, usual Maxwell's equations can  be exploited in a moving inertial frame where sources are at rest. However, the transformation connecting two inertial frames with absolute time synchronization is a Galilean transformation, and  Maxwell's equations do not remain form-invariant with respect to Galilean transformations.
As a result, without changing  synchronization in the lab frame, after the boost we have a much more complicated situation for the electrodynamics of moving sources compared to the usual one.
The main difference consists in the crossed terms $\partial^2/\partial t\partial x$ which arises in d'Alembertian from the non-diagonal component of the metric tensor $g_{01} = v/c$. To get around this difficulty, we observe that metric Eq.(\ref{GGG3}) can always be simplified. The trick needed here is to make a change of the time and spatial  variables. In the new variables Eq.(\ref{GG7})  we obtain metric in the usual Minkowski form. Obviously, transforming to new variables leads to the usual Maxwell's equations and we have standard electrodynamics of moving sources.

A coordinate system endowed with diagonal metric is called, as already said, a Lorentz coordinate system. So, from  an operational point of view, the new  coordinates in the lab frame after the clocks resynchronization are impeccable. However, from the theory of relativity we know that if we wish to assign Lorentz coordinates to an inertial lab frame, the synchronization must be defined in terms of light signals. The following important detail of such synchronization  can hardly be emphasized enough.  If the source of light is in motion, we see that the procedure
for distant clocks synchronizing must be performed by using a moving light source.
The constant value of $c$ for the speed  of light emitted by the moving source destroys the  simultaneity introduced by light signals emitted by the (dipole) source at rest. The coordinates reflecting the constant speed of light $c$ from a moving source
are Lorentz coordinates for that particular source.

Consider now two light sources the 1, 2 say. Suppose that in the lab frame the velocities of 1,2 are $\vec{v}_1$, $\vec{v}_2$ and $\vec{v}_1 \neq \vec{v}_2$. The question now arise how to assign a time coordinate to the lab reference frame. We have a choice between an absolute time coordinate and a Lorentz time coordinate. The most natural choice, from the point of view of connecting to the laboratory reality, is the absolute time synchronization. In this case simultaneity is absolute, and for this we should prepare, for two sources, only one set of synchronized clocks in the lab frame. On the other hand, Maxwell's equations are not form-invariant under Galilean transformations, that is, their form is different on the lab frame. In fact, the use of the absolute time convention, implies the use of much more complicated field equations, and these equations are different for each source.
Now we are in the position to assign Lorentz coordinates.
The only possibility to introduce Lorentz coordinates in this situation consists in introducing individual coordinate systems  (i.e. individual set of clocks) for each source.
It is clear that if operational methods are at hand to fix the coordinates (clock synchronization in the lab frame) for the first source, the same methods can be used to assign values to the coordinates for the second source and these will be two different Lorentz coordinate systems.
It should have been made clear that Lorentz coordinate systems exist only in our mind
and manipulations with non existing clocks are an indispensable prerequisite for the application of the usual Maxwell's equations for  moving light sources.

\subsection{Optical phenomena and the Galilean coordinate transformations}

Light is described by electromagnetic field theory, Maxwell theory. The Maxwell theory meets all requirements of the theory of relativity and, therefore, must accurately describe the properties of such a relativistic object as light. 
In the microscopic approach (i.e. in the approach which based on the way the field behave dynamically) to optical phenomena,  Einstein and absolute time   synchronization conventions give the same result for any convention-invariant phenomena,  and it does not matter which transformation (Galilean or Lorentz) is used.

\subsubsection{The aberration of light}

We are now in the position to understand a  number of interesting optical phenomena in the framework of the electromagnetic field theory, based on the use of the absolute time convention.
For example, consider the effect of light aberration, that is a change in the direction of light propagation ascribed to boosted light sources. We will describe the effect of aberration of light by working only up to the first order  $v/c$. However, even in this simple non relativistic example we are able to demonstrate that Galilean transformations do not leave Maxwell's equations unchanged. When a Galilean transformation of Maxwell's equations is tried, the new terms that have to be put into the electromagnetic field equations lead to the effect of aberration of light. It does not matter which convention  and hence transformation is used to describe the same reality.

The explanation of the effect of aberration of light presented in well-known textbooks is actually based on the use of a Lorentz boost (i.e. of relativistic kinematics) to describe how the direction of a light ray depends on  the velocity of the light source relative to the lab frame. Let us discuss a special case of the aberration of a horizontal light ray. Suppose that a light source, studied in the comoving frame $S'$, radiates a plane wave along the $z$-axis. Now imagine what happens in the lab frame,  where the source  is moving with constant speed $v$ along the $x$-axis. The transformation of observations from the lab frame  with Lorentz coordinates to the co-moving Lorentz frame is described by a transverse  Lorentz boost. On the one hand, the wave equation remains invariant with respect to Lorentz transformations. 
On the other hand, if make a Lorentz boost, we automatically introduce a time transformation $t' = t - xv/c^2$ and the effect of this transformation is just a rotation of the wavefront in the lab frame. This is because the effect of this time transformation is just a dislocation in the timing of processes, which has the effect of rotating the plane of simultaneity on the angle $v/c$ in the first order approximation. In other words, when a uniform translational motion of the source is treated according to Lorentz transformations, the aberration of light effect is described in the language of relativistic kinematics. In fact, the relativity of simultaneity is a  relativistic effect that appears also in the first order in $v/c$.

It should be noted, however, that there is another satisfactory way of explaining  the effect of aberration of light. The explanation consists in using a Galileo boost to describe the  uniform translational motion of the light source in the lab frame. After the Galilean transformation of the wave equation we come to the conclusion that the crossed term described above yields an aberration angle $v/c$. 
It could be said that the crossed term generates anisotropy in the lab frame that is responsible for the change of radiation direction (aberration).
In fact,  in order to eliminate the crossed term in the transformed wave equation, we can make a change of the time variable. After both Galilean coordinate transformation and time shift we obtain the wave equation in "diagonal" form, i.e. without crossed terms. The time shift results in a slope  of the plane of simultaneity. Then, the electromagnetic waves are radiated at the angle $v/c$, yielding the phenomenon of light aberration: the two approaches, treated according to Einstein's or absolute time synchronization conventions give the same result. The choice between these two different approaches is a matter of pragmatics.

\subsubsection{A moving source and stationary mirror}

Because of our usage of Galilean transformations within electrodynamics we have some apparent paradoxes. An analysis of paradoxes leads to a better understanding of the four-dimensional geometrical significance of the concepts of space and time in the theory of relativity.

The peculiarity of the kinematic consequence of using Galilean transformations is that the speed of light emitted by a moving source depends on the relative velocity between source and observer.
A widespread theoretical argument used to support the incorrectness of Galilean transformations is the conclusion that a Galilean transformation of the velocity of light is not consistent with the electron-theoretical explanation of reflection and refraction. 

This idea is a part of the material in well-known books. For instance,
in his famous review  \cite{PA}, Pauli pointed out that if we consider a moving source and  a stationary mirror, the incident light wave with its velocity $c+v$ and a wave scattered by the dipoles of the mirror with their different velocity cannot interfere as required by the electron theory of dispersion since their velocity are different. To quote Pauli \cite{PA}
"[...] it is essential that the spherical waves emitted by the dipoles in the body should interfere with the incident wave. If we now think of the body as at rest, and the light source moving relative to it, then [...] the wave emitted by the dipoles will have velocity different from that of the incident wave. Interference is therefore not possible."   

This conclusion is incorrect. 
It is clear that an incident wave with a certain frequency, no matter what its velocity, excites the electrons of a mirror into oscillations of the same frequency. They then emit radiation with the same frequency. Thus, the incident and scattered wave at any given point have the same frequency and can interfere . The effect of the different velocities is to produce a relative phase which varies with position in space. This, according to well known ideas, affects the velocity and amplitude envelope  of the single wave which results from the superposition of the two separate waves \cite{FO}. 

The following simple analysis confirms these ideas. Let $\exp i(\omega t - kx)$ represent an  incoming wave whose velocity is $\omega/k = c + v$. Similarly, let  $\exp i(\omega t + k'x + \phi)$ represent another  out-coming (scattered radiation) wave of the same amplitude and the same frequency, a different velocity $\omega/k' = c - v$ and different phase.  The superposition of these two waves is represented by  $\exp i(\omega t - kx)  +  \exp i(\omega t + k'x + \phi) = 
	2[\cos[(k+k')x/2 +\phi/2]]\exp i[\omega t - (k-k')x/2 +\phi/2]$. There is the cosine factor representing an amplitude envelope which is stationary in space and whose periodicity is inversely proportional to the difference in the propagating constants $k$ and $k'$ of the two component waves.  This can be written in a simpler form $2[\cos[\omega x/[c(1-v^2/c^2)] +\phi/2]]\exp i[\omega t - xv\omega/[c^2(1-v^2/c^2)]  +\phi/2]$.
	
Suppose that the source at rest is emitting waves at frequency $\omega_0$.	In the lab frame  after the Galilean transformation the velocity of incoming wave is $c+v$. Thus if $\omega_0$ is the natural frequency, the modified frequency would be  $\omega = \omega_0/[1 - v/(c+v)]$.
Therefore the observed in the lab frame frequency is $\omega = \omega_0(1+v/c)$.  The shift in frequency observed in the above situation is the well known  Doppler effect. Our  equation for superposition of two waves now looks like
$2[\cos[\omega_0 x/[c(1-v/c)] +\phi/2]]\exp i[\omega_0(1+v/c) t - xv\omega_0/[c^2(1-v/c)]  +\phi/2]$.

Suppose that an observer in the laboratory performs the standing wave measurement.
We should examine what parts of the measured data depends on the choice of synchronization convention and what parts do not. Clearly, physical meaningful results must be convention-invariant.
We state that time oscillation  has no intrinsic meaning - its meaning only being assigned by a convention. In particular, one can see the connection between the time shift $xv\omega_0/[c^2(1-v/c)]$ in  $\exp i[\omega_0(1+v/c) t - xv\omega_0/[c^2(1-v/c)]  +\phi/2]$ and the issue of distant clock synchrony.
Note that the scale of time (frequency) is also unrecognizable from physical viewpoint.

Suppose we took an ordinary atom, which had a natural frequency $\omega_0$ at rest and we moved it toward the observer in the lab frame at speed $v$. In order to measure the velocity of the atom within the lab frame, the observer first has to  specify frequency (time) standard and  length standard and then has to synchronize distant clock. Let us suppose that the same atom at rest which has natural frequency 
$\omega_0$ is used as frequency standard. If we organize a standing wave by using (dipole) radiation
from atoms at rest with standard frequency, we can use the standing wavelength as a standard of length.

Suppose that distant clocks are synchronized by light signals by using  dipole (atom) radiation source at rest. It is also assumed that light from the source at rest propagate with the same velocity $c$ in all direction in the lab frame. 
Let us go back to our calculations of the speed of light from the moving source when the clocks in the lab frame are synchronized according to the procedure described above i.e. according to the absolute time convention. When  coordinates are assigned in the lab frame, the laboratory observer can directly measure the one-way speed of light. The result he observes is that the speed of light emitted by the moving source is consistent with the Galilean law of addition of velocities. In particular, when the source is moving with velocity $v$ along the $z$-axis, the velocity of light  in the direction parallel to the $z$-axis, is equal to $c+v$ in the positive, and $-c+v$ in the negative orientations. 
The principle of relativity assures that no physical (i.e. convention-invariant) observable can depend on the value of $v$. In particular, the principle of relativity requires that the two-way speed of light is equal to $c$ in any given inertial frame. Our next objective is to understand the results of a measurement of  the  two-way speed of light from the moving  source  described above.

Suppose that the laboratory observer performs a measurement of  the wavelength of the standing wave. Then, when the measured data is analyzed, the laboratory observer finds that the speed of light is equal to $c$. We now give derivation of this interesting and important result. If we analyze the geometry of the situation, we find that from the standing wave measurement we can only extract information  about two-way speed of light. The wavenumber observed in the above situation is $\omega_0/[c(1-v/c)]$.  
So if  $\omega_0/c$ is the wavenumber of  light emitted by the same atom at rest in the lab frame, the observer finds that the wavelength of radiation from moving source (the source moves towards the observer) is decreased by the factor $(1-v/c)$. We see that it is the same factor that we can obtain by assuming that the velocity of light from the moving source in the lab frame is $c$.  
Due to the Galilean vectorial velocities addition, the laboratory observer will measure the same two-way speed of light, irrespective of the source velocity. In other words, the measurement of the two-way speed of light is universal and the laboratory observer actually verifies the principle of relativity.

\section{Synchrotron radiation. Geometry and approximations}

\subsection{Paraxial approximation for the radiation field}

We call $z$ the observation distance along the optical axis of the system, while $\vec{r}$ fixes the transverse position of the observer.  
Using the complex notation, in this and in the following sections we assume, in agreement with Eq. (\ref{ftdef2}), that the temporal dependence of fields with a certain frequency is of the form:

\begin{eqnarray}
	\vec{E} \sim \vec{\bar{E}}(z,\vec{r},\omega) \exp(-i \omega t)~.
	\label{eoft}
\end{eqnarray}
With this choice for the temporal dependence we can describe a plane wave traveling along the positive  $z$-axis with

\begin{eqnarray}
	\vec{E} = \vec{E}_0 \exp\left(\frac{i\omega}{c}z -i \omega t\right)~.
	\label{eoftrav}
\end{eqnarray}
In the following we will always assume that the ultra-relativistic approximation is satisfied, which is the case for SR setups. As a consequence, the paraxial approximation applies too. The paraxial approximation implies a slowly varying envelope of the field with respect to the wavelength. It is therefore convenient to introduce the slowly varying envelope of the transverse field components as

\begin{equation}
	\vec{\widetilde{E}}(z,\vec{r},\omega) = \vec{\bar{E}}(z,\vec{r},\omega) \exp{\left(-i\omega z/c\right)}~. \label{vtilde}
\end{equation}
Introducing angles $\theta_x = x_0/z_0$ and $\theta_y = y_0/z_0$, the transverse components of the envelope of the field in Eq. (\ref{revwied}) in the far zone and in paraxial approximation can be written as

\begin{eqnarray}
	\vec{\widetilde{{E}}}(z_0, \vec{r}_0,\omega) &=& -{i
		\omega e\over{c^2}z_0} \int_{-\infty}^{\infty} dz' {\exp{\left[i
			\Phi_T\right]}}  \left[\left({v_x(z')\over{c}}
	-\theta_x\right){\vec{e_x}}
	+\left({v_y(z')\over{c}}-\theta_y\right){\vec{e_y}}\right] \label{generalfin}
\end{eqnarray}
where the total phase $\Phi_T$ is

\begin{eqnarray}
	&&\Phi_T = \omega \left[{s(z')\over{v}}-{z'\over{c}}\right] \cr &&
	+ \frac{\omega}{2c}\left[z_0 (\theta_x^2+\theta_y^2) - 2 \theta_x x(z') - 2 \theta_y y(z') + z'(\theta_x^2+\theta_y^2)\right]~
	. \label{totph}
\end{eqnarray}
Here $v_x(z')$ and $v_y(z')$ are the horizontal and the vertical components of the transverse velocity of the electron,  $x(z')$ and $y(z')$ specify the transverse position of the electron as a function of the longitudinal position, $\vec{e}_x$ and $\vec{e}_y$ are unit vectors along the transverse coordinate axis. Finally, $s(z')$ is the longitudinal coordinate along the path. The electron is moving with velocity $\vec{v}$, whose magnitude is equal to $v = ds/dt$.

\subsection{Approximation for the electron path}

Let us now discuss the case of the radiation from a single electron with an arbitrary angular deflection $\vec{\eta}$ and an arbitrary offset $\vec{l}$ with respect to a reference orbit defined as the path through the origin of the coordinate system, that is $x(0) = y(0) = 0$.

If the magnetic field in the setup does not depend on the transverse coordinates, i.e. $B = B(z)$, an initial offset $x(0) = l_x$, $y(0) = l_y$ shifts the path of an electron of $\vec{l}$. Similarly, an angular deflection $\vec{\eta} = (\eta_x, \eta_y)$ at $z=0$ tilts the path without modifying it. Cases when the magnetic field of SR sources include focusing elements (or the natural focusing of insertion devices) are out of the scope of this paper. Assuming further that $\eta_x \ll 1$ and $\eta_y \ll 1$, which is typically justified for ultrarelativistic electron beams, one obtains the following approximation for the electron path:

\begin{eqnarray}
&& \vec{r}(z) = \vec{r}_r(z) + \vec{\eta} z + \vec{l} ~, \cr &&
\vec{v}(z) = \vec{v}_r(z) + v \vec{\eta}~,
\label{shiftilt}
\end{eqnarray}
where the subscript `r' refers to the reference path. The pair $(\vec{r}(z),z)$  gives a parametric description of the path of a single electron with offset $\vec{l}$ and deflection $\vec{\eta}$. The curvilinear abscissa on the path can then be written as

\begin{eqnarray}
&& s(z) = \int_0^z dz' \left[1+\left(\frac{dx}{dz'} \right)^2+\left(\frac{dy}{dz'} \right)^2\right]^{1/2} \cr &&
\simeq \int_0^z dz' \left[1 + \frac{1}{2} \left(\frac{dx_r}{dz'} \right)^2+\frac{1}{2}\left(\frac{dy_r}{dz'}\right)^2 + \frac{1}{2} \left(\eta_x^2 + \eta_y^2\right) + \eta_x \frac{d x_r}{dz'} + \eta_y \frac{d y_r}{dz'}\right] \cr &&= s_r(z) + \frac{\eta^2 z}{2} + \vec{r}_r(z) \cdot \vec{\eta}~,
\label{curvabs}
\end{eqnarray}
where we expanded the square root around unity in the first passage, we made use of Eq. (\ref{shiftilt}), and of the fact that the curvilinear abscissa along the reference path is $s_r(z) \simeq z + \int_0^z |d \vec{r}_r/dz'|^2/2$.

We now substitute Eq. (\ref{shiftilt}) and Eq. (\ref{curvabs}) into Eq. (\ref{generalfin}) to obtain:

\begin{eqnarray}
&& \vec{\widetilde{{E}}}(z_0, \vec{r}_0,\omega) = -{i
	\omega e\over{c^2}z_0} \int_{-\infty}^{\infty} dz' {\exp{\left[i
		\Phi_T\right]}} \cr && \times  \left[\left({v_x(z')\over{c}}
-(\theta_x-\eta_x)\right){\vec{e_x}}
+\left({v_y(z')\over{c}}-(\theta_y-\eta_y)\right){\vec{e_y}}\right]
~,\cr && \label{generalfin2}
\end{eqnarray}
where the total phase $\Phi_T$ is

\begin{eqnarray}
&&\Phi_T = \omega \left[\frac{s_r(z')}{v} +\frac{\eta^2 z'}{2 v} + \frac{1}{v}\vec{r}_r(z') \cdot \vec{\eta} -{z'\over{c}}\right] \cr &&
+ \frac{\omega}{2c}\left[z_0 (\theta_x^2+\theta_y^2) - 2 \theta_x x_r(z')  - 2 \theta_x \eta_x z' - 2 \theta_x l_x \right. \cr &&  \left. - 2 \theta_y y(z') - 2 \theta_y \eta_y z' - 2 \theta_y l_y  + z'(\theta_x^2+\theta_y^2)\right]~,
\label{totph2}
\end{eqnarray}
which can be rearranged as

\begin{eqnarray}
&&\Phi_T \simeq
\omega \left[{s_r(z')\over{v}}-{z'\over{c}}\right]  - \frac{\omega}{c} (\theta_x l_x + \theta_y l_y) \cr &&
+ \frac{\omega}{2c}\left[z_0 (\theta_x^2+\theta_y^2) - 2 (\theta_x-\eta_x) x_r(z') \right. \cr && \left. - 2 (\theta_y-\eta_y) y_r(z') + z'\left((\theta_x-\eta_x)^2+(\theta_y-\eta_y)^2\right)\right] ~.\cr &&
\label{totph3}
\end{eqnarray}

\section{Undulator radiation}

\subsection{Existing theory}

\subsubsection{Undulator radiation  from a single electron moving along undulator axis}

Eq. (\ref{generalfin}) can be used to characterize the far field from an electron moving on any path.  
In this section we present a simple derivation of the frequency representation of the radiated field produced by an electron in the planar undulator. The magnetic field on the undulator axis has the form

\begin{eqnarray}
\vec{H}(z) = \vec{e}_y H_w\cos(k_wz) ~,\label{mf}
\end{eqnarray}

The Lorentz force is used to derive the equation of motion of the electron in the presence of a magnetic field. Integration of this equation gives

\begin{eqnarray}
v_x(z) = -{c \theta_s} \sin(k_w z) = -\frac{c \theta_s}{2
	i}\left[\exp(ik_w z)-\exp(-i k_w z) \right]~. \label{vxpl_ap1}
\end{eqnarray}
Here  $k_w=2\pi/\lambda_w$, and $\lambda_w$ is the undulator
period. Moreover, $\theta_s=K/\gamma$, where $K$ is the deflection
parameter defined as

\begin{eqnarray}
K = \frac{e\lambda_w H_w}{2 \pi m c^2}~,\label{Kpar_ap1}
\end{eqnarray}
$m$ being the electron mass at rest and $H_w$ being the maximal
magnetic field of the undulator on axis.

In this case the electron path is given by

\begin{eqnarray}
x(z) = r_w\cos(k_w z)~,\label{oa}
\end{eqnarray}

where $r_w = \theta_s/k_w$ is the oscillation amplitude. 

We write the undulator length as $L = N_w \lambda_w$, where
$N_w$ is the number of undulator periods. With the help of Eq.
(\ref{generalfin}) we obtain an expression, valid in the far zone:

\begin{eqnarray}
&& {\vec{\widetilde{E}}}= {i \omega e\over{c^2 z_0}}
\int_{-L/2}^{L/2} dz' {\exp\left[i
	\Phi_T\right]\exp\left[i\frac{\omega \theta^2 z_0}{2c}\right]}
\left[{K\over{\gamma}} \sin\left(k_w z'\right)\vec{e}_x
+\vec{\theta}\right]~. \cr && \label{undurad_ap1}
\end{eqnarray}
Here

\begin{eqnarray}
&& \Phi_T = \left({\omega \over{2 c \bar{\gamma}_z^2}}+ {\omega
	\theta^2 \over{2  c }}\right) z' -
{K\theta_x\over{\gamma}}{\omega\over{k_w c}}\cos(k_w z') -
{K^2\over{8\gamma^2}} {\omega\over{k_w c}} \sin(2 k_w z')
~,\cr &&\label{phitundu_ap1}
\end{eqnarray}
where the average longitudinal Lorentz factor $\bar{\gamma}_z$ is
defined as

\begin{equation}
\bar{\gamma}_z = \frac{\gamma}{\sqrt{1+K^2/2}}~. \label{bargz_ap1}
\end{equation}
The choice of the integration limits in Eq. (\ref{undurad_ap1})
implies that the reference system has its origin in the center of
the undulator.

Usually, it does not make sense to calculate the intensity
distribution from Eq. (\ref{undurad_ap1}) alone, without extra-terms
(both interfering and not) from the other parts of the electron
path. This means that one should have complete information
about the electron path and calculate extra-terms to be
added to Eq. (\ref{undurad_ap1}) in order to have the total field from
a given setup. Yet, we can find \textit{particular situations} for
which the contribution from Eq. (\ref{undurad_ap1}) is dominant with
respect to others. In this case Eq. (\ref{undurad_ap1}), alone, has
independent physical meaning.

One of these situations is  when the resonance approximation is
valid. This approximation does not replace the paraxial one, based
on $\gamma^2 \gg 1$, but it is used together with it. It takes
advantage of another parameter that is usually large, i.e. the
number of undulator periods $N_w \gg 1$. In this case, the
integral in $dz'$ in Eq. (\ref{undurad_ap1}) exhibits simplifications,
independently of the frequency of interest due to the long
integration range with respect to the scale of the undulator
period.

In all generality, the field in Eq. (\ref{undurad_ap1}) can
be written as

\begin{eqnarray}
&&{\vec{\widetilde{E}}}= \exp\left[i\frac{\omega \theta^2
	z_0}{2c}\right] \frac{i \omega e}{c^2 z_0} \cr && \times \int_{-L/2}^{L/2}
dz'\left\{\frac{K}{2 i \gamma}\left[\exp\left(2 i k_w
z'\right)-1\right]\vec{e}_x +\vec{\theta}\exp\left(i k_w
z'\right)\right\} \cr &&\times \exp\left[i \left(C + {\omega
	\theta^2 \over{2 c }}\right) z' -
{K\theta_x\over{\gamma}}{\omega\over{k_w c}}\cos(k_w z')  -
{K^2\over{8\gamma^2}} {\omega\over{k_w c}} \sin(2 k_w z') \right]
~. \cr &&\label{undurad2_ap1}
\end{eqnarray}
Here $\omega = \omega_r + \Delta \omega$, $C =  k_w \Delta\omega/\omega_r$ and

\begin{eqnarray}
\omega_r = 2 k_w c \bar{\gamma}_z^2~, \label{res_ap1}
\end{eqnarray}
is the fundamental resonance frequency.

Using the Anger-Jacobi expansion:

\begin{equation}
\exp\left[i a \sin(\psi)\right] = \sum_{p=-\infty}^{\infty} J_p(a)
\exp\left[ip\psi\right]~, \label{alfeq}
\end{equation}
where $J_p(\cdot)$ indicates the Bessel function of the first kind
of order $p$, to write the integral in Eq. (\ref{undurad2_ap1}) in a
different way:

\begin{eqnarray}
&&{\vec{\widetilde{E}}}= \exp\left[i\frac{\omega \theta^2
	z_0}{2c}\right] \frac{i \omega e}{c^2 z_0} \sum_{m,n=-\infty}^\infty
J_m(u) J_n(v) \exp\left[\frac{i \pi n}{2}\right] \cr && \times
\int_{-L/2}^{L/2} dz'\exp\left[i \left(C + {\omega \theta^2
	\over{2 c }}\right) z'\right] \cr &&\times \left\{\frac{K}{2 i \gamma}
\left[\exp\left(2 i k_w z'\right)-1\right]\vec{e}_x
+\vec{\theta}\exp\left(i k_w z'\right)\right\}
\exp\left[i (n+2m) k_w z'\right] ~,\cr &&\label{undurad3_ap1}
\end{eqnarray}
where\footnote{Here the parameter $v$ should not be confused with the velocity.}

\begin{equation}
u = - \frac{K^2 \omega}{8 \gamma^2 k_w c}~~~~\mathrm{and}~~~v = -
\frac{K \theta_x \omega}{\gamma k_w c}~. \label{uv}
\end{equation}
Up to now we just re-wrote Eq. (\ref{undurad_ap1}) in a different way.
Eq. (\ref{undurad_ap1}) and Eq. (\ref{undurad3_ap1}) are equivalent. Of
course, definition of $C$ is suited to
investigate frequencies around the fundamental harmonic but no
approximation is taken besides the paraxial approximation.

Whenever

\begin{equation}
C  + \frac{\omega \theta^2}{{2 c}} \ll k_w \label{eqq_ap1} ~,
\end{equation}
the first phase term in $z'$ under the integral sign in Eq.
(\ref{undurad3_ap1}) is varying slowly on the scale of the undulator
period $\lambda_w$. As a result, simplifications arise when $N_w
\gg 1$, because fast oscillating terms in powers of  $\exp[i k_w
z']$ effectively average to zero. When these simplifications are
taken,  resonance approximation is applied, in the sense that one
exploits the large parameter $N_w \gg 1$. This is possible under
condition (\ref{eqq_ap1}). Note that (\ref{eqq_ap1}) restricts the range
of frequencies for positive values of $C$ independently of the
observation angle ${\theta}$, but for any value $C<0$ (i.e. for
wavelengths longer than $\lambdabar_r = c/\omega_r$) there is
always some range of $\theta$ such that Eq. (\ref{eqq_ap1}) can be
applied. Altogether, application of the resonance approximation is
possible for frequencies around $\omega_r$ and lower than
$\omega_r$. Once any frequency is fixed, (\ref{eqq_ap1}) poses
constraints on the observation region where the resonance
approximation applies. Similar reasonings can be done for
frequencies around higher harmonics with a more convenient
definition of the detuning parameter $C$.

Within the resonance approximation we further select frequencies
such that

\begin{eqnarray}
\frac{|\Delta \omega|}{\omega_r} \ll 1~,~~~~ \mathrm{i.e.}~~|C|
\ll k_w ~.\label{resext_ap1}
\end{eqnarray}
Note that this condition on frequencies automatically selects
observation angles of interest $\theta^2 \ll 1/\gamma_z^2$. In
fact, if one considers observation angles outside the range
$\theta^2 \ll 1/\gamma_z^2$, condition (\ref{eqq_ap1}) is not
fulfilled, and the integrand in Eq. (\ref{undurad3_ap1}) exhibits fast
oscillations on the integration scale $L$. As a result, one
obtains zero transverse field, $\vec{\widetilde{E}} = 0$,
with accuracy $1/N_w$. Under the constraint imposed by
(\ref{resext_ap1}), independently of the value of $K$ and for
observation angles of interest $\theta^2 \ll 1/\gamma_z^2$, we
have

\begin{equation}
|v|={K|\theta_x|\over{\gamma}}{\omega\over{k_w c}} =
\left(1+\frac{\Delta \omega}{\omega_r}\right) \frac{2 \sqrt{2}
	K}{\sqrt{2+K^2}} \bar{\gamma}_z |\theta_x| \lesssim
\bar{\gamma}_z |\theta_x| \ll 1~. \label{drop}
\end{equation}
This means that, independently of $K$, $|v| \ll 1$ and we may
expand $J_n(v)$ in Eq. (\ref{undurad3_ap1}) according to $J_n(v)
\simeq [2^{-n}/\Gamma(1+n)]~v^n$, $\Gamma(\cdot)$ being the Euler
gamma function

\begin{eqnarray}
\Gamma(z) = \int_0^\infty dt~t^{z-1} \exp[-t] ~.\label{geule}
\end{eqnarray}
Similar reasonings can be done for frequencies around higher
harmonics with a different definition of the detuning parameter
$C$. However, around odd harmonics, the before-mentioned
expansion, together with the application of the resonance
approximation for $N_w \gg 1$ (fast oscillating terms in powers of
$\exp[i k_w z']$ effectively average to zero), yields
extra-simplifications.

Here we are dealing specifically with the first harmonic.
Therefore, these extra-simplifications apply. We
neglect both the  term in $\cos(k_w z')$ in the phase of Eq. (\ref{undurad2_ap1})
and the term in $\vec{\theta}$ in Eq. (\ref{undurad2_ap1}). First,
non-negligible terms in the expansion of $J_n(v)$ are those for
small values of $n$, since $J_n(v) \sim v^n$, with $|v|\ll 1$. The
value $n=0$ gives a non-negligible contribution $J_0(v) \sim 1$.
Then, since the integration in $d z'$ is performed over a large
number of undulator periods $N_w\gg 1$, all terms of the expansion
in Eq. (\ref{undurad3_ap1}) but those for $m=-1$ and $m=0$ average to
zero due to resonance approximation. Note that surviving
contributions are proportional to $K/\gamma$, and can be traced
back to the term in $\vec{e}_x$ only, while the
term in $\vec{\theta}$ in Eq. (\ref{undurad3_ap1}) averages to zero
for $n=0$. Values $n=\pm 1$ already give negligible contributions.
In fact, $J_{\pm 1}(v) \sim v$. Then, the term in $\vec{e}_x$ in
Eq. (\ref{undurad3_ap1}) is $v$ times the term with $n=0$ and is
immediately negligible, regardless of the values of $m$. The
term in $\vec{\theta}$ would survive averaging when $n=1,
~m=-1$ and when $n=-1, ~m=0$. However, it scales as $\vec{\theta}
v$. Now, using condition (\ref{resext_ap1}) we see that, for
observation angles of interest $\theta^2 \ll 1/\gamma_z^2$,
$|\vec{\theta}|~ |v| \sim (\sqrt{2}~ K~/\sqrt{2+K^2}~)
~\bar{\gamma}_z \theta^2 \ll K/\gamma$. Therefore, the
term in $\vec{\theta}$ is negligible with respect to the term in $\vec{e}_x$
for $n=0$, that scales as $K/\gamma$. All terms corresponding to
larger values of $|n|$ are negligible.

Summing up, all terms of the expansion in Eq. (\ref{alfeq}) but
those for $n=0$ and $m=-1$ or $m=0$ give negligible contribution.
After definition of

\begin{eqnarray}
A_{JJ} = J_0\left(\frac{\omega K^2}{8 k_w c \gamma^2}\right) -
J_1\left(\frac{\omega K^2}{8 k_w c \gamma^2}\right) ~,\label{AJJ}
\end{eqnarray}
that can be calculated at $\omega = \omega_r$ since $|C| \ll k_w$,
we have

\begin{eqnarray}
&&{\vec{\widetilde{E}}}= - \frac{K \omega e }{2
	c^2 z_0 \gamma} A_{JJ} \exp\left[i\frac{\omega \theta^2 z_0}{2c}\right]
\int_{-L/2}^{L/2} dz' \exp\left[i \left(C + {\omega \theta^2
	\over{2  c }}\right) z' \right] \vec{e}_x~,\cr &&\label{undurad5finale}
\end{eqnarray}
yielding the well-known free-space field distribution:

\begin{eqnarray}
&&{\vec{\widetilde{E}}}(z_0, \vec{\theta}) = -\frac{K \omega e
	L  }{2 c^2 z_0 \gamma} A_{JJ}\exp\left[i\frac{\omega \theta^2
	z_0}{2c}\right] \mathrm{sinc}\left[\frac{L}{2}\left(C+\frac{\omega
	\theta^2}{2c} \right)\right] \vec{e}_x~,\cr && \label{generalfin4}
\end{eqnarray}
where $\mathrm{sinc}(\cdot) \equiv \sin(\cdot)/(\cdot)$.
Therefore, the field is horizontally polarized and azimuthal
symmetric.

\subsubsection{An electron with arbitrary angular deflection and offset}

Eq. (\ref{generalfin4})  can be generalized to the case of a particle with a given offset $\vec{l}$ and deflection angle $\vec{\eta}$ with respect to the longitudinal axis, assuming that the magnetic field in the undulator is independent of the transverse coordinate of the particle. Although this can be done using Eq. (\ref{generalfin2}) directly, it is sometimes possible to save time by getting the answer with some trick. For example, in the undulator case one takes advantage of the following geometrical considerations, which are in agreement with rigorous mathematical derivation. First, we consider the effect of an offset $\vec{l}$ on the transverse plane, with respect to the longitudinal axis $z$. Since the magnetic field experienced by the particle does not change, the far-zone field is simply shifted by a quantity $\vec{l}$.  Eq. (\ref{generalfin4}), can be immediately generalized by systematic substitution of the transverse coordinate of observation, $\vec{r}_0$ with $\vec{r}_0 -\vec{l}$. This means that $\vec{\theta}=\vec{r}_0/z_0$ must be substituted by $\vec{\theta} - \vec{l}/z_0$, thus yielding

\begin{eqnarray}
	\widetilde{{E}}\left(z_0,  \vec{l}, \vec{\theta}\right)&=&
	-\frac{K \omega e L} {2 c^2 z_0 \gamma}
	A_{JJ}\exp\left[i\frac{\omega z_0}{2 c}
	\left|\vec{\theta}-\frac{\vec{l}}{z_0}\right|^2\right]
	\mathrm{sinc}\left[\frac{\omega L
		\left|\vec{\theta}-\left({\vec{l}}/{z_0}\right)\right|^2}{4
		c}\right] ~.\label{undurad4bisgg0}
\end{eqnarray}
Let us now discuss the effect of a deflection angle $\vec{\eta}$. Since the magnetic field experienced by the electron is assumed to be independent of its transverse coordinate, the path followed is still sinusoidal, but the effective undulator period is now given by $\lambda_w/\cos(\eta) \simeq (1+\eta^2/2) \lambda_w$. This induces a relative red shift in the resonant wavelength $\Delta \lambda/\lambda \sim \eta^2/2$. In practical cases of interest we may estimate $\eta \sim 1/\gamma$. Then, $\Delta \lambda/\lambda \sim 1/\gamma^2$ should be compared with the relative bandwidth of the resonance, that is $\Delta \lambda/\lambda \sim 1/N_w$, $N_w$ being the number of undulator periods. For example, if $\gamma > 10^3$, the red shift due to the deflection angle can be neglected in all situations of practical relevance. As a result, the introduction of a deflection angle only amounts to a rigid rotation of the entire system. Performing such rotation we should account for the fact that the phase factor in Eq. (\ref{undurad4bisgg0}) is indicative of a spherical wavefront propagating outwards from position $z=0$ and remains thus invariant under rotations. The argument in the $\mathrm{sinc}(\cdot)$ function in Eq. (\ref{undurad4bisgg0}), instead, is modified because the rotation maps the point $(z_0,0,0)$ into the point $(z_0, -\eta_x z_0, -\eta_y z_0)$. As a result, after rotation, Eq. (\ref{undurad4bisgg0}) transforms to

\begin{eqnarray}
	&&\widetilde{{E}}\left(z_0, \vec{\eta}, \vec{l},
	\vec{\theta}\right)= -\frac{K \omega e L A_{JJ}} {2 c^2 z_0 \gamma}
	\exp\left[i\frac{\omega z_0}{2 c}
	\left|\vec{\theta}-\frac{\vec{l}}{z_0}\right|^2\right]
	\mathrm{sinc}\left[\frac{\omega L
		\left|\vec{\theta}-\left({\vec{l}}/{z_0}\right)-\vec{\eta}\right|^2}{4
		c}\right] \cr &&\label{undurad4bisgg00}
\end{eqnarray}
Finally, in the far-zone case, we can always work in the limit for $l/z_0 \ll 1$, that allows one to neglect the term ${\vec{l}}/{z_0}$ in the argument of the $\mathrm{sinc}(\cdot)$ function, as well as the quadratic term in $\omega l^2/(2 c z_0)$ in the phase.  Thus Eq. (\ref{undurad4bisgg00}) can be further simplified, giving the generalization of Eq. (\ref{generalfin4}) in its final form:

\begin{eqnarray}
	\widetilde{{E}}\left(z_0, \vec{\eta}, \vec{l},
	\vec{\theta}\right)&=& -\frac{K \omega e L A_{JJ}} {2 c^2 z_0 \gamma}
	\exp\left[i\frac{\omega}{c}\left( \frac{z_0 \theta^2}{2}-
	\vec{\theta}\cdot\vec{l} \right)\right]
	\mathrm{sinc}\left[\frac{\omega L
		\left|\vec{\theta}-\vec{\eta}\right|^2}{4 c}\right] ~.
	\label{undurad4bisgg}
\end{eqnarray}

It is clear from the above that, according to conventional synchrotron radiation theory, if we consider radiation from one electron at detuning $C$ from resonance, the introduction of a kick only amounts to a rigid rotation of the angular distribution along the new direction of the electron motion. This is plausible, if one keeps in mind that after the kick the electron has the same velocity and emits radiation in the kicked direction owing to the Doppler effect. After such rotation, Eq. (\ref{generalfin4}) transforms into  Eq. (\ref{undurad4bisgg})      

\subsection{Undulator radiation and special theory of relativity}

\subsubsection{Angular-spectral flux radiated in the central cone}

We have seen that in all generality the expression for the undulator field in the far zone  and in the ultrarelativistic (i.e. paraxial) approximation can be written as Eq. (\ref{undurad2_ap1}). 
Within the resonance approximation ($N_w \gg 1$) for the frequencies around the first harmonic it can be simplified  to the well-known expression Eq. (\ref{generalfin4}) where the field is horizontally polarized and azimuthal symmetric. The divergence of this radiation is much smaller compared to the angle $1/\bar{\gamma}_z$. The mathematical reason stems from the fact that the factor $\sin(\cdot)/(\cdot)$ represents the well-known resonance character of the undulator radiation. If we are interested in the angular width of the peak around the observation angle  $\theta = 0$, we can introduce an angular displacement $\Delta\theta$. Taking the first zero of the $\sin(\cdot)/(\cdot)$ function at $C = 0$ we will be able to determine the natural angular width of the radiation for the first harmonic $\theta_c$. The cone with aperture  $\theta_c$ is usually called central cone. It can be found that $\theta_c^2 = 1/(2N_w\bar{\gamma}_z^2) \ll 1/\bar{\gamma}_z^2$.

Now we would like to understand what is 
the characteristic transverse size of the field distribution at the exist of the undulator. 
The radiation from magnetic poles always interferes coherently at zero angle with respect to undulator axis. This interference is constructive within an angle of about $\sqrt{c/(\omega L_w)}$. We can estimate the interference size at the undulator exit as about $\sqrt{cL_w/\omega}$. 
On the other hand, the electron oscillating amplitude is given by $r_w = c\theta_s/k_w = cK/(\gamma k_w)$. It follows that $r_w^2/(cL_w/\omega)  = K^2\omega/(L_wk_w^2\gamma^2) = K^2/[\pi N_w(1 + K^2/2)] \ll 1$, where we use the fact that $\gamma^2 = (1 + K^2/2)\bar{\gamma}_z^2$.
This inequality holds independently of the value of $K$, because $N_w \gg 1$. Thus, the electron oscillating amplitude is always much smaller than the radiation diffraction size at the undulator exit.

We consider the radiation associated with the first order term in the expansion of the  Eq. (\ref{undurad3_ap1}) in power of $v = K \theta_x \omega/(\gamma k_w c)$. 
But in doing so we miss all information about transverse electron trajectory  in the phase factor   Eq. (\ref{phitundu_ap1}) since the term $K\theta_x\omega\cos(k_w z')/(\gamma k_w c)$ is neglected.
In this  approximation the electron orbit scale is always much smaller than the radiation diffraction size and  Eq. (\ref{generalfin4}) gives fields very much in agreement with the dipole radiation theory. So we are satisfied using the non covariant approach when considering the transverse electron motion.

There  are several points to be made about the above result. We have just explained that 
in accounting only  for the radiation in the central cone, we miss all information about the transverse electron motion. To be complete we must add an analysis of the accelerated motion along the $z$-direction (i.e. along the undulator axis).
We assume that the transverse velocity $v_{\perp}(z)$ is small compared to the velocity of light $c$. We consider the small expansion parameter $v_{\perp}/c$, neglecting terms of order $(v_{\perp}/c)^3$, but not of order $(v_{\perp}/c)^2$. In other words we use a second order relativistic approximation for transverse motion.
We should remark that the analysis of the longitudinal motion in the ultrarelativistic approximation is much simpler than in the case of transverse motion. It is easy to see that the acceleration in the constant magnetic field yields an transverse electron velocity $v_{\perp}$ and $\Delta v_z = -v(v_{\perp}/c)^2/2$ parallel to the $z$-axis. If we evaluate the transformations up to the second order $(v_{\perp}/c)^2$, the relativistic correction in the longitudinal motion does not appear. So one should not be surprised to find that, in this approximation,  there is no influence of the difference between the non-covariant and covariant constrained electron trajectories on the undulator radiation in the central cone.

\subsubsection{Influence of the kick}

According to the correct coupling of fields and particles, there is a remarkable prediction of undulator radiation theory concerning to the  undulator radiation from the single electron with  and without kick. 
Namely, when a kick is introduced, there is a red shift in the resonance wavelength of the undulator radiation in the velocity direction. To show this, let us  consider  the covariant treatment, which makes explicit use of Lorentz transformations.

When the kick is introduced, covariant particle tracking predicts a non-zero red shift of the resonance frequency, which arises because in Lorentz coordinates the electron velocity decreases from $v$  to $v - v\theta_k^2/2$ after the kick, while the velocity of light is unvaried and equal to the electrodynamics constant $c$ (see section 4.6 for a detailed derivation).

Now the formula Eq. (\ref{totph3}) is not quite right, because we should have used not the velocity of electron $v$ but $v - v\theta_k^2/2$.    
The shift in the total phase $\Phi_T$  under the integral Eq. (\ref{generalfin2})  can be expressed by the formula $\Delta \Phi_T = \omega\theta_k^2 z'/(2c)$, where we account to that $v \simeq c$ in ultrarelativistic approximation.

Suppose that without kick the electron moves  along the constrained trajectory parallel to the undulator axis. The field which produces this electron in the far zone is given by 
Eq. (\ref{generalfin4}). Referring back to the Eq. (\ref{undurad4bisgg}), we see that the conventional undulator radiation theory gives the following expression for
radiation field after the kick

\begin{eqnarray}
	&&{\vec{\widetilde{E}}} = -\frac{K \omega e
		L  }{2 c^2 z_0 \gamma} A_{JJ}\exp\left[i\frac{\omega \theta^2
		z_0}{2c}\right] \mathrm{sinc}\left[\frac{L}{2}\left(C + \frac{\omega
		\left| \vec{\theta}  -\vec{\theta}_k\right|^2}{2c} \right)\right] \vec{e}_x~.\cr && \label{generalfins}
\end{eqnarray}

The covariant equations say that, when the kick is introduced, the  radiation field in question is given by the formula

\begin{eqnarray}
&&{\vec{\widetilde{E}}} = -\frac{K \omega e
	L  }{2 c^2 z_0 \gamma} A_{JJ}\exp\left[i\frac{\omega \theta^2
	z_0}{2c}\right] \mathrm{sinc}\left[\frac{L}{2}\left(C+ \frac{\omega\theta_k^2}{2c} + \frac{\omega
	\left| \vec{\theta}  -\vec{\theta}_k\right|^2}{2c} \right)\right] \vec{e}_x~,\cr && \label{generalfin6}
\end{eqnarray}

This formula has nearly, but not quite the same form as
Eq. (\ref{generalfins}), the difference consisting in the term $\omega\theta_k^2/(2c)$ in the argument of  $\mathrm{sinc}$ function. Attention must be called to the difference in resonance frequency between the undulator radiation setup with and without  the kick. Remembering the definition of the detuning parameter $C = k_w\Delta\omega/\omega_r$, we can write  
the red shift in resonance frequency as $\Delta\omega/ \omega_r =  - \omega_r\theta_k^2/(2k_w c)$.
With this we also pointed out that the red shift can be written as
$\Delta\omega/ \omega_r =  - \gamma^2\theta_k^2/(1+K^2/2)$. We now see a second order correction $\theta_k^2$ that is, however, multiplied by the large factor $\gamma^2$. 

We are now ready to investigate, more generally, what form the field expression takes under the introduction of a kick. Suppose that, without kick, the electron moves along the trajectory with angle $\vec{\eta}$ with respect to the undulator axis. The field produced by this electron is given by  Eq. (\ref{undurad4bisgg}). We let $\vec{\theta}_k$ be the kick angle of the electron with respect to its initial motion.  The conventional approach gives the following expression for the field after the kick

\begin{eqnarray}
&&{\vec{\widetilde{E}}} = -\frac{K \omega e
	L  }{2 c^2 z_0 \gamma} A_{JJ}\exp\left[i\frac{\omega \theta^2
	z_0}{2c}\right] \mathrm{sinc}\left[\frac{L}{2}\left(C + \frac{\omega
	\left| \vec{\theta}  -\vec{\eta} -\vec{\theta}_k\right|^2}{2c} \right)\right] \vec{e}_x~.\cr && \label{generalfinss}
\end{eqnarray}

In contrast, the covariant approach gives

\begin{eqnarray}
&&{\vec{\widetilde{E}}} = -\frac{K \omega e
	L  }{2 c^2 z_0 \gamma} A_{JJ}\exp\left[i\frac{\omega \theta^2
	z_0}{2c}\right] \mathrm{sinc}\left[\frac{L}{2}\left(C+ \frac{\omega\theta_k^2}{2c} + \frac{\omega
	\left| \vec{\theta}  - \vec{\eta} - \vec{\theta}_k\right|^2}{2c} \right)\right] \vec{e}_x~,\cr && \label{generalfinnn}
\end{eqnarray}

Now this all leads to an interesting situation. According to the conventional theory, the resonance wavelength depends only on the observation angle with respect to the electron velocity direction. 
Equation (\ref{generalfinss}) says that for any kick angle $\vec{\theta}_k$ and for any angle  $\vec{\eta}$ between the undulator axis and  the initial electron velocity direction, the radiation along the velocity direction has no red shift. We would like to emphasize a very important difference between conventional and covariant theory.
The result of the covariant approach Eq. (\ref{generalfinnn}) clearly depends on the absolute value of the kick angle $\theta_k$ and the radiation along the velocity direction has the red shift only when the kick angle has nonzero value.  

We must conclude that 
when we accelerate the electron in the lab frame upstream the undulator, the information about this acceleration is included into the covariant trajectory. Perhaps it is not so puzzling, though, when one remembers that, as well known, after the circular revolution  the electron's interaction with electromagnetic fields depends on the acceleration prehistory which accumulates in the Wigner  (electron spin) rotation.

\subsubsection{Results of experiment}

We now wish to consider an experiment whose results can only be explained on the basis of our corrected undulator radiation theory. We refer to the "beam splitting" experiment at the LCLS \cite{NUHN}. It apparently demonstrated that after a modulated electron beam is kicked on a large angle compared to the divergence of the XFEL radiation, the modulation wavefront is readjusted along the new direction of the motion of the kicked beam. Therefore, coherent radiation from the undulator placed after the kicker is emitted along the kicked direction practically without suppression \footnote{The tuning limit of the kick angle was set by the beamline aperture to $\simeq$ 5 rms of the XFEL radiation divergence, see Fig. 14 in \cite{NUHN}. According to conventional theory, this leads to decrease  the radiation efficiency in the kicked direction by more than three orders of magnitude}.

In the framework of the conventional theory, there is also a second outstanding puzzle concerning the beam splitting experiment at the LCLS. In accordance with conventional undulator radiation theory, if the modulated electron beam is at perfect (undulator) resonance without kick, then after the kick the same modulated beam must be at perfect resonance in the velocity direction. 
However, experimental results clearly show that there is a red shift in the resonance wavelength when the kick is introduced. The maximum power of the coherent radiation is reached when undulator is detuned to be resonant to the lower longitudinal velocity after the kick \cite{NUHN}.    

It should be remarked that any linear superposition of a given radiation field from single electrons conserves single-particle characteristics like parametric dependence on undulator parameters and polarization. Consider a modulated electron beam kicked by a weak dipole field before entering a downstream undulator. Radiation fields generated by this beam can be seen as a linear superposition of fields from individual electrons. Now experimental results clearly show that there is a red shift in the resonance wavelength for coherent undulator radiation when the kick is introduced. It follows that the undulator radiation from the  single electron has red shift when the kick is introduced as well. This argument suggests that results of the beam splitting experiment in reference \cite{NUHN} confirm our  correction for spontaneous undulator emission.
In fact, one of the immediate consequences of our theory is the occurrence of the non-zero red shift of the resonance wavelength when the kick angle has nonzero value.

Our conclusion is in open contrast with the \cite{NA}, that include a kinematical mistake in the description of the undulator resonance condition  for the electron beam with angular deflection between the velocity and the undulator axis.
In the reference \cite{NA} one can read: 
"The FEL radiation wavelength $\lambda_r$ depends on the undulator period $\lambda_u$, and the electron beam Lorentz factor $\gamma$: $\lambda_r = \lambda_u (1+K^2/2 + \gamma^2\phi^2)/(2\gamma^2) ,~~~~ (1)$
$~~~~$ where $\phi$ is the observation angle from the undulator axis. [...]
The dipole corrector in front of the Delta (undulator) is then used to give a kick to the electron beam in the desired circularly polarized photon beam direction. The $K$ value of the Delta (undulator) is also decreased so that resonance condition equation (1) is still satisfied."

Here, we give a reason why this explanation of the red shift is incorrect.
According to the conventional particle tracking, after the beam is kicked there is a trajectory change, while the electron velocity remains as before. The prediction of the conventional undulator radiation theory is that if an electron beam is at perfect undulator resonance without kick, then after the kick the same electron beam must be at perfect resonance in the velocity direction.  

The resonance condition (1) in the reference \cite{NA} corresponds to the  simplest case when the electron beam is moving along the undulator axis. When there is an angle between  the undulator axis and the electron velocity direction, the resonance condition depends only on the observation angle  with respect to the velocity direction. This is not surprising, if one analyzes the situation in the conventional framework  and keeps in mid that after the kick the electron has the same velocity and emits radiation in the velocity direction owing to the Doppler effect.

\section{Synchrotron radiation from bending magnets}

\subsection{Existing theory}

Consider a single relativistic electron moving on a circular orbit. 
It is worth to underline the difference between the geometry which we use and the geometry used in most synchrotron radiation textbooks  for the treatment of bending magnet radiation. The observer in the standard treatment is assumed to be located in a vertical plane tangent to the circular trajectory at the origin, at an angle $\theta$ above the level of the orbit. In other words, in this geometry the $z$ axis is not fixed, but depends on the observer's position. Note that the geometry of the electron motion has a cylindrical symmetry, with the vertical axis going through the center of the circular orbit.  Because of this symmetry, in order to calculate spectral and angular photon distributions, it is not necessary to consider an observer at a more general location. However, since the wavefront is not spherical, this way of proceeding can hardly help to obtain the phase of the field distribution on a plane perpendicular to a fixed $z$ axis.

\subsubsection{Radiation from a single electron moving along an arc of a circle}

We can use Eq. (\ref{generalfin}) to calculate the far zone field of radiation from a relativistic electron moving along an arc of a circle. Assuming a geometry with a fixed $z$ we can write the transverse position of the electron as a function of the curvilinear abscissa $s$ as

\begin{equation}
	\vec{r}(s) = -R\left(1-\cos(s/R)\right) \vec{e_x}
	\label{trmot}
\end{equation}
and

\begin{equation}
	z(s) = R \sin(s/R) \label{zmot}
\end{equation}
where $R$ is the bending radius.

Since the integral in Eq. (\ref{generalfin}) is performed along $z$ we should invert $z(s)$ in Eq. (\ref{zmot}) and find the explicit dependence $s(z)$:

\begin{equation}
	s(z) = R \arcsin(z/R) \simeq z + {z^3\over{6R^2}} \label{sz}
\end{equation}
so that

\begin{equation}
	\vec{r}(z) = - {z^2\over{2 R}} \vec{e_x}~,\label{rpdis}
\end{equation}
where the expansion in Eq. (\ref{sz}) and Eq. (\ref{rpdis}) is justified, once again, in the framework of the paraxial approximation.

%
%

With Eq. (\ref{generalfin}) we obtain the radiation field amplitude in the far zone:

\begin{eqnarray}
	\vec{\widetilde{E}}= {i \omega e\over{c^2 z_0}}
	\int_{-\infty}^{\infty} dz' {e^{i  \Phi_T}} \left({
		z'+R\theta_x\over{R}}\vec{e_x}
	+\theta_y\vec{e_y}\right)~\label{srtwo}
\end{eqnarray}
where

\begin{eqnarray}
	&&\Phi_T = \omega \left[
	\left({\theta_x^2+\theta_y^2\over{2c}}z_0\right)
	+\left({1\over{2\gamma^2c}} +
	{\theta_x^2+\theta_y^2\over{2c}}\right)z' \right.  \cr && \left. +
	\left({\theta_x\over{2Rc}}\right)z'^2 +
	\left(1\over{6R^2c}\right)z'^3\right]~.\label{phh2}
\end{eqnarray}

One can easily reorganize the terms in Eq. (\ref{phh2}) to obtain

\begin{eqnarray}
	&& \Phi_T = \omega\left[
	\left({\theta_x^2+\theta_y^2\over{2c}}z_0\right)-{R\theta_x\over{2c}}\left({1\over{\gamma^2}}
	+{\theta_x^2\over{3}} +\theta_y^2\right) \right. \cr && \left.
	+\left({{1\over{\gamma^2}}+\theta_y^2}\right){\left(z'+R\theta_x\right)\over{2c}}
	+ {\left(z'+R\theta_x\right)^3\over{6 R^2 c
	}}\right]~.\label{phh2b}
\end{eqnarray}
With redefinition of $z'$ as $z' + R \theta_x$ under integral we obtain the final result:

\begin{eqnarray}
	&& \vec{\widetilde{E}}= {i \omega e\over{c^2 z_0}} e^{i\Phi_s}
	e^{i\Phi_0} \int_{-\infty}^{\infty} dz'
	\left({z'\over{R}}\vec{e_x}+\theta_y\vec{e_y}\right) \cr &&
	\times
	\exp\left\{{i\omega\left[{z'\over{2\gamma^2c}}\left(1+\gamma^2\theta_y^2\right)
		+{z'^3\over{6R^2c}}\right]}\right\}~,\label{srtwob}
\end{eqnarray}
where
\begin{equation}
	\Phi_s ={\omega z_0\over{2c}}\left(\theta_x^2+\theta_y^2
	\right)\label{phis}
\end{equation}
and

\begin{equation}
	\Phi_0 = -{\omega R \theta_x\over{2c}}\left( {1\over{\gamma^2}}
	+{\theta_x^2\over{3}} +\theta_y^2 \right)~.\label{phio}
\end{equation}
In standard treatments of bending magnet radiation, the phase term $\exp(i\Phi_0)$ is absent. In fact, the horizontal observation angle $\theta_x$ is always equal to zero. The reason for this is that most textbooks focus on the calculation of the intensity radiated by a single electron in the far zone,  which involves the square modulus of the field amplitude but do not analyze, for instance, situations like source imaging.

\subsubsection{An electron with arbitrary angular deflection and offset}

Up to this point we considered an electron moving along a circular trajectory that lies in the $(x,z)$-plane and  tangent to the $z$ axis. The phase difference in the fields will be determined by the position of the observer position  and by the electron trajectory. Let us now discuss the bending magnet radiation from a single electron with arbitrary angular deflection and offset with respect to the nominal orbit.

The meaning of  horizontal and vertical deflection angles $\eta_x$ and $\eta_y$ is clear once we specify the electron velocity

\begin{eqnarray}
	&&\vec{v}(s) = v\left[-\sin\left({s\over{R}}+\eta_x\right)\cos(\eta_y)
	\vec{e_x} + \sin(\eta_y) \vec{e_y} +
	\cos\left({s\over{R}}+\eta_x\right)\cos(\eta_y) \vec{e_z}
	\right]~,\cr && \label{veloangle}
\end{eqnarray}
so that the trajectory can be expressed as a function of the curvilinear abscissa $s$ as

\begin{eqnarray}
	&&x(s) \vec{e}_x + y(s) \vec{e}_y + z(s) \vec{e}_z = \cr &&  \left[l_x +
	R\cos\left({s\over{R}}+\eta_x\right)\cos(\eta_y) - R\cos(\eta_x)\cos(\eta_y) \right]
	\vec{e_x} \cr && + \left[l_y+s \sin(\eta_y)\right]  \vec{e_y}
	\cr &&+ \left[ R\sin\left({s\over{R}}+\eta_x\right)\cos(\eta_y) -
	R\sin(\eta_x)\sin(\eta_y) \right] \vec{e_z} \cr &&
	\label{trajangle}
\end{eqnarray}
Here we have introduced, also, an arbitrary offset $(l_x,l_y,0)$
in the trajectory. Using Eq. (\ref{trajangle}) an approximated
expression for $s(z)$ can be found:

\begin{equation}
	s(z) = z+ {z^3\over{6 R^2}}+{z^2 \eta_x\over{2 R}} +{z
		\eta_x^2\over{2}}+{z \eta_y^2\over{2}} \label{szangle}
\end{equation}
so that

\begin{equation}
	\vec{v}(z) =   \left(- {v z\over{R}}+ v \eta_x  \right)
	\vec{e_x} + \left(v \eta_y  \right)
	\vec{e_y}~\label{vapprangle}
\end{equation}
and

\begin{equation}
	\vec{r}(z) = \left(- {z^2\over{2R}}+ \eta_x z + l_x\right)
	\vec{e_x} + \left(\eta_y z +l_y \right)
	\vec{e_y}~.\label{trmot2}
\end{equation}
%
%
%

It is evident that the offsets $l_x$ and $l_y$ are always subtracted from $x_0$ and $y_0$  respectively: a shift in the particle trajectory on the vertical plane is equivalent to a shift of the observer in the opposite direction. With this in mind we introduce angles $\bar{\theta}_x = \theta_x -l_x/z_0$ and $\bar{\theta}_y = \theta_y - l_y/z_0$ to obtain

\begin{eqnarray}
	\vec{\widetilde{E}}= {i \omega e\over{c^2 z_0}}
	\int_{-\infty}^{\infty} dz' {e^{i \Phi_T}} \left({
		z'+R(\bar{\theta}_x-\eta_x)\over{R}}\vec{e_x}
	+{(\bar{\theta}_y-\eta_y)}\vec{e_y}\right)~\label{srtwoang}
\end{eqnarray}
and

\begin{eqnarray}
	&& \Phi_T =
	\omega \left({\bar{\theta}_x^2+\bar{\theta}_y^2\over{2c}}z_0 \right)
	+{\omega\over{2c}}\left({1\over{\gamma^2}} +
	\left(\bar{\theta}_x-\eta_x\right)^2  +
	\left(\bar{\theta}_y-\eta_y\right)^2\right)z' \cr && +
	\left({\omega(\bar{\theta}_x-\eta_x)\over{2Rc}}\right)z'^2 +
	\left(\omega\over{6R^2c}\right)z'^3~.\label{phh2ang}
\end{eqnarray}
One can easily reorganize the terms in Eq. (\ref{phh2ang}) to
obtain

\begin{eqnarray}
	&&\Phi_T =
	\omega\left({\bar{\theta}_x^2+\bar{\theta}_y^2\over{2c}}z_0\right)-
	{\omega R(\bar{\theta}_x-\eta_x)\over{2c}} \cr &&\times
	\left({1\over{\gamma^2}} +(\bar{\theta}_y-\eta_y)^2
	+{(\bar{\theta}_x-\eta_x)^2\over{3}}\right) \cr &&
	+\left({{1\over{\gamma^2}}+(\bar{\theta}_y-\eta_y)^2}\right)
	{\omega\left(z'+R(\bar{\theta}_x-\eta_x)\right)\over{2c}} \cr  &&+
	{\omega \left(z'+R (\bar{\theta}_x-\eta_x)\right)^3\over{6 R^2 c
	}}~.\label{phh2angfin}
\end{eqnarray}
Redefinition of $z'$ as $z'+R(\bar{\theta}_x-\eta_x)$ gives the
result

\begin{eqnarray}
	&&\vec{\widetilde{E}}= {i \omega e\over{c^2 z_0}} e^{i \Phi_s} e^{i
		\Phi_0} \int_{-\infty}^{\infty} dz'
	\left({z'\over{R}}\vec{e_x}+(\bar{\theta}_y-\eta_y)\vec{e_y}\right)
	\cr && \times
	\exp\left\{{i\omega\left[{z'\over{2\gamma^2c}}\left(1+\gamma^2
		(\bar{\theta}_y-\eta_y)^2\right)
		+{z'^3\over{6R^2c}}\right]}\right\}~,\label{srtwoang2}
\end{eqnarray}
where
\begin{equation}
	\Phi_s = {\omega z_0
		\over{2c}}\left(\bar{\theta}_x^2+\bar{\theta}_y^2
	\right)\label{phisang}
\end{equation}
and

\begin{equation}
	\Phi_0 = - {\omega R(\bar{\theta}_x-\eta_x)\over{2c}}
	\left({1\over{\gamma^2}} +(\bar{\theta}_y-\eta_y)^2
	+{(\bar{\theta}_x-\eta_x)^2\over{3}}\right)~.\label{phioang}
\end{equation}
In the far zone we can neglect terms in $l_x/z_0$ and $l_y/z_0$, which leads to

\begin{eqnarray}
	&&\vec{\widetilde{E}}= {i \omega e\over{c^2 z_0}} e^{i \Phi_s} e^{i
		\Phi_0} \int_{-\infty}^{\infty} dz'
	\left({z'\over{R}}\vec{e_x}+\left(\theta_y-\eta_y
	\right)\vec{e_y}\right) \cr && \times
	\exp\left\{{i\omega\left[{z'\over{2\gamma^2c}}\left(1+\gamma^2
		\left(\theta_y-\eta_y\right)^2\right)
		+{z'^3\over{6R^2c}}\right]}\right\}~,\label{srtwoang2bis}
\end{eqnarray}
where
\begin{equation}
	\Phi_s = {\omega z_0 \over{2c}}\left(\theta_x^2+\theta_y^2
	\right)\label{phisangbis}
\end{equation}
and

\begin{eqnarray}
	\Phi_o \simeq - {\omega R({\theta_x}-\eta_x)\over{2c}}
	\left({1\over{\gamma^2}} +(\theta_y-\eta_y)^2 
	+{(\theta_x-\eta_x)^2\over{3}}\right)-{\omega\over{c}}(l_x
	\theta_x+l_y\theta_y) ~.\label{phioangbis}
\end{eqnarray}

\subsection{Bending magnet radiation and special theory of relativity}

\subsubsection{Radiation field in space-frequency domain}

Our case of interest is an ultrarelativistic electron accelerating in a circle. 
As already remarked, in conventional (non-covariant) particle tracking the description of the dynamical evolution in the lab frame is based on the use of the absolute time convention. In this case simultaneity is absolute, and we only need one set of synchronized clocks in the lab frame, to be used for the description of the accelerated motion. However, the use of the absolute time convention automatically implies the use of much more complicated field equations, and these equations are different for each value of the particle velocity i.e. for each point along its path.    
This is the reason to prefer the covariant approach within the framework of both dynamics and electrodynamics.

We want to solve the electrodynamics problem based on Maxwell's equations in their usual form. In this case we should analyze the particle evolution within the framework of special relativity, where the problem of assigning Lorentz coordinates to the lab frame in the case of accelerating motion is complicated. 
The only possibility to introduce Lorentz coordinates in this situation consists in introducing individual coordinate systems (i.e. individual rule-clock structure) for each point of the path.

We start by considering an electron moving along a circular trajectory that lies in the $(x,z)$-plane and  tangent to the $z$ axis. Because of cylindrical symmetry, in order to calculate spectral and angular photon distributions, it is not necessary to consider an observer at general location. The observer is assumed to be located in the vertical plane tangent to the circular trajectory at the origin. In ultrarelativistic (paraxial) approximation we evaluate transformations working only up to the order of $v^2_x/c^2$. The restriction to this order provides an essential simplicity of calculations. 
We can interpret manipulation with rule-clock structure in the lab frame simply as a change of variables according to the transformation $x_d = \gamma_x x$, $t_d = t/\gamma_x + \gamma_x xv_x/c^2$. We are dealing with a second order approximation and $\gamma_x = 1 + v_x^2/(2c^2)$.
The overall combination of Galilean transformation and variable changes actually yields to the  transverse Lorentz transformation (see section 3.3.5 for more detail).
Since the Galilean transformation, completed by the introduction of the new variables, is mathematically equivalent to a Lorentz transformation, it obviously follows that transforming to new variables leads to the usual Maxwell's equations. 

In order to keep Lorentz coordinates in the lab frame, as discussed before, we need only to perform a clock resynchronization by introducing the time shift $\Delta t = t_d - t =  - [v_x^2/(2c^2)]t + xv_x/c^2$. The relativistic correction to the particle's offset "$x$" does not appear in this expansion order, but only in order of $v_x^3/c^3$ and $x_d = x$ in our case of interest. Although we have only shown that time shift  in one rather special case, the result is right for any offset and (transverse) velocity direction:  
$\Delta t = t_d - t =  - [|\vec{v}_{\perp}|^2/(2c^2)]z'/c + \vec{r}_{\perp}\cdot \vec{v}_{\perp}/c^2$.
To finish our analysis we need only find a relativistic correction to the longitudinal motion. We remark again that if we evaluate the transformations up to the second order $(v_{\perp}/c)^2$, the relativistic correction in the longitudinal motion does not appear in this approximation.
We have demonstrated the covariant method that can be used for any trajectory - a general way of funding what happens  directly in space-frequency domain and in paraxial approximation.

Let us now see how to apply this covariant  method  to a special situation. 
Let's use  our knowledge of the relativistically correct method for calculating synchrotron radiation emission to  find the photon angular-spectral density distributions from a bending magnet. In  the ultrarelativistic approximation, we have a uniform acceleration of the electron $a = v^2/R = c^2/R$ in the transverse direction. We can, then, write velocity and offset of the electron as follows $v_x = at = az'/v = az'/c$, $x = at^2/2 = az'^2/(2c^2)$. We have now all quantities we wanted. Let us put them all together in relativistic time shift:  $\Delta t =  t_d - t = - a^2z'^3/(2c^5) + a^2z'^3/(2c^5) = 0$. There is no difference! We do not need to use covariant particle tracking for derivation of the  bending magnet radiation. Why should that be? Usually, such a beautiful cancellation is found to stem from a deep underlying principle. Nevertheless, in this case there does not appear to be any such profound implication.  
This is a coincidence. It is because we have deal with uniform acceleration in the transverse direction using a second order (paraxial) approximation when an electron is moving along an arc of a circle. 

This cancellation is not surprising, if one analyzes the general expression for the radiation field from bending magnet in the far zone Eq.(\ref{srtwob}).
In our previous discussion of undulator radiation,  we learned that the relativistic correction  appears only when the transverse electron trajectory is  included in the total phase $\Phi_T$ under the integral Eq.(\ref{generalfin}). Referring back to Eq.(\ref{totph}) for the phase factor $\Phi_T$ , we see that the  term which depends on the transverse position of the electron  can be written as $\exp i(\omega/c)[\theta_x x(z') +  \theta_y y(z')]$. We conclude that the observation angle in the total phase factor under the integral must be related with the contribution of the transverse electron trajectory. Now look at Eq.(\ref{srtwob}). This equation includes only the observation angle $\theta_y$ in the phase factor under the integral. 
This means that the transverse constraint motion of the electron in the bending magnet does not affect synchrotron radiation.
So we are justified using a non-covariant approach for considering the constrained electron motion along the nominal orbit in $(x,z)$-plane.

We point out that the cancellation in relativistic time shift and the independence of the Fraunhofer propagator (to be more precise, in space-frequency domain we are dealing with a paraxial approximation of Green's function of nonhomogeneous Helmholtz equation) on the observation angle $\theta_x$  in the far zone can be regarded as the two sides of the same coin: they are manifestation of the cylindrical symmetry when an electron is moving along an arc of a circle. Because of cylindrical symmetry, in order to calculate spectral and angular photon distributions in the far zone, it is not necessary to consider an observer at a general location. The observer is assumed to be located in the vertical plane tangent to the circular trajectory at the origin. In this case observation angle $\theta_x = 0$ and the observation angle $\theta_y$ is above the level of the orbit. In other words, in this very special geometry the $z$-axis is not fixed, but depends on the observer position. However, this way of proceeding can hardly help to obtain radiation fields in the near zone. Indeed, in the near zone we are dealing with the Fresnel propagator, which obviously depends on the constrained motion of the electron. We use far-zone arguments only to show that there is no influence of the difference between the non-covariant and covariant trajectories on the synchrotron radiation  from bending magnets. The cancellation in the relativistic time shift leads to the same outcome in the near zone as it must be.

\subsubsection{Influence of the kick}

We can check our relativistically correct method against something else we know.
Let us discuss the bending magnet radiation from a single electron with a kick with respect to the nominal orbit in $(x,z)$-plane. In this case, we additionally have a translation along the $y$-axis with constant velocity $v_y = v\theta_k$. We can, then, write the offset of the electron as follows $y = \theta_k z'$. Let's put velocity and offset in the relativistic time shift: $\Delta t = t_d - t = -\theta_k^2z'/(2c)
+ \theta_k^2z'/c = \theta_k^2z'/(2c)$.  So, the shift in the total phase under the integral along the path can be expressed by the formula  $\Delta \Phi_T = \omega\theta_k^2 z'/(2c)$. The result agrees with our red shift calculation in the undulator case when the kick is introduced, as it must be.
Synchrotron radiation from bending magnets is emitted in a broad spectrum and its angular-spectral density distributions are not sensitive to red shift of the critical wavelength.

\section{Synchrotron radiation in the case of particle motion on a helix}

The presence of red shift in bending magnet radiation automatically implies the same problem for conventional cyclotron radiation theory. In fact, the conventional theory predicts that there should be no red shift for radiation emitted by an electron with velocity directed along and across the magnetic lines of force. In the ultrarelativistic limit, thre are  well-known analytical formulas that describe the spectral and angular distribution of cyclotron radiation emitted by an electron moving in a constant magnetic field having a non-relativistic velocity component parallel to the field, and an ultrarelativistic velocity component perpendicular to it. According to the conventional approach, exactly as for the bending magnet case, the angular-spectral distribution of radiation is a function of the total velocity of the particle due, again, to the Doppler effect. At present, relativistic cyclotron radiation results are textbook examples (see e.g. \cite{EC, GI, PE}) and do not require a detail description. We note, however, that cyclotron-synchrotron radiation emission is one of the most important processes in plasma physics and astrophysics and our corrections are very important for a much wider part of physics than that of synchrotron or XFEL sources.

\subsection{Existing theory}

A widely accepted expression for the angular and spectral distributions of radiation from an ultra-relativistic electron on a helical orbit were calculated in \cite{WW,EP}. Let us discuss in some detail the
cyclotron radiation emitted by an  electron moving in constant magnetic field  with a non-relativistic component of the velocity parallel to the direction of the magnetic field, and a ultra-relativistic component perpendicular to it. Here 
we shell only give some final results and discuss their relation with the conventional synchrotron radiation theory from bending magnet. In the case of a uniform translation motion with non-relativistic velocity along the magnetic field direction, the radiation field in the far zone according to \cite{GI}, and using the notation in that reference, is given by

\begin{eqnarray}
	{\vec{\widetilde{E}}}(\chi, \alpha)  &&  \sim
	\Bigg\{ \vec{e}_x
	\left[(\xi^2 + \psi^2) K_{2/3} \left(\frac{\omega}{2\omega_c}
	\left(1+\frac{\psi^2}{\xi^2}\right)^{3/2}\right)\right]\cr   && - i \vec{e}_y \left[(\xi^2 +
	\psi^2)^{1/2} \psi K_{1/3}\left(\frac{\omega}{2\omega_c}
	\left(1+\frac{\psi^2}{\xi^2}\right)^{3/2}\right)\right] \Bigg\}~,  \label{EF}
\end{eqnarray}

where $K_{1/3}$ and $K_{2/3}$ are the modified Bessel functions, $\xi = 1/\gamma$, $\psi = \chi - \alpha$, ($\chi$
is the angle between $\vec{v}$ and $\vec{H}$ and $\alpha$ that between $\vec{n}$ and $\vec{H}$);
the angle $\psi$ is clearly the angular distance between the direction of the electron velocity $\vec{v}$ and the direction of observation $\vec{n}$.
Here the $\omega_c$ is defined by $3eH\gamma^2/(2mc)$.

Actually we have already discussed radiation from an ultrarelativistic electron on a helical orbit in the previous section. Equation   Eq. (\ref{srtwoang2bis}) is the result we worked out above for the bending magnet radiation from a single electron with arbitrary angular deflection with respect to nominal orbit. Eq. ( \ref{EF}) does not look the same as  Eq. (\ref{srtwoang2bis}). It will, however, if we now define the small deflection angle $\eta_y = \pi/2 - \chi$ and the observation angle $\theta_y = \pi/2 - \alpha$ (the observer is also assumed to be located in the vertical plane tangent to the trajectory i.e. $\theta_x, \eta_x = 0$): we get the same result as before 
\footnote{The integrals in Eq. (\ref{srtwoang2bis}) can be expressed in terms of the modified Bessel functions: $\int_{0}^{\infty}x\sin[(3/2)\alpha(x+x^3/3)]dx = (1/\sqrt{3})K_{2/3}(\alpha)$,
$\int_{0}^{\infty}x\cos[(3/2)\alpha(x+x^3/3)]dx = (1/\sqrt{3})K_{1/3}(\alpha)$. Then, making the necessary variable changes, the formula  reduces to Eq. ( \ref{EF}) }.
It is clear from Eq. ( \ref{EF}) that if we consider the radiation from an electron with relativistic factor $\gamma$ moving on circular orbit, the introduction of the kick only amounts to a rigid rotation of the angular distribution along the new direction of the electron motion.

\subsection{Radiation for a helical motion and special theory of relativity}

The angular spectral distribution in  Eq. ( \ref{EF}) was recovered in treatments that make no explicit use of the theory of relativity  \cite{WW,EP}. When there is a motion along the field, that is $\eta \neq 0$, the calculation leading to Eq. ( \ref{EF}) is rather elaborate (we performed these calculations in section 9.1.2). It is therefore desirable to have an independent derivation. This was carried out in \cite{OS}. The simplest way of analyzing the radiation for an ultrarelativistic helical motion  makes use of the theory of relativity and involves practically no calculations.

The reference frame $S'$ in which the electron moves in circular motion can be transformed to a frame $S$ in which the electron proceeds following a helical trajectory. In \cite{OS} it was shown that Eq. ( \ref{EF}) holds, indeed, in the frame $S$ for a particle whose velocity is $(v_x,v_y,v_z)  = (v\sin \chi\sin\phi,v\cos\chi, v\sin \chi\cos\phi )$. The Lorentz transformation, which leads to the value $v_y = v\cos\chi$ for the $y$-component of the velocity yields  
$(v_x,v_y,v_z)  = ( v'\sin\phi'/\gamma_y,v_y, v'\cos\phi'/\gamma_y )$, where $\gamma_y = \sqrt{1-v^2_y/c^2}$, $v'$ is the velocity of the electron in the frame $S'$ and the phase angle $\phi' = \phi$ is invariant. This means that, in order to end up in $S$ with a transverse  velocity $v_{\perp} = v\sin \chi$ ( to the magnetic field direction), one must start in $S'$ with $v' = \gamma_yv\sin \chi$. In the ultrarelativistic approximation $\gamma_{\perp}^2 = 1/(1-v^2_{\perp}/c^2) \gg 1$, and one finds the simple result $v = v'$, so that a Lorentz boost with non-relativistic velocity $v_y$ leads to a rotation of the particle velocity $\vec{v}$ of the angle $\eta = \pi/2 - \chi \simeq v_y/c$.  If one transforms the radiation field for a particle in a circular motion in the system $S'$ and neglects second order terms $v_y^2/c^2 \ll 1$ in observation angle and frequency, one obtains the result that the effect of a boost amounts to a rigid rotation of the angular-spectral distribution of the radiation emitted by the electron moving with velocity $v$ on a circle that is, once more,  Eq. ( \ref{EF}).
So the way for computing the radiation in the case of uniform translation is simple.
One describes a complicated situation by finding a reference system where the analysis is already done (radiation in the case of circular motion) and  transforms back to the old reference frame. 

From above argument, one could naively expect that according to the theory of relativity there should be no red shift for the radiation emitted by an electron with velocity directed along and across the magnetic lines of force. But when the situation is described as we have done it here, there does not seem to be any paradox at all; it comes out quite naturally that the covariant way of analyzing the radiation for helical motion  considered in \cite{OS}  is based on the Lorentz transformation. In other words, within the lab frame the Lorentz coordinates are automatically enforced. It assumed  that in the Lorentz lab frame the electron proceeds following a helical trajectory with velocity $v$. This is employed as initial condition. In \cite{OS} it is correctly demonstrated that in  the ultrarelativistic approximation a Lorentz boost along the field direction with non relativistic velocity $v_y$ leads to the circular motion of the electron with the same velocity $v$. Thus the boost will leave the radiation properties unchanged. 

Now what about the value of the electron velocity on a helical orbit in the Lorentz lab frame? How this velocity is defined in \cite{OS}? It is  generally believed that $\vec{x}(t) = \vec{x}(t)_{cov}$ and this is the reason why in \cite{OS} there is no distinction between the two (non covariant and covariant) approaches to describe the electron motion on a helix. 
If we will keep the Lorentz coordinate system in the lab frame downstream  of the kicker, we will find that the covariant velocity on the helical orbit after the kick decreases from $v$ to $v - v\theta^2_k/2$  and the covariant way of analyzing the radiation for a helical motion considered in  \cite{OS} will leads  to a red shift in the critical wavelength, as it must be.

We may also point out that
there are two different ways (from the viewpoint of initial conditions) to organize the same uniform translation along the magnetic field direction in the Lorentz lab  frame. 
Suppose that an electron moves, initially, at  ultrarelativistic velocity $v$ parallel to the $z$- axis upstream a uniform magnetic field (i.e. bending magnet) directed along the $y$-axis. In other words, we start by considering  an electron moving along a circular trajectory that lies in the $(x,z)$-plane.  Then we rotate $\vec{H}$ in the $(y,z)$-plane by  angle $\eta = \pi/2 - \chi$. We consider a situation in which the electron is in uniform motion  with velocity $v\eta$ along the magnetic field direction. It is clear that if we consider the radiation from an electron moving on a circular orbit, the introduction of the magnetic field vector rotation will leave the radiation properties unchanged.
Now we consider another situation in which there is no bending magnet rotation, but there is a kicker upstream the bending magnet. When the kick in $y$ direction is introduced, there is a red shift of the critical wavelength which arise because the electron velocity decreases from $v$ to $v - v\theta_k^2/2$ after the kick.

The difference between these two situations, ending with a final uniform translation along the magnetic fields direction is very interesting.
It comes about as the result of the difference between two Lorentz coordinate systems in the lab frame. By trying to accelerate the electron upstream the bending magnet we have changed Lorentz coordinates for that particular source.
We know that in order to keep a Lorentz coordinates system in the lab frame after the kick we need to perform a clock resynchronization. So we should expect the electron velocity to be changed.
Now the difference between the two setups 
is understandable. When we do not perturb the electron motion upstream of the bending magnet, no clock resynchronization takes place, while when we do perturb the motion, clock resynchronization is introduced.

\section{How to solve problems involving many trajectory kicks}

We shell now discuss the situation where there are $n$ arbitrary spaced kickers, all different from one another in terms of the rotation angle introduced.  Let us consider how we may apply covariant particle tracking in this circumstance, and try to understand what is happening when we have for example an undulator downstream of the kicker setup. One might say that this is getting ridiculous. If one wants to calculate the radiation from the undulator one should take into account all kicks in  the electron trajectory, from the generation of the electron. However, this situation is not surprising, if one analyzes the general expression for the radiation field from a single electron Eq.(\ref{revwied}).
In fact, we should note that, in general, one needs to know the entire history of the electron from $t' = -\infty$ to $t' = \infty$ since the integration in  Eq.(\ref{revwied}) is performed between these limits. 
However, this statement should be interpreted physically, depending on the situation under study: integration should in fact be performed from and up to times when the electron does not contribute to the field anymore. 

We should pointed out that it is the electrodynamics theory, which ultimately decides what part of the particle trajectory is important for calculating undulator radiation  and what part can be neglected. The most important, general statement concerning the relevant  part of the particle trajectory, is that it must be calculated according to the covariant method (if one wants to use the usual Maxwell's equations).

Let us consider the ultrarelativistic assumption $1/\gamma^2 \ll 1$, which is verified for synchrotron radiation setups. In general, the introduction of a small parameter in any theory brings simplifications.  
The ultrarelativistic approximation implies a paraxial approximation and 
Eq.(\ref{revwied}) can be simplified to  Eq.(\ref{generalfin}).
Suppose that we take a situation in which the rotation angle  of the first  bending magnet upstream of the undulator is much larger than $1/\gamma$. In other words, we now consider an electron moving along a standard synchrotron radiation setup. The electron enters the setup via a bending magnet, passes through a straight section, an undulator, and another straight section. Finally, it leaves the setup via another bend.     
Note that, although the integration in Eq.(\ref{generalfin}) is performed from $-\infty$ to $\infty$, the only (edge) part of the trajectory into the bending magnets contributing to the integral is of order of the radiation formation length $d_s$. 
Mathematically, it is reflected in the fact that $\Phi_T(z')$ in Eq.(\ref{generalfin})
exhibits more and more rapid oscillations as $z'$ becomes larger than the formation length. 
At the critical wavelength the formation length is simply of order of $\rho/\gamma$, 
$\rho$ being the radius of the bend. That simply corresponds to an orbiting angular interval $\Delta\theta \simeq 1/\gamma$. Typically, the critical wavelength of the radiation from a bending magnet in synchrotron radiation source is about 0.1 nm and the formation length in this case is only  few millimeters.

Note that for ultrarelativistic systems in general, the formation length is always much longer than the radiation wavelength. This counterintuitive result follows from the fact  that for ultrarelativistic systems one cannot localize sources of radiation  within a macroscopic part of the trajectory.   
The formation length can be considered as the longitudinal size of a single electron source. It does not make sense at all to talk about the position where electromagnetic signals are emitted within the formation length. This means that, as concerns the radiative process in the bending magnet, we cannot distinguish between radiation emitted at point $A$ and radiation emitted at point $B$ when the distance between these two points is shorter than the formation length $d_s$. Let us now consider the case of a straight section of length $L$ inserted between the bending magnet and the undulator. One can still use the same reasoning considered for the bend to define a region of the trajectory where it does not make sense to distinguish between different points. As in bending magnet case, the observer sees  a time compressed motion of the source and  in the case of straight motion the apparent time corresponds to an apparent  distance $\lambdabar \gamma^2$.  At the critical wavelength the bending magnet formation length $d_s \simeq \rho/\gamma$ is simply order of the straight line formation length  $\lambdabar \gamma^2$.

Intuitively,  bending magnets act like switchers for the ultrarelativistic electron trajectory. 
We consider the case when switchers are presented in the form of bending magnets, but other setups can be considered where switchers have different physical realizations. 
The only feature that these different realizations must have in common, by definition of switcher, is that the switching process must depends exponentially on the distance from the beginning of the process. Then, a characteristic length $d_s$ can be associated to any switcher.
Consider, for example, a plasma accelerator where an electron is accelerated with high-gradient fields. In this case it is the accelerator itself that switches on the relativistic electron trajectory, since acceleration in the GeV range takes place within a few millimeters only. In the X-ray range the acceleration distance $d_a$ is  shorter than the formation length $\lambdabar \gamma^2$ for the following straight section. In this particular case length $d_a$ plays the role of the characteristic length of the switcher $d_s$, which switch on the ultrarelativistic electron trajectory.

Let us now return to our consideration of the standard synchrotron radiation setup and let us analyze the radiation process in an insertion device (undulator). We have actually the "creation" of the relativistic electron within a distance of order  $\lambdabar \gamma^2$  from the very beginning of the straight section upstream the undulator. It is assumed that the length of the straight section $L$ is much longer than the formation length $\lambdabar \gamma^2$ that is clearly always the case in  the X-ray range. When the switching distance $d_s \lesssim \lambdabar \gamma^2 \ll L$, the nature of the switcher is not important for  describing the radiation from the undulator installed within the straight section. 

Downstream of the switcher we have a uniformly moving electron.
The fields associated to an  electron with a constant velocity exhibit an interesting behavior when the speed of the charge approaches that of light. Namely, in the space-frequency domain there is an equivalence of the fields of a relativistic electron and those of a beam of electromagnetic radiation.  
In fact, for a rapidly moving electron we have  nearly equal transverse and mutually perpendicular electric and magnetic fields. These are indistinguishable  from the fields of a beam of radiation. 
This virtual radiation beam has a macroscopic transverse size of order $\lambdabar\gamma$ \footnote{
An ultrarelativistic electron at synchrotron radiation facilities at nanometer-wavelength scale (in the space-frequency domain) has indeed a macroscopic  transverse size of order of 1 $\mu$m}.  
At the exit of the switcher we have a "naked" (or "field-free")  electron i.e. an electron that is not accompanied by  virtual radiation fields.  There is a process of formation of the "field-dressed" electron (i.e. the formation of the fields from a fast moving charge)  within the distance of order  $\lambdabar \gamma^2$  from the very beginning of the straight section downstream of the switcher.

The electron trajectory being divided into two essentially different parts: before and after the switcher. If we accelerate the electron in the lab frame upstream of the switcher, the information about this acceleration is included into the first part of the covariant trajectory. But this acceleration prehistory  (together with the fields of the ultrarelativistic electron) is washed out during the switching process and at the entrance of the straight section we have a "naked" electron.

We start with the description of the field formation process along the straight section downstream of the switcher, based on the covariant approach. First of all we have to synchronize distant clocks within the lab frame. The synchronization procedure that follows is the usual Einstein synchronization procedure. It is assumed that in the Lorentz lab frame the electron proceeds following a rectilinear trajectory with velocity $v$. This assumption is used as initial condition. Then we can analyze situation downstream the switcher by using the usual Maxwell's equations. 

When one analyzes the process of "field-dressed" electron formation from the viewpoint of the non covariant approach, one assumes the same initial conditions  (rectilinear trajectory with velocity $v$) for the electron motion. Then one solves the electrodynamics problem of  fields formation by using the usual Maxwell's equations. We already mentioned that the type of clock synchronization which results in time coordinate $t$ in an electron trajectory $\vec{x}(t)$ is never discussed in accelerator physics.
However, we know that the usual Maxwell's equations are only valid  in the Lorentz frame. The non covariant approach is  obviously  based on a definite synchronization assumption, but this is actually a hidden assumption. In other words, within the lab frame the Lorentz coordinates are then automatically enforced. 

So one should not be surprised to find that in this simple case of rectilinear motion (i.e. in the situation when we have only deal  with the description initial conditions) there is no difference between covariant and non covariant calculations of the initial conditions at the undulator entrance. 

Because of the characteristics of undulator radiation, in order to calculate the radiation field within the central cone, we only need to account for the longitudinal accelerated motion. So we are satisfied using a non covariant approach for considering the constrained motion along the undulator.  
We conclude that it does not matter which approach is used to describe the standard synchrotron radiation setup. The two approaches, treated according to Einstein's or absolute time synchronization conventions give the same result for the radiation within the central cone.

Let us now see what happens with a weak dipole magnet (a kicker), which is installed in the straight section upstream of the undulator and is characterized by a small kick angle $(\gamma\theta_k)^2 \ll 1$. What do we expect for the undulator radiation? At first glance the situation is similar to the switcher setup and the electron trajectory is again divided into two parts: before and after the kicker. The most important difference, however, is that  electrodynamics now dictates that both trajectories are important for the calculation of the undulator radiation. When the electron passes through the kicker there is no synchrotron radiation (to be more precise, in this case radiation is indistinguishable from the self-electromagnetic fields of the electron), washing out the virtual radiation fields like in the switcher case. We expect that an electron that passes through a kicker is  still "field-dressed", but we have an electron whose fields has been perturbed, and now include information about the acceleration.

According to the conventional theory, as usual for Newtonian kinematics, the Galilean vectorial law of addition of velocities is actually used. Non-covariant particle dynamics shows that the direction of the electron trajectory changes after the kick, while its speed remains unvaried. In contrast, covariant particle tracking, which is based on the use of Lorentz coordinates, yields different results for the trajectory of the electron. The electron speed decreases from $v$ to $v(1 - \theta^2_k/2)$.  This result is at odds with the prediction from non-covariant particle tracking, because Einstein's addition law  for non-parallel velocities is used to calculate the electron trajectory.

According to the conventional algorithm for solving electrodynamics field equations, which deals with the usual Maxwell's equations, and particle trajectories calculated by using non-covariant particle tracking, the undulator radiation along the velocity direction has no red shift of resonance frequency for any kick angle $\theta_k$. 

According to the correct coupling of fields and particles, there is a remarkable  prediction of synchrotron radiation theory  concerning the setup described above. Namely, there is a red shift of  the  resonance frequency of the undulator radiation in the kicked direction. To show this, let us first consider the covariant treatment, which makes explicit use of Lorentz transformations.
When the kick is introduced, covariant particle tracking predicts a non-zero red shift of the resonance frequency, which arises because in Lorentz coordinates the electron velocity decreases from $v$ to $v - v\theta_k^2/2$, while the velocity of light is unvaried and equal to the electrodynamics constant $c$.
The red shift in the resonance frequency can be expressed by the formula $\Delta\omega_{r}/\omega_{r} = - \gamma^2\theta_k^2/(1+K^2/2)$.

It should be note, however, that there is another satisfactory way of explaining the red shift.  
We can reinterpret this result with the help of a non-covariant treatment, which deals with non- covariant particle trajectories, and with Galilean transformations of the electromagnetic field equations. According to non-covariant particle tracking the electron velocity is unvaried. However, Maxwell's equations do not remain invariant with respect to Galilean transformation,  and the velocity of light has increased from $c$, without kick, to $c(1 + \theta^2/2)$ with kick.  The reason for the velocity of light being different from the electrodynamics constant $c$ is due to the fact that, according to the absolute time convention, the clocks after the kick are not resynchronized.
Now everything fits together, and our calculations  show that covariant and non-covariant treatments (at the correct coupling fields and particles) give the same result for the red shift prediction, which is obviously convention-invariant and  has direct objective meaning.

One way to demonstrate incompatibility between the standard approach to relativistic electrodynamics, which deals with the usual Maxwell's equations, and particle trajectories calculated by using non-covariant particle tracking, is to make a direct laboratory test of synchrotron radiation theory. In other words, we are stating here that, despite the many measurements done during decades, synchrotron radiation theory is not an experimentally well-confirmed theory.

We have already pointed out that results of the beam splitting experiment in reference \cite{NUHN} confirm our  correction for spontaneous undulator emission. These measurements clearly show that there is a red shift in the resonance wavelength when the kick is introduced \cite{NUHN}.    
The potential for exploiting synchrotron radiation sources in order to confirm the predictions of  corrected synchrotron radiation theory, is analyzed in \cite{OURS2}. The emittance of the electron beam in new generation synchrotron radiation sources is small enough, so that one can neglect finite electron beam size and angular divergence  in the soft X-ray wavelength range, and such synchrotron radiation source can be examined under the approximation of a filament electron beam. This allows us to take advantage of analytical presentations for single electron synchrotron radiation fields.  The spontaneous radiation pulse goes through a monochromator filter and its energy is subsequently measured by a detector. The proposed experimental procedure is relatively simple, because is based on relative  measurements in the velocity direction with and without transverse kick. Such a measurement is critical, in the sense that the prediction of conventional theory is the absence of red shift, and has never been performed to our knowledge.

\section{Summary}

\subsection{Covariant particle tracking in a constant magnetic field}

The study of relativistic particle motion in a constant magnetic field according to usual accelerator engineering, is intimately connected with the old (Newtonian) kinematics: the Galilean vectorial law of addition of velocities is actually used. However, Maxwell's equations are not covariant under   Galilean transformations. We cannot take one kinematics for one part of physical phenomena and the other kinematics for the other part, namely Galilean transformations for mechanics and Lorentz transformations for electrodynamics. We must decide which part must be retained and which must be modified. We demonstrated that there is no principle difficulty with the non-covariant approach in mechanics and electrodynamics. It is perfectly satisfactory. It does not matter which transformation is used to describe the same reality. Nevertheless, there is a reason to prefer the covariant approach within the framework of both mechanics and electrodynamics. As we have seen, in fact, the choice of the non-covariant approach also implies the use of much more complicated (anisotropic) electromagnetic field equations.  

The Lorentz transformations give rise to non-Galilean transformation rules for  velocities.
According to the covariant approach, the Einstein addition law  for non-parallel velocities is used to calculate the electron trajectory in a constant magnetic field.    
It is not surprising that there is a difference between covariant and non-covariant  velocities, in particular $v_{cov} < v$.
But non-covariant and covariant approaches produce the same particle's three-momentum.  The point is that both approaches describe correctly the same  physical reality and curvature radius of the trajectory in a given magnetic field and consequently the three-momentum has an objective meaning, i.e. it is convention-invariant. In contrast to this, the velocity of the particle has objective meaning to within a certain accuracy because of the finiteness of velocity of light. 

Authors of textbooks are dramatically mistaken in their belief about the usual momentum-velocity relation. The covariant equation of motion 
tells us that force is the rate of change of momentum, but it does not tells  us how the momentum varies with speed.
The usual equations for a particle motion in the three-dimensional space are not a mathematical result, derived from the covariant four-dimensional dynamics equation. In these equations the additional restriction has already been imposed: it is implicit in the assumption that we are working in the  three-dimensional momentum representation $\vec{p} = m\vec{v}\gamma$.

We therefore discovered that for a  motion along a curved trajectory, the usual momentum-velocity relation does not hold. Summarizing, we can state that for a rectilinear the motion covariant velocity transformation (made according to Einstein's addition velocity rule) is consistent with the covariant three-momentum transformation,  and the usual momentum-velocity relation holds. But this result was incorrectly extended to curved trajectories.

We emphasize the difference between the notion of path and trajectory in three-dimensional space.
The conventional nature of trajectory in relativistic dynamics should not be confused with the notion of path. The trajectory of a particle conveys more information about its motion, because every position is described additionally by the corresponding time instant. The path is rather a purely geometrical notion. The path  has an exact objective meaning  i.e. it is convention-invariant. In contrast to this, consistently with conventionality of the value of velocity, the trajectory of a particle is convention-dependent and has no exact objective meaning.

Attempts to solve the dynamics equation  in manifestly covariant form  in the case of constant magnetic field can be found in literature. The trajectory which was found does not include relativistic kinematics effects. Therefore, it cannot be identified with the covariant trajectory even if, at first glance, it appears to be derived following covariant prescription.
It is generally believed that the usual momentum-velocity relation holds for the arbitrary world-line $x(\tau)$. We state that this incorrect and that the four-velocity can not be decomposed into  $u = (c\gamma, \vec{v}\gamma)$  when we dealing with a particle accelerating along the curved trajectory in the Lorentz lab frame. The presentation of the time component as  the simple relation $d\tau = dt'/\gamma$ between proper time and coordinate time is based on the hidden assumption that the type of clock synchronization, which provides the time coordinate $t$ in the lab frame, is based on the use of the absolute time convention.

The theory of relativity shows that the unusual momentum-velocity relation discussed above has to do with the acceleration along curved trajectories. It is, in fact, a relativistic effect which has no a non-covariant analogue. In this case there is a difference between  covariant and non-covariant  particle trajectories. One can see that this essential point has never received attention by the physical community. Only the solution of the dynamics equations in covariant form gives the correct coupling between the usual Maxwell's equations and particle trajectories in the lab frame. 
A closer analysis of the concept of velocity, i.e. a discussion of the methods by which a time coordinate can actually be assigned in the lab frame, opens up the possibility of a description of such physical phenomena as radiation from a relativistic electron accelerating along a curved trajectory in accordance with the theory of relativity.

\subsection{Relativity and XFELs}

\subsubsection{Relativistic kinematics effects and ultrarelativistic asymptotics}

The appearance of relativistic effects in radiation phenomena does not depend on  a large  speed of the radiation sources.  Lorentz transformations always give rise to relativistic kinematic  effects and no matter how small ratio $v/c$ may be. According to the covariant approach,
the various relativistic kinematics effects turn up in successive orders of approximation. 

In lowest (first) order. - relativity of simultaneity. 

In the next (second) order. - time dilation, Lorentz contraction, and Wigner rotation. 

In still higher order. - relativistic correction in the law of composition of velocities.

These relativistic kinematics effects give rise to convention-invariant relativistic radiation effects. In particular, the relativity of simultaneity is responsible for aberrations to the first order of $v/c$. In the second order, one gets the transverse Doppler effect. According to the classical theory there should be no change in frequency. From the relativity theory, the difference arises from the time dilation, and is of order $(v/c)^2$.

For an arbitrary parameter $v/c$ covariant calculations of the radiation process is very difficult. There are, however, circumstances in which  calculations can be greatly simplified. As example of such circumstance is a synchrotron radiation setup.
Similar to the non-relativistic asymptote, the ultrarelativistic asymptote also  provides the essential simplicity of the covariant calculation. The reason is that the ultrarelativistic assumption implies the paraxial approximation. Since the formation length  of the radiation is much longer than the wavelength, the radiation is emitted at small angles of order $1/\gamma$ or even smaller, and we can therefore enforce the small angle approximation. We assume that the transverse velocity  is small compared to the velocity of light. In other words, we use a second order relativistic approximation for the transverse motion.
Instead of small (total) velocity parameter $(v/c)$ in the non-relativistic case, we use a small transverse velocity parameter  $(v_{\perp}/c)$.
The next step is to analyze the longitudinal motion, following the same method. We should remark that the analysis of the longitudinal motion in a synchrotron radiation setup is very simple. If we evaluate the transformations up to  second order  $(v_{\perp}/c)^2$, the relativistic correction in the longitudinal motion does not appear in this approximation. 

According to covariant approach, the various relativistic kinematics effects concerning to the synchrotron radiation setup, turn up in successive orders of approximation.

In the first order $(v_{\perp}/c)$. - relativity of simultaneity. Wigner rotation, which in the ultrarelativistic approximation appears in the first order already, and results directly from the relativity of simultaneity.

In the second order $(v_{\perp}/c)^2$. - time dilation. Relativistic correction in law of composition of velocities, which already appears in the second order, and results directly from the time dilation.

\subsubsection{Effect of aberration of light in XFELs}

The Wigner rotation effect plays an essential role only in the description of extended (macroscopic) relativistic objects. But up to 21 st century there were no macroscopic objects possessing relativistic velocities, and there was a general belief that only microscopic particles  in experiments can travel at velocities close to that of light. The 2010s saw a rapid development of new laser light sources in the X-ray wavelength range. An X-ray free electron laser (XFEL) is an example where improvements in accelerator technology makes it possible to develop ultrarelativistic macroscopic objects with an internal structure (modulated electron bunches), and the first order kinematics term  $(v_{\perp}/c)$ plays an essential role in their description. 
We demonstrated that relativistic kinematics enters XFEL physics in a most fundamental way through the Wigner rotation of the modulation wavefront, which,  in ultrarelativistic approximation, is closely associated to the relativity of simultaneity.

There are several cases where the first order relativistic effect can occur in XFELs, mainly through the introduction of an angular trajectory kick. 
It is generally understood that a transverse kick does not change the orientation of a modulation wavefront, and hence suppresses  the radiation emitted in the direction of the electron motion. 
We have shown that the covariant approach within the framework of both mechanics and electrodynamics 
predicts an effect in complete contrast to the conventional  treatment. Namely, in the ultrarelativistic limit, the wavefront of modulation, that is a plane of simultaneity,  is always perpendicular to  the electron beam velocity. As a result, the Maxwell's equations predict strong emission of coherent undulator radiation  from the  modulated electron beam in the kicked direction.

It is possible to present intuitive arguments to explain why a modulated electron beam after the kick radiates in the kicked direction. Consider, downstream of the kicker, a Lorentz reference frame $S'$ moving with uniform motion at speed $v_{\perp}$  relative to the lab frame $S$. A setup in the  inertial frame $S'$ downstream of the kicker reproduces the situation upstream of the kicker.  In the previous sections we transformed the source (modulated electron beam) to the lab frame and after this we calculated radiation emitted by such a source. Now, it would be interesting to show that there is another possibility. The emitted radiation can be calculated in the moving  Lorentz frame $S'$. The next question is, what is the change in the radiation beam direction which is viewed from the lab frame? The direction of the wavefront of the light wave depends essentially on the velocity of the light source relative to the observer, a phenomenon commonly known as aberration. Aberration of light is a shift of the direction of an incident beam of light due to the motion of the source relative to the observer. An elementary explanation of this effect is well-known. This phenomenon is fully understandable in terms of transformation of velocities between different (inertial) reference frames both in Einstein's kinematics  and in old (Newtonian) kinematics treatments.

The rule for computing aberration effect is simple. One takes the velocity of light with respect to the source and adds it vectorially to the velocity of the source with respect to observer. The direction of the resulting vector is the apparent direction of the light source as measured at the observer position.  Application of this rule for the case when the angle of aberration is at its maximum, i.e. when the direction of the observer's motion is perpendicular to the direction of the source radiation, results in an angle of aberration of $v_{\perp}/c$ radians. For this result to hold, it is important that the transverse speed of the observer $v_{\perp}$ is very much smaller than the speed of light $c$. Since all the velocities mentioned in the rule are relative velocities, the rule confirms to the principle of relativity. The radiation in the kicked direction can be quantitatively explained with the help of the rule described above for calculating the angle of aberration.  We find that in ultrarelativistic approximation $v \to c$ and the aberration angle $v_{\perp}/c$ coincides  with the kicked angle of the electron beam  $\theta_k = v_{\perp}/v$.

\subsubsection{Relativistic kinematics effects and existing XFEL theory}

The usual XFEL theory based on the use  of old Newtonian kinematics for particle dynamics and the  Einstein's kinematics for the electrodynamics. In fact, the usual theoretical treatment of relativistic particle dynamics involves only a corrected Newton's second law and is based on the use Galilean transformations. 
For rectilinear motion of the modulated electron beam, non-covariant and covariant approaches produce the same trajectories, and Maxwell's equations are compatible with the result of conventional particle tracking. 
However, one of the consequences of non-commutativity of non-collinear Lorentz boosts is a difference between covariant and non covariant particle trajectories in a constant magnetic field. 
We conclude that previous theoretical and experimental results in XFEL physics should be reexamined in light of the pointed difference between conventional and covariant particle tracking. 

In a typical configuration for an XFEL, the orbit of a modulated electron beam is controlled to avoid 
large excursions from the undulator axis. All existing XFEL codes are based on  a model in which
the modulated electron beam moves only along the undulator axis.
However, random errors in the focusing system can cause angular trajectory errors (or "kicks"). Analysis of the trajectory errors on the XFEL amplification process showed that any XFEL undulator magnetic field must satisfy stringent requirements. However, semi-analytical studies of this critical aspect in the design of a XFEL sources are based on an incorrect coupling of fields and particles. The pleasant surprise is that  the tolerances predicted  are more stringent than they need be according to the corrected XFEL theory. This can be considered as one of the reason for the exceptional progress in XFEL developments over last decade.

Let us now move on to consider the predictions of the existing XFEL theory in the case of non-collinear electron beam motion. As well-known result of conventional particle tracking states that after an electron beam is kicked by a weak dipole magnet there is a change in the trajectory of the electron beam, while the orientation of the modulated wavefront remains as before. In other words, the kick results in a difference between the directions of the electron motion and the normal to the modulation wavefront (i.e. in a wavefront tilt). In existing XFEL theory the wavefront tilt is considered as real. According to this belief, there are many physical effects that can be understood in therms of wavefront tilt. Let us consider one example. One finds some papers (see e.g. \cite{TKS, ML2}) which say that a wavefront tilt leads to significant degradation of the electron beam modulation in XFELs.

First, suppose that modulation wavefront is perpendicular to the beam velocity $v$.
The effect of betatron oscillations, which can influence the operation of the XFEL, has its origin in an additional longitudinal velocity spread. Particles with equal energies, but with different betatron angles, have different longitudinal velocities. In other words, on top of the longitudinal velocity spread due to the energy spread, there is an additional source of velocity spread. To estimate the importance of the last effect, we should calculate the dispersion of the longitudinal velocities due to both effects. The deviation of the longitudinal velocity from nominal value is $\Delta v_z = v\Delta \gamma/\gamma^3 - v\Delta \theta^2/2$. 
The finite angular spread of the electron beam results in a difference in time when each electron arrives at the same longitudinal position, and this spoils the phase coherence. This is so called normal debunching effect.

From the viewpoint of the  existing XFEL theory, the time difference is enhanced by the kick angle $\theta_k$. In this case, according to conventional (non-covariant) particle tracking, the angle of  wavefront tilt is $\theta_{tilt} = \theta_k$. It is a widespread belief that the wavefront tilt has physical meaning, and that the deviation of the longitudinal velocity component (i.e. velocity component which is perpendicular to the modulation wavefront within the framework of Galilean kinematics) is now given by the expression  $\Delta v_z =  - v|\Delta \vec{\theta} + \vec{\theta}_k|^2/2$.  
If such picture is correct, the crossed term $v\vec{\theta}_k\cdot\Delta \vec{\theta}$ leads to a significant degradation of the modulation amplitude. This mechanism is called smearing of modulation and should be distinguished from the normal debunching.

Many experts would like to think that any debunching process obviously has objective meaning. The theory of relativity says, however, that normal debunching has objective meaning, but smearing effects not exist at all. The explanation of the new debunching mechanism  clearly demonstrates the essential dependence of the smearing effect on the choice of the coordinate system in the four-dimensional space, which from the physical point of view is meaningless.
In ultrarelativistic asymptotics the wavefront tilt has no exact objective meaning since, due to finiteness of the velocity of light, we cannot specify any experimental method by which this tilt could be ascertained. The angle of wavefront tilt depends on the choice of a procedure for clock synchronization in the lab frame, as a result of which it can be given any preassigned values within the interval $(0,\theta_k)$. For instance, in the ultrarelativistic asymptote, the orientation of the modulation wavefront is always perpendicular to the electron beam velocity (i.e. $\theta_{tilt} = 0$) when the evolution of the modulated electron beam is treated using Lorentz coordinates. No physical effects may depends on an arbitrary constant or an arbitrary function
\footnote{It has been claimed in the recent paper \cite{ML2} that accounting for the quadrupole lattice in the baseline XFEL undulator  it is possible to obtain a mechanism for the modulation wavefront to tilt forward, towards the new direction of propagation. On the basis of these claims, \cite{ML2}  has even concluded that the problem is solved and, therefore, that the conventional approach requires neither revision nor replacement. However, the paper \cite{ML2} is incorrect and misleading. The new mechanism of wavefront rotation depends on the choice of a coordinate system, and  therefore it has no physical meaning. This wavefront rotation effect is completely analogous to the smearing effect. For instance, it is easy to see that when the evolution of the modulated electron beam is treated by using Lorentz coordinates, the wavefront is always perpendicular to the velocity and therefore new mechanism of wavefront rotation is not a real phenomenon}.

The tilt of the modulation wavefront is not a real observable effect. Indeed, if we  couple particle system with electromagnetic fields in accordance with the principle of relativity, we find that  coherent undulator radiation from the modulated electron beam is always emitted in the kicked direction, independently of the system of coordinates. It is not difficult to see this using a Lorentz coordinate system where Maxwell's equations are valid and the modulation wavefront is always perpendicular to the beam velocity. In Maxwell's electrodynamics, coherent radiation is always emitted in the direction of the normal to the modulation wavefront. Indeed, we may consider the amplitude of the beam radiated as a whole to be the resultant of radiated spherical waves. This is because Maxwell's theory has no intrinsic anisotropy. The electrons lying on the plane of simultaneity gives rise to spherical radiated wavelets, and these combine according to Huygens'principle to form what is effectively a radiated wave.

We can derive the same results for observables (the direction of radiation propagation  has obviously an exact objective meaning) with the help of Galilean transformations. According to this old kinematics, the orientation of the modulation wavefront is unvaried. However, Maxwell's equations do not remain invariant with respect to Galilean transformations and the choice of the old kinematics implies the use of anisotropic field equations. In particular, the wave equation for radiated spherical wavelets transforms into Eq.(\ref{GGT2}). The main difference consists in the anisotropic crossed term,  which is of order $v_{\perp}/c$. In this case the secondary waves (wavelets) are not spherical, but they are all equal as a consequence of homogeneity. As a result, the wavefront remains plane but the direction of propagation is not perpendicular to the wavefront. In other words, the radiation beam motion and the radiation wavefront (phase front) normal have different directions. Then, the  Huygens'construction shows that the radiated wave propagates in the kicked direction with the wavefront tilt $\theta_k$.

Now let us understand physically why the new debunching mechanism does not exist in framework of Galilean kinematics. In this old kinematics the crossed term $v\vec{\theta}_k\cdot\Delta \vec{\theta}$ leads to a degradation of modulation amplitude in the forward direction. Our Galilean transformed electrodynamics says, however, that by making a measurement on the coherent radiation, one can observe only radiation in the kicked direction. But the crossed term is absent in the expression for the deviation of the velocity component  along the kicked direction. It comes out quite naturally that the smearing effect is not a real phenomenon.

The two (covariant and non-covariant) approaches give the same result for real observable effects. The choice between two different approaches is a matter of pragmatics. However, we would like to emphasize a difference in the conceptual background between these two approaches. The non-covariant approach gives additionally a physical insight into the particular laws of nature it deals with. For instance, the dynamical line of arguments explains the radiation in the kicked direction is based on the structure of the electromagnetic field equations. In the covariant approach  the dynamics, based on the electromagnetic field equations, is actually hidden in the language of relativistic kinematics (Wigner rotation).   

The existing XFEL theory based on the use of the absolute time convention (i.e. old kinematics)  for particle dynamics. The understanding of Galilean transformations in terms of the theory of relativity  has always represented a tough challenge to the physicists who meet those new concepts for the first time. The aim of this final note about existing XFEL theory
is to give a new proof of the conflict between conventional particle tracking and Maxwell's electrodynamics. This new proof is perhaps simpler than the ones we have given before.  The  purpose is to show how one can demonstrate in a simple way that the conventional theory is absolutely incapable of correctly describing the distribution of the electromagnetic fields  from a fast moving modulated electron beam downstream the kicker.

In the case of Maxwell's electrodynamics, the fields of the modulated electron beam moving with a constant velocity exhibit an interesting behavior when the velocity of charges approaches that of light. In the space-time domain there is an equivalence of the fields of a relativistic modulated electron beam and those of a laser-like radiation beam.  In fact, for a rapidly moving  modulated electron beam we have  nearly equal transverse and mutually perpendicular electric and magnetic fields. These are indistinguishable  from the fields of a laser beam. According to Maxwell's equations, the wavefront of the laser beam is always perpendicular to the propagation direction \footnote{Within the deep asymptotic region when the transverse size of the modulated electron beam $\sigma \ll \lambdabar\gamma$ the Ginzburg-Frank formula can be applied.  In this asymptotic region one has no more electron beam emittance effect, and radiation can be considered as virtual radiation from a filament electron beam (with no transverse dimensions).  However, in XFEL practice we only deal with the deep asymptotic region where $\sigma \gg  \lambdabar\gamma$. Then, it can be seen that the field distribution in the space-time domain is essentially a convolution in the space domain between the transverse charge distribution of the electron beam and the field spread function described by the Ginzburg-Frank formula. Assuming a Gaussian (azimuthally-symmetric) transverse density distribution of the electron beam we obtain the
radially polarized virtual radiation beam}.

This is indeed the case for virtual laser-like radiation beam  in the region upstream the kicker. 
In the old kinematics case, the kick results in a difference between the directions of the electron motion and the normal to the modulation wavefront, i.e. the kick results in a modulation wavefront tilt. Now let us see what happens with a virtual radiation beam. What do we expect for radiation wavefront orientation after the kick? In existing literature theoretical analysis is presented, of an XFEL driven by an electron beam with wavefront tilt, and this analysis is based on the exploitation  of usual Maxwell's equations and standard simulation codes. 

At first glance Maxwell's electrodynamics dictates that the wavefront of the radiation beam must be always perpendicular to the propagation direction.
The most important thing, however, is that the old kinematics now says that
the wavefront of virtual radiation beam  remains as before, but the direction of propagation is not perpendicular to the radiation beam wavefront. In other words, the radiation beam motion and the radiation wavefront normal have different directions.  So one should not be surprised to find that the virtual radiation beam (which is indistinguishable from the real radiation laser-like beam in ultrarelativistic asymptote) propagates in the kicked direction with the wavefront tilt $\theta_k$.
This is the prediction of conventional XFEL theory and is obviously absurd from the viewpoint of  Maxwell's electrodynamics. Therefore, something is fundamentally, powerfully, and absolutely  wrong. 
The difficulty above is a part of the continual problem of XFEL physics, which started with 
coherent undulator radiation from an ultrarelativistic modulated electron beam in the kicked direction, and now has been focused on the wavefront tilt of the self-electromagnetic fields of the  modulated electron beam.

Now let us return to the virtual laser-like radiation beam with wavefront tilt. We have already remarked that the usual study of a modulated electron beam motion in a magnetic field of weak dipole magnet is intimately connected with the old kinematics. 
It does not matter which kinematics and hence transformation is used to describe the same reality. What matter is that, once fixed, such kinematics should be applied and kept in a consistent way in both dynamics and electrodynamics.  
We can interpret the wavefront tilt of a virtual radiation beam with the help of the non-covariant treatment, which deals with non-covariant particle tracking, and Galilean transformations of electromagnetic field equations. The choice of the old kinematics implies the use of anisotropic field equations. As a result, the virtual radiation beam motion and virtual radiation wavefront normal have different directions. Using only a kicker setup (i.e. without undulator radiation setup) we demonstrated that in conventional XFEL theory the description of the dynamical evolution in the lab frame is based on the use of the absolute time convention. In this non-covariant particle tracking, time differ from space and particle's trajectories can be seen from the lab frame view as the result of Galilean boosts that track the motion of the modulated electron beam through the kicker setup.

\subsection{Relativity and synchrotron radiation sources}

The first order kinematics term $(v_{\perp}/c)$ plays an essential role only in the description of the coherent radiation from the modulated electron beam. In a storage ring the distribution of the longitudinal position of the electrons in a bunch is essentially uncorrelated.
In this case, the radiated fields due to different electrons are also uncorrelated and the average power radiated is a simple sum of the radiated power from individual electrons; that is we sum intensities, not fields. 
A motion of the single ultrarelativistic electron in a constant magnetic field, according to the theory of relativity, influences the kinematics terms of the second order $(v_{\perp}/c)^2$ only. 

The relativistic correction to the synchrotron radiation emission from a single electron appears if and only if the transverse electron trajectory is involved in the solution of electrodynamics equations. If we analyze  the general expression for the synchrotron radiation field in the far zone, we find that the  term which depends on the transverse position of the electron  can be written as $\exp i(\omega/c)[\theta_x x(z') +  \theta_y y(z')]$. This is simply the exponent in the Fraunhofer propagator. We conclude that the observation angle in the Fraunhofer phase factor under the integral must be related with the contribution of the transverse electron trajectory. If we consider the limit as the observation angles go to zero, we find that the transverse electron trajectory
$x(z'), y(z')$ does not affect synchrotron radiation emission.

In a bending magnet we have an electron which is moving along an arc of a circle. Suppose that trajectory lies in the $(x,z)$-plane. Note that the geometry of the electron motion has a cylindrical symmetry. Because of this symmetry, in order to calculate spectral and angular photon distributions, it is not necessary to consider an observer at  arbitrary observation angle $\theta_x$.
The observer in the standard treatment is assumed to be located in a vertical plane tangent to the circular trajectory at the origin, at an angle $\theta_y$ above the level of the orbit. In other words, in this geometry $\theta_x = 0$ and  the $z$ axis is not fixed, but depends on the observer position. 
This means that transverse constrained motion of the electron in the bending magnet does not affect the synchrotron radiation. So, all we have to do is project the motion on the $z$-axis and we are satisfied using conventional approach for the description of the bending magnet radiation. 
This is because we deal with  cylindrical symmetry when an electron is moving along an arc of a circle.

We now move on to consider another situation, a very practical one. To generate specific synchrotron radiation characteristics, radiation is often produced from special insertion devices called undulators.
The resonance approximation, that can always be applied in the case of undulator radiation setups, yields  simplifications of the theory.  This approximation does not replace the paraxial one, but it is used together with it. It takes advantage of another parameter that is usually large, i.e. number of undulator periods $N_w \gg 1$. In this approximation, all undulator radiation is emitted within an angle much smaller than $1/\gamma$. This automatically selects observation angles of interest. In fact, if we consider observation angles outside the diffraction angle, we obtain zero intensity with accuracy $1/N_w$. In working out the corresponding formula for the radiation field in the far zone using the limitation for the observation angles described above, we find that observation angles in the Fraunhofer phase factor can be taken to be zero and that the transverse constrained electron trajectory does not affect the undulator radiation. So, we are satisfied using the conventional approach for describing the undulator radiation into the central cone, that is the practical situation of interest. This practically means that the relativistic kinematics effects (similar to the bending magnet radiation) are only important  in  the prediction of the initial conditions at the radiator entrance.

We would like to make a historical note. The difference between covariant and non-covariant particle trajectories was never understood. So, accelerator physicists did not appreciate that there was a contribution to the synchrotron radiation from relativistic kinematics effects. They thought only in terms of old (Newtonian)  kinematics that was not compatible with Maxwell's equations.  At this point, a reasonable question arises: since storage rings are designed without accounting for the relativistic kinematics effects, how can they actually operate?  In fact, electron dynamics in storage ring is greatly influenced by the emission of radiation. Due to synchrotron radiation, electron motion becomes dumped. However,
dumping is counterbalanced in storage ring by quantum effects. These two radiation effects determine transverse electron beam size, energy spread and bunch length. 

This example deals with a situation where electron beam kinetics is determined by the emission of synchrotron radiation from bending magnets. 
However, because of the cylindrical symmetry, covariant and non-covariant solutions for the electron motion along an arc of a circle yield similar properties of synchrotron radiation except the following  modifications. First,  relativistic corrections are important only for bending magnet edge radiation. But the influence of this effect on the electron beam kinetics can be roughly estimated as the ratio of the radiation formation length (which is typically  a few millimeters)   to bending magnet length. This practically means that such difference is not important in the prediction of storage ring parameters. Second, the covariant approach predicts a non-zero red shift of the critical frequency, which arises when  there are perturbations of the electron motion in the vertical direction. But synchrotron radiation from bending magnets is emitted within a wide range of frequencies, and the output intensity (in contrast to the undulator case) is not sensitive on the red shift.

\section{Conclusions}

When we have discovered that during the motion along a curved trajectory, usual momentum-velocity relation does not hold, we have suddenly connected our theory to an enormous practical development. With this radically new factor in the XFEL theory, new optimum XFEL design will have to be created. 
We must, however, leave that subject to the accelerator engineers who are interested in working out the details of particular applications. Our paper only supplies the base for such design - the basic principles for the description of the radiation from a relativistic electron
in accordance with the principle of relativity.

\section{Acknowledgements}

We acknowledge many useful discussions with Gianluca Geloni and Vitaly Kocharyan. We are also idebted to Gianluca Geloni for carefully reading this manuscript, as well as for his continuous advice during its development.


\begin{thebibliography}{99}


\bibitem{M} C. Moeller, "The Theory of relativity", Clarendon, 1952

\bibitem{TKS} T. Tanaka, H. Kitamura and T. Shintake, Nucl. Instr. and Meth.  A 528, 172 (2004)

\bibitem{BH} P. Baxevanis, Z. Huang, and G. Stupakov Phys. Rev. ST AB 20, 040703 (2017)

\bibitem{ML} J. MacArthur, et al., 'Coherent undulator radiation from a kiked electron beam', in Proceedings of the 2017 FEL Conference, Santa Fe, USA, WEP048 (2018).

\bibitem{ML2} J. MacArthur, et al., 'Microbunching rotation and coherent undulator radiation from a kicked bwam' Proceedings of IPAC2018, Vancuver, Canada, A06 Free Electron Laser THPMK082


\bibitem{L} Y. Li et al., Phys. Rev. ST AB 13, 080705 (2010)

\bibitem{NUHN} H.-D. Nuhn et al., `Commissioning of the Delta polarizing undulator at LCLS', in Proceedings of the 2015 FEL Conference, Daejeon, South Korea, WED01 (2015).


\bibitem{OURS1} G. Geloni, V. Kocharyan and E. Saldin, "Misconception Regarding Conventional Coupling of Fields and Particles in XFEL Codes" DESY 16-017 (2016).

\bibitem{OURS2} G. Geloni, V. Kocharyan and E. Saldin, "A Critical experimental Test of Synchrotron Radiation Theory with 3rd Generation Light Source" DESY 16-079 (2016).


\bibitem{OURS3} G. Geloni, V. Kocharyan and E. Saldin, "Evidence of Wigner Rotation Phenomena in the Beam Splitting Experiment at the LCLS" DESY 16-128 (2016).


\bibitem{OURS4} G. Geloni, V. Kocharyan and E. Saldin, "On the Coupling of Fields and Particles in Accelerator and Plasma Physics" DESY 16-194 (2016).

\bibitem{OURS5}  G. Geloni, V. Kocharyan and E. Saldin, "Radiation by Moving Charges" DESY 17-047 (2017)

\bibitem{OURS6}  G. Geloni, V. Kocharyan and E. Saldin,"On Radiation Emission from a Microbunched Beam with Wavefront Tilt and its Experimental Observation" DESY 17-093 (2017)


\bibitem{OURS7}  G. Geloni, V. Kocharyan and E. Saldin,"Relativity and Accelerator Engineering" DESY 17-143 (2017)


\bibitem{WI} E. Wigner, Ann. Math. 40, 149, (1939)

\bibitem{WI1} E. Wigner, Z. Phys. 124, 665 (1948)

\bibitem{WI2} E. Wigner, Rev. Mod. Phys. 29, 255 (1957)



\bibitem{Rit} V. Ritus, Phys. Usp. 50, 95-101 (2007)

\bibitem{MA} G. Malykin, Phys. Usp. 49, 37 (2006)  

\bibitem{IV} T. Ivezic, Phys. Scr. 82(2010)055007

\bibitem{LL} L. Landau and E. Lifshitz, "The Classical Theory of Fields" Pergamon, Oxford, 1975



\bibitem{RM} J. Rafelski "Relativity Matter" Springer International Publishing, 2017


\bibitem{GF} E. Gourgoulhon "Special Relativity in General Frames" Springer-Verlag Berlin Heidelberg, 2013

\bibitem{CT} B. Kosyakov "Introduction to the Classical theory of Particles and Fields" Springer-Verlag Berlin, 2007

\bibitem{FA} W. Furry, American Journal of Physics, 23, 517 (1955)


\bibitem{JA} J. Awrejcewicz, "Classical mechanics" Springer, 2012

\bibitem{SCHE} F. Scheck, Classical field theory" Springer-Verlag (2012)


\bibitem{NA} A. Lutman, et al, Nature Photonics 10, 468 (2016)




\bibitem{PA} W. Pauli, "Theory of Relativity" Pergamon Press, 1958

\bibitem{FO} J. Fox, American Journal of Physics, 33,1, 1965 


\bibitem{JACK} J. Jackson, "Classical Electrodynamics", 3rd ed., Wiley, New York (1999)


\bibitem{EC} G. Ecker, Theory of Fully Ionized Plasma" Academic Press, 1972

\bibitem{GI} V. Ginzburg, "Application of Electrodynamics in Theoretical Physics and Astrophysics" Gordon and Breach Science Publisher, 1989

\bibitem{PE} A. Peratt, "Physics of the Plasma Universe", 2015

\bibitem{WW} K. Westfold, Astrophysical Journal 130, 241

\bibitem{EP} R. Epstein and P. Feldman, Astrophysical Journl 150, 109

\bibitem{OS} L. Oster, Phys. Rev. 121 p 961 (1961)






\end{thebibliography}
\end{document}